%% file: article.tex
\documentclass[final,3p,times,twocolumn]{elsarticle}
\usepackage{amssymb}
\usepackage{amsmath}
\usepackage{dblfloatfix}
\usepackage{xcolor}
\usepackage{listings}
\usepackage[section]{placeins}
\usepackage[hidelinks]{hyperref}
\usepackage{xfrac}
\usepackage{listings}
\usepackage{multirow}
\usepackage{rotating}
\usepackage{cancel}
\definecolor{commentgreen}{RGB}{2,112,10}
\newcounter{bla}

\journal{Computer Physics Communications}
\makeatletter
\def\input@path{{./images/}}
\makeatother

\graphicspath{{./images/}}

\begin{document}

\begin{frontmatter}



\title{equilibrium-c: A Lightweight Modern Equilibrium Chemistry Calculator for Hypersonic Flow Applications}


\author[a]{Nicholas N. Gibbons\corref{author}}

\cortext[author] {Corresponding author.\\\textit{E-mail address:} n.gibbons@uq.edu.au}
\address[a]{Centre for Hypersonics, School of Mechanical \& Mining Engineering, The University of Queensland}

% This needs some rework

\begin{abstract}
%A submitted program is expected to satisfy the following criteria: it must be of benefit to other physicists, or be an exemplar of good programming practice, or illustrate new or novel programming techniques which are of importance to computational physics community; it should be implemented in a language and executable on hardware that is widely available and well documented; it should meet accepted standards for scientific programming; it should be adequately documented and, where appropriate, supplied with a separate User Manual, which together with the manuscript should make clear the structure, functionality, installation, and operation of the program.
equilibrium-c (eqc) is a program for computing the composition of gas mixtures in chemical equilibrium. In typical usage, the program is given a known thermodynamic state, such as fixed temperature and pressure, as well as an initial composition of gaseous species, and computes the final composition in the limit of a large amount of time relative to the reaction speeds. eqc includes a database of thermodynamic properties taken from the literature, a set of core routines written the C programming language to solve the equilibrium problems, and a Python wrapper layer to organise the solution process and interface with user code. Dependencies are extremely minimal, and the API is designed to be easily embedded in multi-physics codes that solve problems in fluid dynamics, combustion, and chemical processing. In this paper, I first introduce the equations of chemical equilibrium, then spend some time discussing their numerical solution, and finally present a series of example problems, with an emphasis on verification and validation of the solver.
\end{abstract}


\end{frontmatter}

% \linenumbers

% All CPiP articles must contain the following
% PROGRAM SUMMARY.

\newpage

{\bf PROGRAM SUMMARY}
  %Delete as appropriate.

\begin{small}
\noindent
{\em Program Title:} equilibrium-c                                          \\
{\em CPC Library link to program files:} (to be added by Technical Editor) \\
{\em Developer's repository link:}\\ \href{https://github.com/uqngibbo/equilibrium-c}{github.com/uqngibbo/equilibrium-c}\\
{\em Code Ocean capsule:} (to be added by Technical Editor)\\
{\em Licensing provisions:} MIT \\
{\em Programming language:} C, Python                            \\
%{\em Supplementary material:}   \\
  % Fill in if necessary, otherwise leave out.
{\em Nature of problem:}\\
equilibrium-c solves the reacting thermochemistry equations to find the equilibrium composition for a given thermodynamic state. The code is open-source and has been developed with a minimal dependencies philosophy. equilibrium-c be used on its own via a python interface or embedded into a more elaborate multi-physics code that may require chemical equilibrium as a submodule.\\
{\em Solution method:}\\
The equations are solved using a modified Newton's method for multidimensional nonlinear equations, with analytic derivatives for the Jacobian. These modifications consist of an underrelaxation factor for each update, as well as careful handling of small species and a novel convergence metric.\\
{\em Additional comments:}\\
Emphasis is placed, in the paper, on reacting supersonic flow, owing to the nature of the author's primary research interest. However, the code is applicable to a diverse range of applications in research and beyond. Readers who are interested in pursuing these applications are encouraged to contact the author via the corresponding email address.
\end{small}

\newpage

\section{Introduction}
\label{sec:intro}
\begin{figure}[h]
    \scriptsize
    \centering
    \def\svgwidth{0.80\columnwidth}
    \input{equilibrium.pdf_tex}
    \caption{Chemically reacting flow passing through a shockwave.}
    \label{equilibrium}
\end{figure}
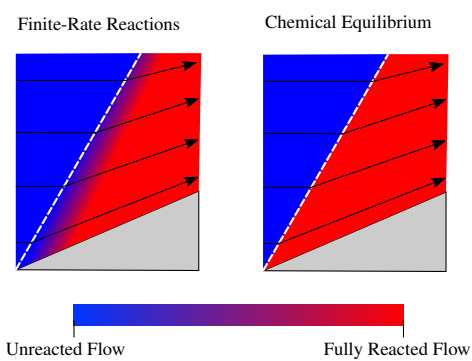

\noindent Many research problems involving rockets, high-speed aeroplanes, and spacecraft reentering the atmosphere encounter fluid flows with chemical reactions. These reactions affect the dynamics and thermochemistry of the flow, and must be considered in modelling calculations in order to compute accurate values of drag, heat transfer, or radiation that are needed for engineering design. In some situations, considerable time and complexity may be saved in these calculations by assuming the gas is in chemical equilibrium: The state that the gas would reach in the limit of a large time relative to the reaction time constants, as in figure \ref{equilibrium}.

The equilibrium composition of a gas depends only on its thermodynamic state, which means a dedicated computer program for computing it would be valuable across a wide range of applications, from hypersonic wind-tunnel design to research in combustion and surface chemistry. In addition, computer programs designed for scientific research have a range of desiderata beyond their functional capabilities: they should be open source and freely available, have good documentation that explains their structure and algorithmic design, they should be verified and validated, preferably using automated tests that can be run by a user, and they should be easy to install and have minimal dependencies.
One of the earliest efforts to provide a program for equilibrium calculations was STANJAN \cite{reynolds_stanjan86}, a Fortran-77 code that introduced the Method of Element Potentials. This method considers the equilibrium calculation as a constrained non-linear optimisation problem using Lagrange multipliers, a formulation that can handle complex mixtures and a wide range of initial conditions, and for these reasons it was adopted by most subsequent programs. Some further development of the method was introduced in Steven Pope's \texttt{ceq} \cite{pope_ceq04}, written in Fortran-90, which implemented a procedure called ``Gibbs function continuation'' that aimed to improve convergence in the case of linear constraints on the final species composition. Element potentials were also adopted in NASA's Chemical Equilibrium with Applications (CEA) \cite{nasacea_I}, a widely used Fortran-77 code and associated database developed at NASA's Lewis Research Centre, with a most recent release (CEAv2) from 2002. CEA has found widespread use in the aerospace research community, as in addition to solving chemical equilibrium problems, it also provides a range of application modes for solving problems such as rocket nozzles or shockwaves. More recent work in this space includes the equilibrium solver in the C++ combustion toolkit Cantera\cite{cantera}, and a Python toolkit called EQTK \cite{dirks2007} which was developed for dilute liquid phase calculations needed for DNA and RNA research.

This brief review demonstrates that there is a broad interest in chemical equilibrium solvers for enabling various fields of applied research, but also serves to highlight some of the problems with the existing programs. STANJAN, ceq, and CEA are essentially unmaintained, and their use of old-school FORTRAN and text-based input files make them difficult for modern programs to interface with. Cantera is modern and reasonably well maintained, though its large size and project complexity means that linking it to existing code is also difficult. EQTK is lighter in weight and has good design and documentation, though its intended field of application is dilute liquid chemistry, and thus it does not use the Element Potential method that was developed to handle the very hot gas phase flows encountered in hypersonics.

To try and address these shortcomings, this paper introduces a new program for the calculation of chemical equilibrium problems, called \texttt{equilibrium-c} or eqc for short. eqc is a re-implementation of the same equations solved by CEA, using an updated numerical method and a focus on the non-functional desiderata of a good research code. The program is open-sourced under the permissive MIT license and can be freely downloaded from the project's \href{https://github.com/uqngibbo/equilibrium-c}{github} page. It requires minimal dependencies (basic development libraries, a C compiler, and python3-numpy), works on both Linux and Windows, and can be compiled into a dynamic library for easy embedding in high performance multi-physics codes. It comes with a dozen unit tests which are continuously integrated whenever new changes are made, but can also be run by any user to verify that the implementation is functioning as advertised. Finally, it is documented in this paper and in a number of example scripts, some of which serve to verify the code against experimental data.

%The first and second sections of this paper describe the equations that are solved by eqc to compute chemical equilibrium under various thermodynamic constraints. The next part describes the numerical algorithm used to solve these equations, and the final part describes a handful of example problems, with an emphasis on verification and validation.

\section{Thermodynamics of Chemical Equilibrium}
\noindent The program described in this paper follows the element potential method by \citet{nasacea_I}, in which we seek an equilibrium state by minimising the value of a certain thermodynamic potential. For a problem at fixed temperature and pressure, the relevant potential would be the Gibbs energy $ G = U + p V - T S$, or for a fixed internal energy and volume problem, the potential would be the Helmholtz energy $ F = U - T S$. These potentials change as the composition of the gas is varied, even if the rest of the thermodynamic state is kept fixed, which allows the equilibrium state to be obtained via constrained minimisation, with the constraints provided by the initial composition and the requirement of elemental conservation.

The basic theory of this process is presented in many textbooks, which would ordinarily preclude a derivation in a publication such as this one; however, the author has found that almost always this material is presented in a way that is hand-wavy or otherwise unsatisfying. \citet{nasacea_I}, for example, offer no explanation whatsoever for their formulation. \citet{turns}, one of the best selling undergraduate textbooks on combustion theory, merely introduce the derivatives of the Gibbs energy as a mysterious ``chemical potential'', as if such a definition were an explanation for why the method works. A slightly better approach is given \citet{andersonhypersonics}, who introduces an analytic form of the second law of thermodynamics that includes a term for non-equilibrium or irreversible entropy change.

\begin{equation}
dS = \frac{p dV + dE}{T} + dS_{irrev}
\label{irrev_2nd_law}
\end{equation}

Putting aside, for the moment, any curiosity about where this equation comes from, we can proceed to rearrange it to isolate the irreversible entropy change, and then introduce differential changes in the pressure $V dp$ and $S dT$.

\begin{equation*}
T dS - p dV - dE = dS_{irrev}
\end{equation*}

\begin{equation*}
T dS + S dT - S dT - p dV + V dp - V dp - dE = dS_{irrev}
\end{equation*}

\begin{equation*}
d (TS) + S dT - d(pV) - V dp - dE = dS_{irrev}
\end{equation*}

\begin{equation*}
d (TS - pV - E) + S dT - V dp = dS_{irrev}
\end{equation*}

The differential term in brackets is minus the Gibbs energy, and assuming a constant pressure and temperature process, we seem to be entitled to assume $dp$ and $dT$ are equal to zero. These simplifications lead to the final expression as follows.

\begin{equation}
-dG = dS_{irrev}
\end{equation}
\vspace{0mm}

This derivation appears to show that the change in the Gibbs energy for a gas mixture at constant pressure and temperature is equal to the change in irreversible entropy. Anderson asserts that these irreversibilities ``are caused chemical nonequilibrium during the process'', but ``irreversible'' is just thermodynamics jargon for ``non-equilibrium'', so this explanation offers no insight into why this is the case or where equation \ref{irrev_2nd_law} actually comes from.\\

To understand why minimising the Gibbs energy results in chemical equilibrium, or indeed why chemistry has to do with thermodynamics in the first place, we must start with the principles of statistical mechanics that underpin the more familiar laws of classical thermodynamics. To begin with, we follow chapter 11 of \citet{andersonhypersonics} and consider a box of uniform gas molecules with volume $V$, without the complexity of reactions for the moment. This gas will consist of a large number of individual particles with different energies, which are constantly colliding with each other and exchanging energy through various means, though these energies can be added up to get a total energy $E$ which is conserved in the melee of collisions.

\begin{figure}[h]
    \scriptsize
    \centering
    \def\svgwidth{0.99\columnwidth}
    \input{box.pdf_tex}
    \caption{Illustration of particles and energy levels in an ideal gas.}
    \label{box}
\end{figure}
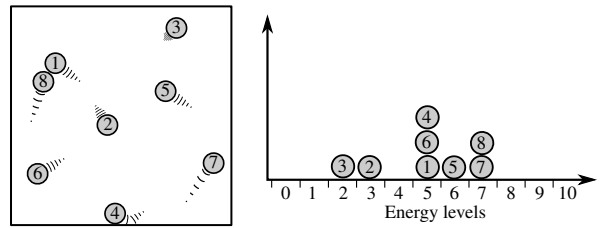

To develop a statistical model for the distribution of particle energies, we need one major starting assumption, which is that the energy states available to the particles are \emph{quantized}. This means that there are a finite (but very large) number of different energy states, and that these states are occupied by individual particles for brief moments of time as they rattle around the box. A schematic depiction of this is shown in figure \ref{box}.\\

In this figure, the energy levels are shown with a certain number of particles occupying them, for example there are three particles occupying the 5th energy level ($N_5=3$). The complete distribution of these \emph{occupation numbers} $N_j$ is of interest to us for many reasons, since they can be summed over to give the total number of particles, as well as the total energy.

\begin{equation*}
N = \sum N_j ~~~~~~~~~ E = \sum \varepsilon_j N_j
\end{equation*}

Where $\varepsilon_j$ is the amount of energy in the jth energy level. A given distribution of $N_j$ is called a \emph{macrostate}, for example, the macrostate shown in figure \ref{box} is $N=[0,0,1,1,0,3,1,2,0,...]$. If we were to take particle 4 and move it down one energy level we would get a different macrostate, namely $N=[0,0,1,1,1,2,1,2,0,...]$. Importantly, if we were to take particle 4 and \emph{swap} it with particle 3, we would not change the overall set of occupation numbers, meaning that the macrostate would be the same. This implies that there may be multiple ways of arranging a set of particles into a given macrostate. Indeed, a quick calculation shows that there are 8!/3!/2!=3360 ways of arranging the 8 particles into the macrostate in figure \ref{box}. These individual arrangements are called \emph{microstates}, and though they are physically identical in terms of energy, the number of available microstates affect the probability of each macrostate showing up the random dance of molecular collisions.

To further understand this point, an analogy may be helpful. Consider rolling a pair of six-sided dice and adding the results together, as is done in many board games. A pair of 1s gives a total of two, and a pair of 6s gives a total of twelve, the two rarest totals, each with only one permutation of individual rolls that can make it. A total of three however, can be achieved with a 1 and a 2, or a 2 and a 1, making it twice as likely. The number of permutations increases as the totals increase up to seven, which can be made with 1\&6, 2\&5, 3\&4, 4\&3, 5\&2, 6\&1, making it much more likely to appear than the one or the twelve, as can be seen in table \ref{dice2d6}.
\begin{table}[h]
\centering
\begin{tabular}{cc|cccccc}
        &   & \multicolumn{6}{c}{Roll 1}\\
        &   & 1 & 2 & 3 & 4 & 5 & 6 \\
\hline
\multirow{6}{*}[0.25ex]{\hspace*{0.8em}\turnbox{90}{Roll 2}}
        & 1 & \textcolor{red}{2} & \textcolor{orange}{3} & \textcolor{orange}{4} & \textcolor{green}{5}  & \textcolor{green}{6}   & \textcolor{blue}{7}   \\
        & 2 & \textcolor{orange}{3} & \textcolor{orange}{4} & \textcolor{green}{5} & \textcolor{green}{6}  & \textcolor{blue}{7}   & \textcolor{green}{8}   \\
        & 3 & \textcolor{orange}{4} & \textcolor{green}{5} & \textcolor{green}{6} & \textcolor{blue}{7}  & \textcolor{green}{8}   & \textcolor{green}{9}   \\
        & 4 & \textcolor{green}{5} & \textcolor{green}{6} & \textcolor{blue}{7} & \textcolor{green}{8}  & \textcolor{green}{9}   & \textcolor{orange}{10}   \\
        & 5 & \textcolor{green}{6} & \textcolor{blue}{7} & \textcolor{green}{8} & \textcolor{green}{9}  & \textcolor{orange}{10}  & \textcolor{orange}{11}   \\
        & 6 & \textcolor{blue}{7} & \textcolor{green}{8} & \textcolor{green}{9} & \textcolor{orange}{10} & \textcolor{orange}{11}  & \textcolor{red}{12}   \\
\end{tabular}
\label{dice2d6}
\caption{Totals (macrostates) resulting from a pair of six-sided dice rolls.}
\end{table}

In this analogy, the totals are the macrostates, and the individual dice roll combinations are the microstates. Since the seven has the most microstates it is the most likely total to appear in a random pair of rolls, and we might say it is the most probable macrostate.

This analogy extends to the particles of gas in the box. Some sets of occupation numbers will have more microstates associated with them than others, and the most likely macrostate is of interest for predicting the behaviour of the collective. \citet{andersonhypersonics} provides an expression for the number of microstates $W$ involved in any given macrostate or set of occupation numbers $N_j$.

\begin{equation}
W = \prod_j \frac{(N_j + g_j - 1)!}{(g_j-1)!N_j!}
\label{W1}
\end{equation}

This equation introduces the degeneracy, or statistical weight of an energy level $g_j$. Degeneracies arise due to the fact that multiple distinct quantum states may have the same energy, such as the angular momentum of a particle in different directions. To extend our analogy from before, imagine instead of rolling a pair of six-sided dice with exactly one of each number on the faces, we were instead to roll a pair of twenty sided dice with some duplicates of the numerals 1-6 on them. When counting the microstates then, we would have to account for the number of duplicates to determine the most likely total.

Equation \ref{W1} can be used to determine the most likely macrostate via mathematical optimisation. First, we choose to work with the logarithm of $W$ rather than the actual value, as its maximum will occur at the same place, and the corresponding sum of terms is easier to deal with that a product.

\begin{equation*}
ln(W) = \sum_j ln(N_j + g_j - 1)! - ln(g_j-1)! - ln(N_j)!
\end{equation*}

Next we assume that the energy levels can be lumped into a succession of quasi-levels with similar energy levels. This leads to each state having a large number of occupying particles and a large number of degeneracies, $N_j >> 1$ and $g_j >> 1$. 

\begin{equation*}
ln(W) = \sum_j ln(N_j + g_j)! - ln(g_j)! - ln(N_j)!
\end{equation*}

We can then employ Stirling's approximation to remove the factorials: $ln(a!) = a ln(a) - a$.

\begin{equation*}
ln(W) = \sum_j \left[N_j ln\left(1+\frac{g_j}{N_j}\right) + g_j ln\left(\frac{N_j}{g_j} + 1 \right) \right]
\end{equation*}

This equation can be maximised to determine the most probable macrostate, the set of $N_j$'s with the most number of microstates. To do this we again follow \citet{andersonhypersonics}, who employs the method of Lagrange multipliers to maximise subject to a constrained number of total particles $N$ and total energy $E$.

\begin{equation*}
N = \sum_j N_j
\end{equation*}

\begin{equation*}
E = \sum_j \varepsilon_j N_j
\end{equation*}

Skipping the intermediate steps, the result of the minimisation is as follows, introducing Lagrange multipliers $\alpha$ and $\beta$.

\begin{equation*}
N^*_j = \frac{g_j}{e^\alpha e^{\beta \varepsilon_j}}
\end{equation*}

$\alpha$ can be removed from this expression by introducing the constraint of total $N$, and it can be shown that $\beta$ is the inverse of the quantity normally referred to as the temperature (or in physical units, the Boltzmann constant $k$ times the temperature). This leads to an expression for the most probable macrostate $N^*_j$:

\begin{equation}
N^*_j = N \frac{g_j e^{-\varepsilon_j/k T}}{\sum_i g_i e^{-\varepsilon_i/k T}}
\label{mostprobable_N}
\end{equation}

This is the famous Boltzmann distribution. The Boltzmann distribution is interesting because although it underpins much of classical thermodynamics, there is no fundamental law of physics that compels a system to exist or remain in it. As we have seen it is simply one configuration out of countless others, though it is the one with the most microstates that can manifest it.

In order to understand why this distribution is so important, consider again the rolling of a pair of six-sided dice, but this time we should think of the average of the two rolls, which is merely the total divided by two. The most likely average of a given pair is three-and-a-half, corresponding to the sum of seven discussed earlier, though a circumspect gambler would not be terribly surprised to find either a one (from a pair of 1's) or a six (from a pair of 6's) should arise from time to time.

\begin{figure}[h]
    \scriptsize
    \centering
    \def\svgwidth{0.99\columnwidth}
    \input{dice_pdfs.pdf_tex}
    \caption{Discrete PDFs of multiple dice rolls, normalised across domain and range.}
    \label{dice_pdfs}
\end{figure}
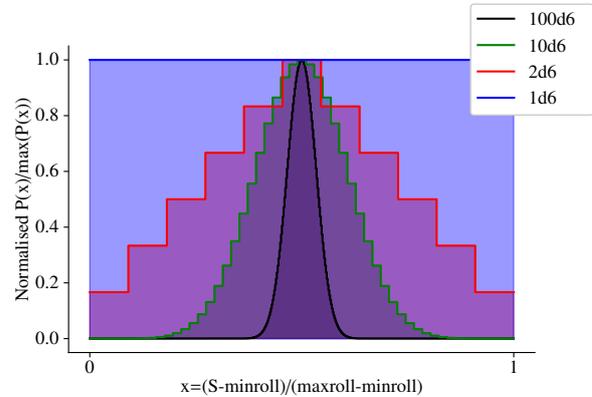

Consider however, what happens to the average as the number of dice rolls increases. With three dice there are now $6^3=216$ total permutations, but still only one way of rolling an average of one, i.e. three 1's, so extreme values have become less likely, and for the same reason, the intermediate values have become more likely. The general formula for the number of permutations associated with a given total when rolling $n$ six sided dice is given in \ref{dice_formula}, which has been used to generate figure \ref{dice_pdfs}, in which the discrete probability functions of different $n$ values are plotted with both axes normalised.

This figure demonstrates an important property of additive probabilistic systems that is related to the Law of Large Numbers. This property is that, as the number of degrees of freedom (dice, in this example) increases, the PDF contracts about the mean value, making extreme values less and less likely and the near-mean values more and more likely. With one thousand dice, the chances of getting the most probable macrostate (a total of exactly 3500 for a mean of 3.5) is 0.738 \%, but the chances of getting \emph{approximately} the macrostate, say 3.5 $\pm$ 0.1 is $\approx$ 93.6 \%. With a million dice rolls, the chances of being within half a percent of the mean are 99.99997 \%. 

The number of particles making up a hot gas is, of course, much larger even than this. And so when we do thermodynamics, we can bet that the gas will be in the mostly likely macrostate, or any of the huge number of nearly identical nearby ones. Thanks to the astronomical number of particles in the system, and Law of Large numbers, this bet will be right every time.\\

Having understood why the most probable macrostate is so important, we now want to connect the statistical mechanics expressions considered so far into the more familiar ones of classical thermodynamics. To compute the properties of a system at the most probable macrostate (i.e. at thermodynamic equilibrium), we must evaluate $ln(W)$ using the expression for $N^{*}_j$. (Recall that $N^{*}_j$ is the set of occupation numbers that define the most probable macrostate). This results in $ln(W^{max})$, the log of the number of microstates associated with the most probable macrostate. $ln(W^{max})$ is a number which has a very special place in thermodynamics, we call it the entropy.

\begin{equation}
S = k_B ~ ln(W^{max})
\end{equation}

Combining equations \ref{mostprobable_N} and \ref{W1} gives the following expression for $ln(W^{max})$.

\begin{equation*}
ln(W^{max}) = \sum_j N_j^{*} \left[ln \frac{g_j}{N_j} + 1 \right]
\end{equation*}

Substituting equation \ref{mostprobable_N} results in a simple expression with the summation over states removed. 

\begin{equation}
ln(W^{max}) = N \left[ln \frac{Q}{N} + 1 \right] + \frac{E}{kT}
\label{Wmax2}
\end{equation}

This introduces the partition function $Q$, which is defined as $Q = \sum_j g_j e^{- \varepsilon_j/k T}$. The specifics of the partition function are important for actually evaluating a statistical mechanics problem, but for the purposes of this derivation we can just note that $Q$ is a function of both the volume and the temperature $Q(V,T)$. Simplifying equation \ref{Wmax2} further, we can swap $W^{max}$ for $S$ and get an expression that is very close to the standard thermodynamics-style definition of entropy.

\begin{equation}
S = k N \left[ln \frac{Q(V,T)}{N} + 1 \right] + \frac{E(V,T)}{T}
\label{entropy_Q}
\end{equation}

This expression is a function of two independent variables, though exactly which two can be chosen for convenience. In the following section, we will be interested in the derivative of this expression, which will allow us to track the entropy as the properties of the system change. Considering $T$ and $V$ as independent variables, the infinitesimal difference in entropy is:

\begin{equation*}
d S = \frac{\partial S}{\partial T}\Bigr|_{\substack{V}} d T
              + \frac{\partial S}{\partial V}\Bigr|_{\substack{T}} d V
\end{equation*}

Considering first term, it can be shown that (see \ref{S_wrt_T}).

\begin{equation}
\frac{\partial S}{\partial T}\Bigr|_{\substack{V}} = \frac{\partial E}{\partial T}\Bigr|_{\substack{V}} \frac{1}{T}
\end{equation}

The $V$ derivative of the entropy is slightly more involved. It involves the pressure $p$, which in statistical mechanics is defined as minus the derivative of the energy with respect to volume, at constant entropy.

\begin{equation}
p = - \frac{\partial E}{\partial V}\Bigr|_{\substack{S}}
\end{equation}

This, rather abstract, definition is a kind of generalised pressure that reduces to the familiar quantity in the typical case. It can be used to show (see \ref{S_wrt_V}) that:
\begin{equation}
\frac{\partial S}{\partial V}\Bigr|_{\substack{T}} = \frac{p}{T} + \frac{\partial E}{\partial V}\Bigr|_{\substack{T}} \frac{1}{T}
\end{equation}

Putting these equations together we end up with a familiar expression for the second law of thermodynamics.
\begin{equation*}
d S = \frac{1}{T} \left[p dV + \frac{\partial E}{\partial V}\Bigr|_{\substack{T}} dV + \frac{\partial E}{\partial T}\Bigr|_{\substack{V}} dT \right]
\end{equation*}

\begin{equation}
d S = \frac{p dV + dE}{T}
\label{dS1}
\end{equation}

We are now in a position to understand exactly what this expression means. Since we have used $N_j^{*}$ to derive it, $d S$ apparently refers to the change in equilibrium entropy, the extra microstates that get added to the most probable macrostate due to slow changes in the system  volume and temperature. If these changes are done rapidly, the system may wobble away from this equilibrium state temporarily, but in time it will eventually arrive there.\\

So far we have been considering a single species gas, to try and get a clear picture of what thermodynamic equilibrium is and how it is calculated. In order to derive the chemical equilibrium equations we will need to consider a mixture of species, but thankfully the equations are just a linear combination of those already established. Again following \citet{andersonhypersonics}, we can derive the following expression for the number of microstates associated with a specific macrostate, where the $s$ index runs over the different chemical species in the mixture.

\begin{equation*}
ln(W) = \sum_s \sum_j N^s_j \left[ln \frac{g^s_j}{N^s_j} + 1 \right]
\end{equation*}

The most probable macrostate is then.

\begin{equation}
N^{s^*}_j = \frac{N^s}{Q^s} g^s_j e^{-\varepsilon^s_j/k T}
\label{mostprobable_N_state}
\end{equation}

And $ln(W^{max})$ is therefore.

\begin{equation}
ln(W^{max}) = \sum_s N^s \left[ln \frac{Q^s}{N^s} + 1 \right] + \frac{E^s}{kT}
\label{multispecies_w}
\end{equation}

It is not immediately obvious from this expression, but at this point we have run into a problem which was not present for the single species gas. Previously, we considered the total number of particles $N$ to be a known constant, introduced into the minimisation process as a constraint via a Lagrange multiplier. But for the multispecies case, what is known is not the individual species totals $N^s$, but the total number of atoms of each element in the initial mixture. For example, in a simple binary dissociation reaction $A + B \rightleftharpoons AB$, the constraint would be formulated as.

\begin{equation*}
\begin{split}
N^A + N^{AB} = \mathcal{N}_A\\
N^B + N^{AB} = \mathcal{N}_B
\end{split}
\end{equation*}

It is as if we have maximised the individual species macrostates, but not the macrostate considering the entire mixture collectively. Nonetheless, the same logic from before applies: The state of thermodynamic equilibrium occurs at the macrostate with the most microstates, and that means we need to perform another maximisation, this time using equation \ref{multispecies_w}, to find the set of $N_s$ that maximise the total $W$. The equilibrium condition therefore occurs at the stationary point of $W$, computed using $N_s$ as independent variables and with the thermodynamic state fixed.

\begin{equation}
0 = \sum_s \frac{\partial ln(W^{max})}{\partial N_s}\Bigr|_{\substack{E,V}} d N_s
\label{dw1}
\end{equation}

At this point we could go ahead and solve equation \ref{dw1} for an equilibrium state at fixed volume and total energy. However, this kind of thermodynamic state is unusual, and not terribly helpful for gas dynamics calculations, which usually use intensive variables with more physical meaning such as the temperature and pressure. We will therefore introduce a helpful trick that makes the problem more tractable. Consider the following equation for the differential change in $ln W$, which considers changes to both the thermodynamic state and composition.

\begin{equation*}
\begin{split}
d ln W^{max} =
          \frac{\partial ln W^{max}}{\partial V}\Bigr|_{\substack{E,N_s}} dV
 +        \frac{\partial ln W^{max}}{\partial E}\Bigr|_{\substack{V,N_s}} dE \\
 + \sum_s \frac{\partial ln W^{max}}{\partial N_s}\Bigr|_{\substack{E,V}} dN_s
\end{split}
\end{equation*}

The first two terms on the right hand side are the change in disorder affected by the changing thermodynamic state. The final term is the change in disorder associated with chemical change, and looking at equation \ref{dw1}, we can see that the condition for equilibrium is that it is zero. Rearranging to isolate this term we get.

\begin{equation}
d ln W^{max} 
 -        \frac{\partial ln W^{max}}{\partial V} dV
 -        \frac{\partial ln W^{max}}{\partial E} dE
 = \sum_s \frac{\partial ln W^{max}}{\partial N_s} dN_s
\label{dw2}
\end{equation}

Note carefully the conceptual interpretation of this equation. The right hand side is the disorder associated with chemical nonequilibrium, which is the part we are interested in solving. But the left hand side is the difference between the total disorder change and the disorder change attributed solely to the thermodynamic state change. By finding a composition where this difference is zero, we can determine the equilibrium state without ever having to actually differentiate $ln(W^{max})$ with respect to the composition, which is how a chemical equilibrium solver can do its work using only standard thermodynamic quantities. More formally, this amounts to solving for the following expression:

\begin{equation}
d ln W^{max} 
 -        \frac{\partial ln W^{max}}{\partial V} dV
 -        \frac{\partial ln W^{max}}{\partial E} dE
 = 0
\label{dw3}
\end{equation}

We can assume that the individual species are in their respective Boltzmann distributions, and so $ln W^{max}$ is really $S/k$, the total entropy divided by the Boltzmann constant. 

\begin{equation*}
d S 
 -        \frac{\partial S}{\partial V} dV
 -        \frac{\partial S}{\partial E} dE
 = 0
\end{equation*}

The rest of the derivation is the same as for the single species gas, and we end up with a slightly rearranged copy of equation \ref{dS1}.

\begin{equation}
d S - \frac{p dV + dE}{T} = 0
\label{dS2}
\end{equation}

The derivation given by given by \citet{andersonhypersonics} begins at this point, starting with the second law expression expression for entropy change and asserting that the right hand is zero in a reversible or equilibrium process. We are now in a position to understand what this means: Equation \ref{dS2} is also the stationary point of the mixture entropy with respect to composition. And that is the condition of chemical equilibrium.\\

To complete our derivation, we must show that minimising the Gibbs energy reduces to equation \ref{dS2}. The Gibbs energy is a thermodynamic potential that measures the maximum amount of work that can be extracted from a closed system at constant temperature and pressure, defined as follows. 

\begin{equation}
G = E + p V - T S
\end{equation}

$G$ can be expressed as sum of individual species, using the number of particles (in moles) of each species $N_s$ and the species energy and entropy per mole $E_s$ and $S_s$.

\begin{equation*}
G = \sum_s E_s N_s + p V - \sum_s T S_s N_s
\end{equation*}

Using the ideal gas law $p V = N R_u T$, and noting that $N=\sum_s N_s$, the $p V$ term can also be removed.

\begin{equation*}
G = \sum_s E_s N_s + R_u T \sum_s N_s  - \sum_s T S_s N_s
\end{equation*}

\begin{equation}
G = \sum_s \left[ E_s + R_u T - T S_s \right] N_s
\label{sum_species_G}
\end{equation}

The bracketed term in equation \ref{sum_species_G} is the Gibbs energy per mole of species $s$, and can be defined as:

\begin{equation}
G_s \equiv  E_s + R_u T - T S_s
\label{species_G}
\end{equation}

Leading to a tidy formula:

\begin{equation}
G = \sum_s G_s N_s
\end{equation}

We are interested in a stationary point of $G$, the point at which $dG$ is equal to zero in every direction in species space.

\begin{equation}
d G = \sum_s \frac{\partial G}{\partial N_s} d N_s = 0
\label{differential_G}
\end{equation}

Evaluating the partial derivatives of G with respect to $N_s$ is actually a little bit involved, because of the entropy $S_s$ depending on the complete set of of $N_s$'s. The full derivation of this expression is given in \ref{dGdNs_appendix}, but the end result is a surprisingly tidy expression where all of the extra terms from the product rule cancel to get:

\begin{equation*}
\frac{\partial G}{\partial N_s} = G_s
\end{equation*}

Continuing with equation \ref{differential_G}, we expand using the definition of $G_s$, equation \ref{species_G} to get:

\begin{equation*}
 dG = \sum_s \left[E_s + R_u T - T S_s\right] d N_s = 0
\end{equation*}

Note that, since $E=N_s E_s$, $dE = E_s dN_s$, and likewise $dS = S_s dN_s$. Slightly more complicated is the following expression for $d V$ coming again from the ideal gas law:

\begin{equation*}
\begin{split}
V = \frac{R_u T}{p} \sum_s N_s\\
\therefore ~ dV = \frac{R_u T}{p} \sum_s d N_s
\end{split}
\end{equation*}

These expressions allow us to remove the sums over $dN_s$ as follows.

\begin{equation*}
dG = \sum_s E_s dNs + R_u T \sum_s d N_s - T \sum_s S_s d N_s = 0
\end{equation*}

\begin{equation*}
0 = dE + p dV - T dS
\end{equation*}

\begin{equation*}
0 = \frac{dE + p dV}{T} - dS
\end{equation*}

This is almost, but not quite the same as equation \ref{dS2}. The final step is to flip the signs by multiplying the expression by negative one, to get:

\begin{equation}
0 = dS - \frac{dE + p dV}{T}
\end{equation}

Incidentally, this sign flip is why minimising $G$ corresponds to maximising disorder, and the equivalence of the two expressions is why minimising $G$ is the same as finding the state of equilibrium.

\section{Governing Equations}
\label{governing_equations}

\noindent The methodology used by eqc to solve for chemical equilibrium is based off of \citet{nasacea_I}. Here we present the a derivation of the equations for solving a constrained pressure and temperature problem, with other problem types given in the Appendices. We will be interested in minimising the Gibbs Energy $G$, but we will begin by dividing it by the total mixture mass $\rho V$, to get the Gibbs energy per unit mass $g$. This quantity is more convenient to work with, and will have its stationary point at the same place because the total mass $\rho V$ is constant.

\begin{equation*}
g = \frac{G}{\rho V} = \frac{N_s}{\rho V} G_s = n_s G_s
\end{equation*}

This expression introduces the unusual intensive quantity $n_s$, the number of moles of species $s$ present \emph{per unit mixture mass}. As far as the author is aware this quantity does not have a standardised name, however the term \emph{specific molarity} seems appropriate. Its relationship to the more familiar measures of composition such as molar concentration $c_s$, mass fraction $Y_s$, and mole fraction $X_s$ are as follows.

\begin{equation*}
n_s = \frac{N_s}{\rho V} ~~~ : ~~~ c_s = \rho n_s ~~~ : ~~~ Y_s = M_s n_s ~~~ : ~~~ X_s = \frac{n_s}{n}
\end{equation*}

This last equation uses the sum of the specific molarities $n=\sum_s n_s$, which is also notable as being the inverse of of the mixture molar mass $n = 1/M$.

We will need the Gibbs energy in its explicit form, with all of the terms expanded out. Equation \ref{species_G} is the definition of $G_s$, which gives the following.

\begin{equation*}
g = \sum_s n_s G_s =  \sum_s n_s \left[ E_s + R_u T - T S_s \right]
\end{equation*}

Tabulated values of the internal energy per mole $E_s$ are available from the NASA Lewis thermodynamic database, which have been computed at a standard pressure of $100,000$ Pa or 1 BAR. This standard state is sometimes written as $E^\circ_s$, however, we will exclusively by considering thermally perfect gases where $E_s$ has no dependency on pressure, so $E_s=E^\circ_s$ in general. The same is unfortunately not true of the entropy $S_s$, which does vary with pressure. The molar entropy of species $s$ at an arbitrary pressure can be written as follows, using the proof in \ref{appendix_entropy_derivation}.

\begin{equation}
S_s = S^\circ_s - R_u ln \left(\frac{n_s}{n}\right) - R_u ln \left(\frac{p}{p^\circ}\right)
\label{species_entropy}
\end{equation}

$S^\circ_s$ can be computed using the Lewis tables, which assume $p^\circ=100,000$ Pa. The fully expanded expression for minimisation is then as follows. 

\begin{equation*}
g = \sum_s n_s \left[ E_s + R_u T - T S^\circ_s + R_u T ln \left(\frac{n_s}{n}\right) + R_u T ln \left(\frac{p}{p^\circ}\right) \right]
\end{equation*}

Note that the first three terms in the bracket of this equation are constant throughout the minimisation process, and are, on a closer inspection, just the Gibbs energy at the standard conditions. To save some ink and space then, our final step will be to define $G_s^\circ \equiv E_s + R_u T - T S^\circ_s$, and then we can write the objective function to be minimised as follows.

\begin{equation}
g = \sum_s n_s \left[ G_s^\circ + R_u T ln \left(\frac{n_s}{n}\right) + R_u T ln \left(\frac{p}{p^\circ}\right) \right]
\label{pt_objective_function}
\end{equation}

Minimising this expression is a non-linear constrained optimisation problem in $n_s$. The constraints are the available elements in some initial composition $n_s^0$, which can be expressed using a matrix of elemental constituents $a_{is}$, the number of atoms of element $j$ in species $s$. As an example, consider the dissociation of $CO_2$.

\begin{equation*}
CO_2 \rightleftharpoons CO + \frac{1}{2} O_2
\end{equation*}

This problem contains only two atomic elements, $C$, and $O$, and the constraint matrix would be as follows.\\

\begin{figure}[h]
    \centering
    \def\svgwidth{0.65\columnwidth}
    \input{nuclear_matrix2.pdf_tex}
    \label{nuclear_matrix}
\end{figure}

Given an initial composition $n_s^0$ then, the constraint equation can be written.

\begin{equation}
a_{js} n_s - a_{js} n^0_s = 0
\end{equation}

Following \citet{nasacea_I}, the minimisation process proceeds using the method of Lagrange multipliers, which introduces Lagrange multipliers $\lambda_j$, one for each atomic element in the mixture, and for each row in the matrix $a_{js}$. Defining the Lagrangian $\mathcal{L}$ as:

\begin{equation}
\begin{split}
\mathcal{L} = &\sum_s n_s \left[ G_s^\circ + R_u T ln \left(\frac{n_s}{n}\right) + R_u T ln \left(\frac{p}{p^\circ}\right) \right]\\
             + &\sum_j \lambda_j \left( \sum_s a_{js} n_s - a_{js} n^0_s \right)
\end{split}
\label{lagrangian}
\end{equation}
\vspace{2mm}

Solving the optimisation problem then becomes a matter of seeking the set of $n_s$ and $\lambda_j$ that satisfy $\partial \mathcal{L}/ \partial n_s=0$ and  $\partial \mathcal{L}/ \partial \lambda_j=0$ simultaneously. This set of equations can be derived by differentiating equation \ref{lagrangian} with respect to each unknown. This derivation is given in \ref{pt_lagrangian_appendix}, with the final results shown in equations \ref{Gequation_s} - \ref{nequation} below.

\begin{equation}
\begin{split}
\frac{G_s^\circ}{R_u T} + ln \left(\frac{n_s}{n}\right) + ln \left(\frac{p}{p^\circ}\right) + \sum_j \lambda_j \frac{a_{js}}{R_u T} = 0\\
~~ (s=0,...,n_s)
\end{split}
\label{Gequation_s}
\end{equation}

\begin{equation}
\sum_s a_{js} n_s - a_{js} n^0_s = 0 ~~~ ~~~ (j=0,...,n_e)
\label{aequation}
\end{equation}

We also introduce the apparently redundant linear equation for $n$, the sum of the unknown molarities. This is because we will want to treat $n$ as a separate unknown to be solved along with the individual species ones, which simplifies the resulting expressions and adds some robustness to the numerical method for solving them. 
\begin{equation}
\sum_s n_s - n = 0 ~~~ ~~~ (j=0,...,n_e)
\label{nequation}
\end{equation}

%These equations are solved using a multidimensional Newton's method with a number of special modifications needed for handling the numerical issues involved with solving chemical equilibrium. The core of the method is a gradient descent algorithm that uses a matrix of first derivatives that are computed analytically.

\section{Numerical Method}

\noindent The basic algorithm used to solve equations \ref{Gequation_s}-\ref{aequation} is a generalisation of the familiar Newton's method. For a single nonlinear equation $f(x)=0$, this method begins by taking an initial guess of the unknown variable $x_0$, then computes the gradient of the objective function $f$ at that point $d f/ d x(x_0)$, and finally descends to the x-axis in a straight line to find a new guess $x_1$. This process is repeated until an error criterion drops below a specified tolerance, at which point the value $x_n$ should be extremely close to the actual root.\\

When solving a coupled system of nonlinear equations, a multidimensional generalisation of Newton's method exists that can be employed in an analogous way. Figure \ref{newton_2d} visualises one step of this process, using a pair of concave example functions $f_1(x,y)$ and $f_2(x,y)$ that are plotted in blue and red respectively.

\begin{figure}[h]
    \scriptsize
    \centering
    \def\svgwidth{0.99\columnwidth}
    \input{newton_2d.pdf_tex}
    \label{newton_2d}
\end{figure}

A given guess for the independent variables $x_0$ and $y_0$ produces a pair of errors $f_1(x_0, y_0)$ and $f_2(x_0, y_0)$ which are represented by the blue and red circular markers. The root of the two equations is point where both functions are equal to zero and equal to each other, which is shown in the figure as the green star-shaped marker.

To update an initial guess in 2D we first differentiate both functions at the point $x_0$, $y_0$, which produces a tangent plane for each function that is centred on each of the coloured markers. Each of these planes cuts through the z=0 plane in a line rather than a point, and any position along each line would be a better guess for the zero of that function, though not necessarily a better guess for the entire coupled system of equations. What the multidimensional Newton's method actually does is choose the point where the two lines intersect, marked with a green circle in figure \ref{newton_2d}, as the next iteration $x_1$,$y_1$, which will tend to be closer to the actual root for sufficiently well behaved functions. The method can be formalised in any number of dimensions by considering the first order Taylor series expansion of a multivariate function $f$, one of a set of functions we would like to simultaneously solve.

\begin{equation*}
df = \frac{\partial f}{\partial x_0} dx_0 + \frac{\partial f}{\partial x_1} dx_1 + ... \frac{\partial f}{\partial x_n} dx_n
\end{equation*}

Geometrically, this expression defines a tangent plane to the function at a given point $\mathbf{x}$, by proscribing a linear variation $df$ for an arbitrary movement in the dependant variables $\mathbf{dx}$. The dotted lines in figure \ref{newton_2d} are obtained by descending all the way to the x-y axis, i.e. $df = -f(x_0, x_1, x_2, ...)$.

\begin{equation*}
-f(x_0, x_1, x_2, ...) = \frac{\partial f}{\partial x_0} dx_0 + \frac{\partial f}{\partial x_1} dx_1 + ... \frac{\partial f}{\partial x_n} dx_n
\end{equation*}

For a collection of m functions, we are interested in the point where the dotted lines meet, which defines a linear system.

\begin{equation*}
\begin{split}
-f_0(x_0, x_1, x_2, ...) = \frac{\partial f_0}{\partial x_0} dx_0 + \frac{\partial f_0}{\partial x_1} dx_1 + ... \frac{\partial f_0}{\partial x_n} dx_n\\
-f_1(x_0, x_1, x_2, ...) = \frac{\partial f_1}{\partial x_0} dx_0 + \frac{\partial f_1}{\partial x_1} dx_1 + ... \frac{\partial f_1}{\partial x_n} dx_n\\
\vdots \\
-f_m(x_0, x_1, x_2, ...) = \frac{\partial f_m}{\partial x_0} dx_0 + \frac{\partial f_m}{\partial x_1} dx_1 + ... \frac{\partial f_m}{\partial x_n} dx_n\\
\end{split}
\end{equation*}

Abstracting this system of equations into matrix form produces the familiar expression for the multidimensional Newton's method that can be found in many textbooks, where the matrix of partial derivations is often given the label $J$ for Jacobian.

\begin{equation}
-\mathbf{F}(\mathbf{x}_i) = J(\mathbf{x}) \Delta \mathbf{x}_i
\label{multinewton}
\end{equation}

In this case the vector $F$ is the combination of equations \ref{Gequation_s} and \ref{aequation} stacked on top of each other, and $\mathbf{x}_i$  is the vector of unknowns we are trying to solve for. Clearly a random guess at this vector will not be a solution of equations \ref{Gequation_s} and \ref{aequation}, which is to say that it will give a non-zero remainder on the right-hand side. However, by starting with such a guess and iterating over $i$ using equation \ref{multinewton}, these remainders can be reduced with each iteration until values of $x_i$ are correct to within a specified precision. A slight complicating factor is that the magnitudes of the molarities can vary dramatically between species, so we actually want to solve for the natural logarithms of the composition $ln(n_s)$ and $ln(n)$. This affects the Jacobian matrix $J$, which is made of the following first derivatives.

\begin{equation*}
\frac{\partial F_s}{\partial ln(n_s)} = 1 ~~~ \frac{\partial F_s}{\partial ln(n)} = -1 ~~~ \frac{\partial F_s}{\partial \lambda_j} = \frac{a_{js}}{R_u T}
\end{equation*}

\begin{equation*}
\frac{\partial F_j}{\partial ln(n_s)} = a_{js} n_s  ~~~ \frac{\partial F_j}{\partial ln(n)} = 0 ~~~ \frac{\partial F_j}{\partial \lambda_k} = 0
\end{equation*}

\begin{equation*}
\frac{\partial F_n}{\partial ln(n_s)} = n_s  ~~~ \frac{\partial F_n}{\partial ln(n)} = -n ~~~ \frac{\partial F_n}{\partial \lambda_k} = 0
\end{equation*}
\vspace{3mm}

In these equations, the expression $F_s$ is used to represent one instance of equation \label{Gequation_s}, the expression $F_j$ used to represent an instance of equation \ref{aequation}, and the expression $F_n$ means equation \ref{nequation}. These are the set of coupled nonlinear equations we wish to solve. \citet{nasacea_I} then actually perform the matrix multiplication by $J$ analytically to get what they call the Gibbs iteration equations, beginning with $F_s$.

\begin{equation*}
\Delta ln(n_s) - \Delta ln(n) + \sum_j a_{sj} \frac{\Delta \lambda_j}{R_u T} = -\frac{G_s}{R_u T} - \sum_j \frac{\lambda_j a_{js}}{R_u T}
\end{equation*}
\vspace{3mm}

Before we get to the rest of the Gibbs iteration equations, this one deserves some attention. The first thing to note about it is that the Lagrange multipliers appear linearly in the update equation, as well as linearly in the original Lagrangian. Since a Newton's method will always solve a linear set of equations in one step, we can exempt the Lagrange multipliers from the iteration process by guessing that they are zero at the beginning of each step, and recognising that $\Delta \lambda_j$ is the correct and final value of the Lagrange multiplier needed to implement the constraints. A second point to note is that Lagrange multipliers are arbitrary non-physical quantities, that can always be redefined for our convenience.

Following \citet{nasacea_I} then, we set $\lambda_j=0$ and redefine $-\Delta \lambda_j/R_u T \equiv \pi_j$. This gives \citet{nasacea_I}'s equation 2.18, and the other Gibbs equations are straightforwardly derived in the same manner.\footnote{\citet{nasacea_I} contains another equation that is not considered here, namely 2.19 which sums over the non-gaseous species and is required for dealing with multi-phase flow. Since these are rarely of interest in normal hypersonics research, eqc does not consider condensed species.}

\begin{equation}
\Delta ln(n_s) - \Delta ln(n) - \sum_j a_{sj} \pi_j = -\frac{G_s}{R_u T} ~~~ ~~~ (s=0,...,n_s)
\label{gibbs_Fs}
\end{equation}

\begin{equation}
\sum_s a_{sj} n_s \Delta ln(n_s) = \sum_s a_{js} n^0_s - a_{js} n_s  ~~~ ~~~ (j=0,...,n_e)
\label{gibbs_Fj}
\end{equation}

\begin{equation}
\sum_s n_s \Delta ln(n_s) - n \Delta ln(n) = n - \sum_s n_s
\label{gibbs_Fn}
\end{equation}
\vspace{3mm}

Equation \ref{gibbs_Fs} provides one expression for each species involved with the calculation, equation \ref{gibbs_Fj} one for each chemical element, and equation \ref{gibbs_Fn} one more to implement the total $n$ constraint. These equations would be sufficient to solve for the equilibrium composition at fixed temperature and pressure, however, a keen observer may have noticed that equation \ref{gibbs_Fs} contains only a single unknown $\Delta ln(n_s)$  which could be computed directly if the $\Delta ln(n)$ and $\pi_j$ were known. \citet{nasacea_I} suggests exploiting this structure by using equation \ref{gibbs_Fs} to get an expression for each $\Delta ln(n_s)$, and substituting them into the other two expressions to get what they call the Reduced Gibbs Iteration equations.

\begin{equation}
\begin{split}
\sum_i \sum_s a_{js} a_{is} n_s \pi_i + \sum_s a_{js} n_s \Delta ln(n) =\\
\sum_s a_{js} n^0_s - a_{js} n_s + \sum_s \frac{a_{js} n_s G_s}{R_u T}\\ ~~~ ~~~ (j=0,...,n_e)
\end{split}
\label{reduced_gibbs_Fj}
\end{equation}

\begin{equation}
\sum_i \sum_s a_{is} n_s \pi_i + \left( \sum_s ns - n \right) \Delta ln(n) = n - \sum_s n_s + \sum_s \frac{n_s G_s}{R_u T}
\label{reduced_gibbs_Fn}
\end{equation}

These are the actual equations solved by eqc to get the equilibrium composition at fixed temperature and pressure. There are $n_e+1$ of them, where $n_e$ is the number of chemical elements involved in the reactions, which is a major reduction of the original system which had an additional one for each species. Having solved for $\Delta ln(n)$ and the Lagrange multipliers $\pi_j$, equation \ref{gibbs_Fs} is used to directly obtain $\Delta ln(n_s)$, which is used to update the specific molarities, and a new iteration begins. This process is continued until the system of equations convergences, or perhaps, does not.

%At each iteration a root mean squared quantity is computed, using the change in total molarity ($\Delta ln(n)$), the change in species molarities ($\Delta ln(n_s)$) and the error in the nuclear constraints specified by the $a_{js}$ matrix as follows.

%\begin{equation}
%\epsilon = \sqrt{\frac{(\Delta ln(n))^2 + \sum_{s} (\Delta ln(n_s))^2 + \sum_j (b_j - b^0_j)^2}{1 + N_s + N_e}}
%\end{equation}

%Where $N_s$ is the number of species and $N_e$ the number of chemical elements in the problem. The quantity $\epsilon$ is a scalar residual which tracks level of convergence of the system, and by default the iteration is continued until the residual converges to a value of $\epsilon < 1 \times 10^{-10}$, or perhaps, does not.

\subsection{Numerical Convergence and Robustness}
\label{convergence}
%Figure \ref{flowchart} shows the basic steps in the numerical algorithm used to solve the system of nonlinear equations required for chemical equilibrium. Step one is the contruction of a matrix form of \ref{reduced_gibbs_Fj} - \ref{reduced_gibbs_Fn}, which be written as $J \Delta x = F$, where $J$ the Jacobian matrix of derivatives, $F$ is the residual vector of remainders, and $\Delta x$ the change in unknown variables $ln(n)$ and $pi_j$. Starting with a guess for the species molarities $ln(n_s)$, the matrix problem is solved using Gauss-Jordan elimination with partial pivoting to compute $\Delta ln(n)$ and $pi_j$, and then equation \ref{gibbs_Fs} is used to directly compute $\Delta ln(n_s)$. With the changes known, an iteration can be performed to calculate new tentative values for $ln(n)$ and $ln(n_s)$, until convergence is reached.

\noindent The basic algorithm described so far is capable of solving many chemical equilibrium problems, but early testing of the solver revealed many cases where convergence failed, either due to numerical instability in the algorithm or because of oscillation in the solution that eventually hit the maximum step limit. This subsection describes a handful additions to the method that were needed for robust convergence, and taken together allow eqc to solve a comprehensive range of equilibrium problems, including combustion for propulsion applications, high temperature ionisation for re-entry strength shockwaves, and facility calculations in reflected shock tunnels and expansion tubes.\\

The first addition is similar to a fix suggested by \citet{nasacea_I}: an under-relaxation factor on the unknown variable updates. In pressure-temperature problems, this consists of limiting the updates to $n$ and $n_s$ by a factor $\Lambda$, which is defined using a fraction of the current guess of the total specific molarity $n$.

\begin{equation}
\Lambda = min \left(1.0, \frac{1}{2} \frac{|ln(n)|}{|\Delta ln(n_s)|} \right)
\end{equation}
\vspace{2mm}

The fraction $1/2$ is an adjustable factor that seems to work well for most purposes. The equations for updating the unknown molarities from iteration $i$ to $i+1$ are then:

\begin{equation}
ln(n_s)^{i+1} =  ln(n_s)^i ~ + ~ \Lambda ~ \Delta ln(n_s)
\label{lnns_update_eqn}
\end{equation}

\begin{equation}
ln(n)^{i+1} =  ln(n)^{i} ~ + ~ \Lambda ~ \Delta ln(n)
\end{equation}
\vspace{1mm}

For problems where the gas temperature is also an unknown, such as fixed density and internal energy ($\rho e$) problems (see \ref{rhoe_equations} and \ref{ps_equations}), $\Lambda$ is also applied to the temperature update, though it is computed using $|ln(T)|$ instead of $|ln(n)|$. 

\begin{equation}
T^{i+1} =  exp \left( ln(T)^{i} ~ + ~ \Lambda ~ \Delta ln(T) \right)
\end{equation}
\vspace{1mm}

Notice that in these update equations, the actual first class unknown is $ln(n_s)$. This is important enough to deserves some extended discussion.\\
\newpage

The number one cause of numerical difficulties in solving the equilibrium equations is that, very often, one more species compositions can become extremely small. This can lead to the matrix in the Newton method becoming ill-conditioned and difficult to solve. Solving for the logarithms of the compositions fixes this problem at the cost of introducing a new one, which is that at precisely $n_s=0$, the logarithm is undefined.\\\footnote{This bit of numerical trivia actually has some important physics underlying it. The logarithms in the equations come from the entropy, and recall that the entropy has to do with the number of ways that the particles in a system can be arranged for a given set of occupation numbers. If there are no particles present of one species, then how many ways can they be arranged? Clearly the answer is that the question makes no sense, and yet this creates a problem in the equations much further downstream.} 

\citet{nasacea_I} deal with this problem by automatically detecting when a species's molarity has dropped below a trace value and then removing it from the calculation, but this adds complexity and can cause convergence problems when a species's equilibrium concentration is near the trace limit, since it may oscillate between being present and absent. In this work I have developed the following algorithm for ensuring convergence.\\

\begin{enumerate}
\item At the beginning of the calculation, the elemental constraint vector $X_0$ is used to set an initial guess for the molarities, with a floor to prevent $n_s$ being zero.
\begin{lstlisting}[language=C,basicstyle=\small\ttfamily]
for (s=0; s<nsp; s++){
    ns[s] = fmax(X0[s]/M0, n*1e-4);
}
\end{lstlisting}

\vspace{5mm}
\item Once set to known nonzero values, the species molarities are used to compute the logarithmic species molarities.
\begin{lstlisting}[language=C]
for (s=0; s<nsp; s++){
    lnns[s] = log(ns[s]);
}
\end{lstlisting}
This is the only time that the log function is called. For the rest of the algorithm, \texttt{lnns} is treated as the first class unknown, from which $n_s$ may be computed, but never the other way around.

\vspace{5mm}
\item At the end of each iteration of the Newton's method, $\Delta ln(n_s)$ is computed using equation \ref{lnns_update_eqn}, and then $n_s$ is computed using the exponential function.
\begin{lstlisting}[language=C]
for (s=0; s<nsp; s++){
    ns[s] = exp(lnns[s]);
}
\end{lstlisting}
At around \texttt{lnns}$=-746$, this call to the exponential function will underflow, giving $n_s=0.0$. However, as long as the logarithmic value is still defined, the floating point math will correctly give $n_s \times ln(n_s)=0.0$ in the various places in the algorithm where it is needed. In practice, I have found that the calculations always converge before the values of \texttt{lnns} get anywhere close to underflowing. 

\vspace{5mm}
\item Continue iterating until the right hand side of the nonlinear equations is close enough to zero.
\end{enumerate}

This final step also requires some discussion. In general, a Newton's method can be terminated in one of two possible ways. Either one can monitor the independent variables being solved for, in this case $\Delta ln(n_s)$, and stop iterating when they stop changing, or one can monitor the remaining error in the nonlinear equations, and stop iterating when the remainder on the right hand side is sufficiently close to zero for all of them. Both of these methods, if implemented naively, fail frequently when solving the chemical equilibrium equations implemented in this work.\\

The problem with monitoring $\Delta ln(n_s)$ is illustrated in the following example, which comes from the cold air example script include in the code's repository. Turning on verbose debugging to get extra information, we consider the following output, from iteration 11.

{\scriptsize
%\begin{lstlisting}[language=C]
\begin{verbatim}
iter  9: [34.55] 26.50 8.05 0.00 0.00 0.00  (3.549e-11)
 sp: N2  lnns:  3.2771 dlnns: -0.000000  lambda:  1.0000
 sp: O2  lnns:  2.0857 dlnns: -0.000000  lambda:  1.0000
 sp: N   lnns: -34.958 dlnns: -146.8798  lambda:  0.0180
 sp: O   lnns: -34.958 dlnns: -57.75229  lambda:  0.0460
 sp: NO  lnns: -32.422 dlnns: -0.120385  lambda:  1.0000
\end{verbatim}
%\end{lstlisting}
}

At room temperature, the dissociated species N and O are present in negligible quantities, \texttt{lnns} $\approx -34.9$, which corresponds to a mole fraction of about 1e-17. Note however, that $\Delta ln(n_s)$ for these species is still large, so large in fact, that the relaxation factor \texttt{lambda} is attempting to limit it. If the solver is allowed to continue, it takes a total of 65 iterations for \texttt{lnns[3]} to reach its converged value of -179.18 or a mole fraction of $\approx$ 1.8e-78, which is a ridiculous level of precision for such an insignificant value. Largely because of this problem, the code does not monitor $\Delta ln(n_s)$ to check for convergence.\\

A typically better way to terminate a Newton's method is by looking at the right-hand side of the equations being solved, a vector quantity sometimes called the \emph{residual}. For a fixed pressure/temperature problem, the residuals are given by equations \ref{Gequation_s}, \ref{aequation}, and \ref{nequation}, reproduced below.

\begin{equation*}
\begin{split}
\frac{G_s^\circ}{R_u T} + ln \left(\frac{n_s}{n}\right) + ln \left(\frac{p}{p^\circ}\right) + \sum_j \lambda_j \frac{a_{js}}{R_u T} = F_s = 0\\
~~ (s=0,...,n_s)
\end{split}
\end{equation*}
\begin{equation*}
\sum_s a_{js} n_s - a_{js} n^0_s = F_j = 0 ~~~ ~~~ (j=0,...,n_e)
\end{equation*}
\begin{equation*}
\sum_s n_s - n = F_n = 0 ~~~ ~~~ (j=0,...,n_e)
\end{equation*}
\vspace{4mm}

We can monitor the convergence by looking at the L2 norm of the remainder on the right hand side, defined as follows.

\begin{equation*}
\varepsilon = \sqrt{\sum_s F_s^2 + \sum_j F_j^2 + F_n^2}
\end{equation*}
\vspace{2mm}

This almost, but not quite works. Note that the equation for $F_s$ has a logarithm in it, which causes the exact same problem as before; the residual can be large even when the molarity $n_s$ is insignificant, indeed at $n_s=0$ the equations become singular.

To track convergence, eqc actually does use this second method, where we track the L2 norm of the residual and stop iterating when it becomes small. However, to avoid the aforementioned problem, when computing the L2 norm of the residual, eqc actually uses $n_s F_s$ when considering equation \ref{Gequation_s}. This is a new function which has its zero in the same place as $F_s$, but is not singular as $n_s \rightarrow 0$ and has better behaviour when $n_s$ is small. The actual quantity that is monitored for convergence then is given by equation \ref{L2_residual_norm}, where a tolerance of $\varepsilon<1\times 10^{-11}$ is the default behaviour. 

\begin{equation}
\varepsilon = \sqrt{\sum_s (n_s F_s)^2 + \sum_j F_j^2 + F_n^2}
\label{L2_residual_norm}
\end{equation}
\begin{figure*}[b]
    \scriptsize
    \centering
    \def\svgwidth{0.99\textwidth}
    \input{co2.pdf_tex}
    \caption{Comparison of eqc numerical solutions with CO2 dissociation problem from \citet{turns}, Chapter 2.}
    \label{co2_diss}
\end{figure*}
\newpage

\section{Example Problems}
\noindent This section introduces some example problems to verify and validate the solver. First up is a single reaction problem taken from a popular combustion textbook \cite{turns}, in this case the dissociation of carbon dioxide.
\begin{equation*}
CO_2 \rightleftharpoons CO + \frac{1}{2} O_2
\end{equation*}

\subsection{Carbon Dioxide Textbook Problem}

\noindent An attraction of a studying a single reaction is that the equilibrium problem has an analytic solution that the numerical solver can be compared to. Deriving this solution begins with the equation for the Gibbs energy, written in terms of the mole densities $N_s$, the species partial pressures $p_s$, and the Gibbs energy per mole of species $s$.

\begin{equation*}
G = \sum_s N_s G_s = \sum_s N_s \left[G^{\circ}_s + R_u T ln\left(\frac{p_s}{p^{\circ}} \right) \right]
\end{equation*}
\vspace{1mm}

At a fixed temperature and pressure, the equilibrium condition is given by $dG=0$, which amounts to solving for the following equation.

\begin{equation}
0 =  \sum_s \left[G^{\circ}_s + R_u T ln\left(\frac{p_s}{p^{\circ}} \right) \right] d N_s
\label{equilibrium_condition_2}
\end{equation}
\vspace{2mm}

Rather than separate this expression into a set of nonlinear equations as before, we can take advantage of the single reaction to replace the three species composition unknowns with a single unknown quantity $\alpha$, which is a kind of reaction progress variable.

\begin{equation*}
(1-\alpha) CO_2 \rightleftharpoons \alpha CO + \frac{\alpha}{2} O_2
\end{equation*}
\vspace{2mm}

At $\alpha=0$ the mixture is entirely CO$_2$, while at $\alpha=1$ the mixture is entirely dissociated, consisting of CO and O$_2$ in a 2:1 ratio. The single reaction also allows us to package up the thermochemistry into a single value $K_p$, the famous equilibrium constant, which is in fact not a constant, but a function of temperature and pressure.

\begin{equation*}
K_p \equiv \exp\left[\frac{G^{\circ}_{CO_2} - G^{\circ}_{CO} - \frac{1}{2}G^{\circ}_{O_2}}{R_u T}\right]
\end{equation*}
\vspace{2mm}

The equilibrium condition is given by a cubic polynomial which is derived in \ref{co2_derivation}, with the final answer shown in equation \ref{alpha_cubic}.

\begin{equation}
\alpha^3 \left( 1 - \frac{p}{p^{\circ}}\frac{1}{K_p^2} \right) - 3 \alpha + 2 = 0
\label{alpha_cubic}
\end{equation}
\vspace{2mm}

Equation \ref{alpha_cubic} has one real and two imaginary solutions which can be solved for analytically, specifically the real solution gives the equilibrium value of $\alpha$, and then the mole fractions using equation \ref{X_from_alpha}. A python script, included in the eqc repository, has been written that solves equation \ref{alpha_cubic} using the thermodynamic tables from \citet{turns} to compute $G^\circ_s$ for each species, which ensures a completely separate comparison to eqc with no shared code or methods.

Figure \ref{co2_diss} shows the output of this script compared with curves from eqc, solving the equilibrium composition over a series of fixed temperature and pressures. These results show that the code agrees well with the independent analytic solution.
\newpage

\subsection{Equilibrium Air Chemistry}
\noindent With minor algorithmic differences, the equations that are solved by eqc are the same as those described in \cite{nasacea_I}, and the thermodynamic tables used by both codes are identical. This means for any problem that both codes can solve, the two should give the same answers to within a range of numerical error associated with the Newton's method convergence tolerance. This section performs such test using a  commonly encountered gas mixture for modelling hypersonic wind tunnels, a five species air mixture at known temperature and pressure.\\

A minimal working example the eqc code, written in the Python programming language is shown below. As mentioned in the preamble to this paper, the core equilibrium equations of eqc are written in the C programming language, but they are accessed by the user through a Python layer that handles various IO and memory allocation tasks. This separation of responsibilities is convenient from a programming perspective, and leads to a nice user interface.\\

{\footnotesize
\begin{lstlisting}[language=Python]
from numpy import array
from pyeq import EqCalculator

spnames = ['N2', 'O2', 'N', 'O', 'NO']
eq = EqCalculator(spnames)

T = 2500.0
p = 0.1*101.35e3
Xs0 = array([0.76, 0.23, 0.0, 0.0, 0.0])
Xs1 = eq.pt(p, T, Xs0, 0)
\end{lstlisting}}
\vspace{4mm}

Results from CEA can be obtained from the online webpage \url{cearun.grc.nasa.gov}, which gives an interface to the CEA solver that accepts the same inputs as above. Results from the two solvers are compared below.

{\scriptsize
\begin{verbatim}
Mole Fractions:
       N2         O2         N           O         NO
----------------------------------------------------------
eqc: 0.747849  0.209004  7.93101e-07  0.0207964  0.0223493
CEA: 0.74785   0.20900   7.93200e-07  0.020799   0.022349
\end{verbatim}}

Although the agreement between the two codes is good, the numbers are not precisely identical, particularly the atomic species which differ in the 4th and 5th significant figure. Since the two codes are nominally solving the same equations with the same thermodynamic data, this discrepancy might just be a difference in convergence, or it could be indicative of a problem with one of both codes.\\

Is there any way that we can independently verify the state arrived at is the correct mathematical solution to the equations? There is. Recall that the mathematical foundation of both codes is solving a constrained minimisation problem, formulated using the method of Lagrange multipliers. Recall that the Lagrangian, equation \ref{lagrangian}, is as follows.

\begin{equation*}
\begin{split}
\mathcal{L} = &\sum_s n_s \left[ G_s^\circ + R_u T ln \left(\frac{n_s}{n}\right) + R_u T ln \left(\frac{p}{p^\circ}\right) \right]\\
             + &\sum_j \lambda_j \left( \sum_s a_{js} n_s - a_{js} n^0_s \right)
\end{split}
\end{equation*}

Note that this equation is not actually used directly anywhere in the code. Instead we took derivatives of it and then differentiated again to get an analytic Newton's method. However, if all of the algebra and code has been implemented correctly, the solution we eventually arrive at should be a local minimum of this function.\\

We can take advantage of this fact to verify the implementation in the follow way: Firstly, take the equilibrium state computed above and put it into the full Lagrangian which has been implemented in the C code, purely for the purpose of this exercise. Then, slightly perturb one of the species compositions and compute $L$ again, and do this for all of the species in the calculation. This gives a vector of differences that can be used to compute the derivatives of $L$ with respect to $n_s$, which should be zero or near zero if we are indeed at the equilibrium state.

The results of this calculation are shown below, along with the L2 norm of the derivatives, which is on the order of $0.01$. Given that the actual Lagrangian is on the order of $1 \times 10^{10}$, and knowing that finite differences have a small amount of numerical error, this is a convincing demonstration that the code has found the stationary point, and hence the correct equilibrium composition.

\begin{verbatim}
dLdn_N2  = 0
dLdn_O2  = 0.0051049762
dLdn_N   = 0
dLdn_O   = 0
dLdn_NO  = 0.047740311
---------------------
|dLdns|:  0.04801247861015218
\end{verbatim}
\newpage

%\footnote{An important point that is not shown in this direct comparison is that eqc is much faster than CEA when solving a large number of problems, by approximately two orders of magnitude. In this kind of comparison, CEA is hurt by its use of plain-text input and output files, which must be written to persistent file space and read back one at a time, a very slow process by modern processor standards. In contrast eqc works by reading batches of jobs from numpy arrays directly, and can be embedded in any program with access to the C language. This arrangement also allows the code to be run in parallel using distributed or shared memory parallelism, such as when dealing with very large datasets, or in coupling to a computational fluid dynamics (CFD) flow solver.}

\subsection{Equilibrium Normal Shock with Ionisation}
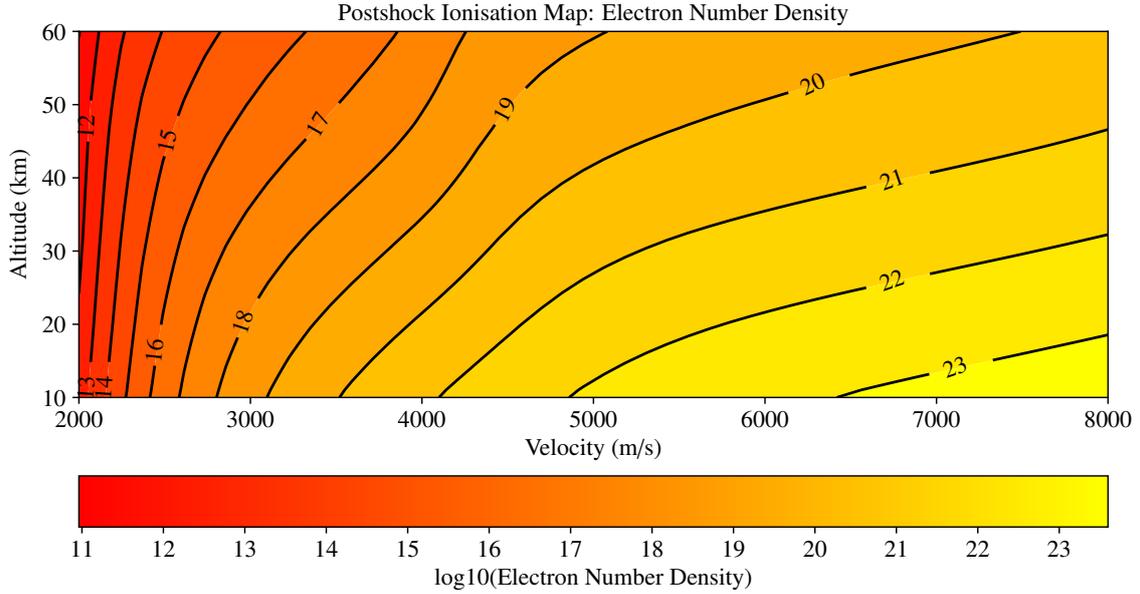
\begin{figure*}[t]
    \centering
    \small
    \def\svgwidth{0.90\textwidth}
    \input{end.pdf_tex}
    \caption{Post-normal shock electron number density, computed vs. Altitude and Flight velocity}
    \label{end}
\end{figure*}

\noindent One of the most common uses for equilibrium chemistry in hypersonics research is computing the properties of shockwaves, particular for hypersonic flight in the atmosphere, where significant dissociation of the Nitrogen and Oxygen molecules is observed behind the shock. This dissociation cools the shocked gas and alters the shock propagation speed, affecting everything from hypersonic wind tunnel operation to flight vehicle heating and transition to turbulence. This section demonstrates an application code that solves for the equilibrium postshock gas state using eqc, designed to be robust at very high shock speeds, high enough to cause ionisation of the postshock gas.

Normal shock waves are described by the Rankine-Hugoniot jump conditions, a set of equations that describe equality of flux between two gas states on either side of the shock.
\begin{equation}
\begin{split}
\rho_1 v_1 & = \rho_2 v_2 \\
\rho_1 v_1^2 + p_1 & = \rho_2 v_2^2 + p_2 \\
h_1  + \frac{1}{2} v_1^2 & = h_2 + \frac{1}{2} v_2^2
\end{split}
\label{jump}
\end{equation} 
With no reactions, these equations can be solved analytically to get a postshock state from a preshock mach number $M_2$, a formula which is present in many textbooks. With dissociation however, the equations must be solved numerically, with a fixed density/internal energy equilibrium problem solved at each step of the outer iteration loop. Although it is possible to use a multi-equation non-linear solver to solve the equations in \ref{jump} separately, this turns out to be a bad idea. Instead, an improved method has been derived which reduces the problem to a single nonlinear root-finding problem, beginning by rearranging the Rankine-Hugoniot equations to compute state 2 assuming that the velocity $v_2$ is known.
\begin{equation}
\begin{split}
\rho_2 & = \frac{\rho_1 v_1}{v_2} \\
p_2 & = (\rho_1 v_1^2 + p_1) - \rho_2 v_2^2\\
e_2 & = \frac{(h_1  + 1/2 v_1^2) - 1/2 \rho_2 v_2^3 - p_2 v_2}{\rho v_1}
\end{split}
\end{equation} 

With the density $\rho_2$ and internal energy $e_2$ computed from the guess $v_2$, eqc can be called in $\rho e$ mode to compute a postshock temperature $T_2$ and composition $X_2$. However, this composition is not certain to obey the jump conditions in equations \ref{jump}, in fact for any guess of $v_2$ a postshock composition can be computed, which makes the problem seem ill-posed. The solution to this situation is to introduce the perfect gas equation of state, which relates the pressure to the density, composition and temperature.
\begin{equation}
p_{eos} = \rho_2 \sum_s \frac{R_u}{M_s X^s_2} T_2
\end{equation}

This pressure $p_{eos}$ will, in general, disagree with the pressure computed using the jump conditions $p_2$, so a numerical method can be used to vary the guess $v_2$ until the two pressures agree, which will yield the correct postshock gas state. In this example, the single-equation Newton's method from scipy is used to solve for $(p_2 - p_{eos})/p_2=0$, yielding a robust normal shock calculation procedure that is well behaved even at high velocities. A comparison to CEA run in SHOCK mode for a 6 km/s, partially ionised air condition at 300 K and 600 Pa, is shown below.

{\small
\begin{verbatim}
Preshock Gas State (76.7% N2, 23.3% O2 by mol.):
 p=600 Pa  T=300 K  v= 6000 m/s

Postshock Gas State:
           eqc       CEA        (% diff)
-------------------------------------------
 v (m/s)|  474.5      474.8      (0.05 %)
 T (K)  |  6564.6     6568.5     (0.06 %)
 p (Pa) |  231418.8   234470.0   (1.32 %)
 X N2   |  4.260e-01  4.261e-01  (0.04 %)
 X O    |  3.249e-01  3.249e-01  (0.00 %)
 X N    |  2.396e-01  2.394e-01  (0.09 %)
 X NO   |  8.549e-03  8.590e-03  (0.48 %)
 X e-   |  3.570e-04  3.576e-04  (0.17 %)
 X NO+  |  3.295e-04  3.299e-04  (0.14 %)
 X O2   |  2.591e-04  2.610e-04  (0.74 %)
 X O+   |  1.437e-05  1.439e-05  (0.12 %)
 X N+   |  1.015e-05  1.017e-05  (0.13 %)
 X N2+  |  4.004e-06  4.020e-06  (0.41 %)
 X O-   |  1.004e-06  1.016e-06  (1.17 %)
\end{verbatim}}

The ionisation fraction associated with this condition is 0.07 \%, and the electron number density is $9 \times 10^{20}$. Normal shock analysis is frequently used to predict postshock conditions at various altitudes and flight velocities, to investigate issues of radio blackout, chemical modelling, and electromagnetic/fluid coupling effects. In figure \ref{end}, we can see the results of running the normal shock script over a wide range of altitude and velocities map. The results show good behaviour down to $\approx$ 2000 m/s or about Mach 6, where the ionisation is essentially negligible, though the code is still performing well in this difficult region. 
\newpage

\subsection{Reflected Shock Tunnel Analysis}
\begin{figure}[h]
\footnotesize
\centering
\def\svgwidth{0.90\columnwidth}
\input{nozzle_exit_state.pdf_tex}
\label{nozzle_exit_state}
\caption{Station numbering for flow states in reflected shock tunnel analysis.}
\end{figure}
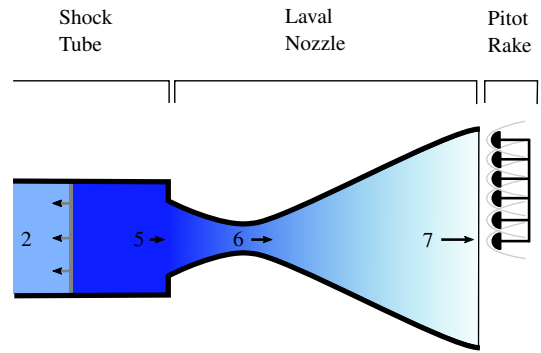

\noindent A workhorse of the experimental hypersonics community is the Reflected Shock Tunnel, an apparatus for generating a brief burst of flow that can actually match the enthalpy and Reynolds numbers experienced during flight. Reflected Shock Tunnels work by creating a strong shockwave or `incident shock' that travels down tube of test gas, compressing and heating it before reflecting off a solid boundary at the far end. This reflected shock further heats and compresses the test gas, while also rupturing a plastic diaphragm that allows the compressed test gas to flow into a Laval nozzle where it expands to supersonic or hypersonic speeds. A schematic depiction of this process is shown in figure \ref{nozzle_exit_state}.\\

Experimentalists measure the velocity of the incident shock and the pressure of the gas feeding into the nozzle, but the flow coming out of the nozzle cannot be measured directly, and is typically reconstructed using some kind of computer program that takes these measurements as inputs. Since temperatures in the shocked gas are high, including equilibrium chemistry in these calculations is critical to getting accurate results. This section presents an application script that uses eqc to perform such a calculation, comparing to known experimental data and to another program.\\

The only additional theoretical component needed for the analysis is an equation for the shock reflection, which can be obtained from the Rankine-Hugoniot equations (equations \ref{jump}). Note that these equations are in a shock centered reference frame: they assume that the shock is stationary and the incoming flow is moving at $v_1$. In the case of the incident shock in a shock tube, the test gas is stationary and the shock velocity is known, so the transformation is simply that $v_1$ is the shock velocity. However, in the reflected state it is $v_2$ that is zero and we must solve the velocity of the shock (at thus the reference frame) itself. We do this by guessing the reflected shock velocity $v_r$, and computing the shock-stationary frame $v_2'$ as follows:

\begin{equation} % This needs a picture to explain what v_1 is in this context
v_2' = v_1 + v_r
\end{equation}

The reflected shock stationary reference frame jump conditions are then the same as before, except the Newton's method is adjusting $v_r$ until the two pressures $p_{eos}$ and $p_2$ match.

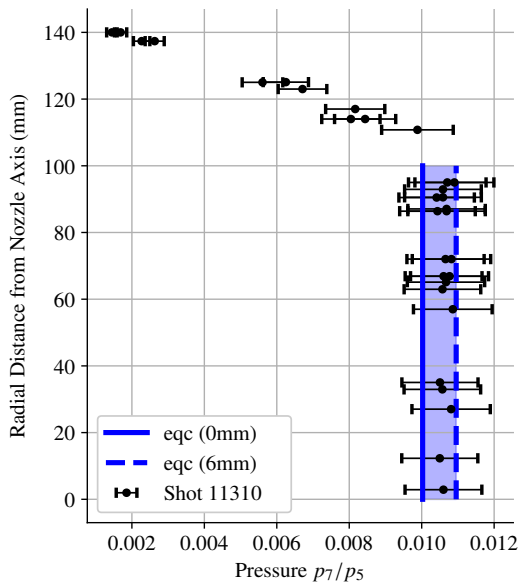
\begin{figure}[h]
\footnotesize
\centering
\def\svgwidth{0.90\columnwidth}
\input{nozzle_results.pdf_tex}
\label{nozzle_results}
\caption{eqc calculations of test number 11310 compared to the measurements from \cite{chan_nozzles18}.}
\end{figure}

In this section the code has been compared to an experiment conducted in the T4 Reflected Shock Tunnel at the University of Queensland, published in \cite{chan_nozzles18}. In particular, I have simulated shot number 11310, a so-called `survey' shot in which the tunnel is fired at a rake of pressure sensors located just downstream of the nozzle. These sensors measure the pitot pressure, computed by passing the flow coming out of the nozzle through a shockwave and then isentropically compressing it to zero velocity. Figure \ref{nozzle_results} shows a schematic of the experimental setup and a comparison to experimental data taken from \cite{chan_nozzles18} figure 11a. 

The solid blue line is the ratio of pitot to stagnation pressure computed using the actual area of T4's Mach 7 expansion nozzle, however, \cite{chan_nozzles18} note that the formation of a viscous boundary layer along the nozzle wall effectively reduces this area, estimating an $\approx$ 6mm displacement thickness due to this effect. The dotted blue line is a calculation with eqc that accounts for this effect by reducing the effective exit radius by this amount. Between these two calculations, the results agree nicely with the experimental data.

\section{Conclusions}
\noindent This article began with a review of existing codes for solving chemical equilibrium problems. Despite finding a number of codes in various states of publication and availability, it noted that no program exists that is simultaneously well-documented, open-source, modern, and lightweight, all important factors for a contemporary research code. The paper then introduced a new code, equilibrium-c, that aims to provide chemical equilibrium solutions while also satisfying these more nebulous criteria.\\

On the documentation front, the paper itself provides much of the important detail. We began with a derivation of the equations of thermochemical equilibrium, a bit of fundamental bookkeeping that is surprisingly absent from the modern literature. This section lead into the numerical discretisation of the equations, describing novel algorithms, developed here for the first time, to improve convergence prevent a complex calculation from going off its rails. The final part of the paper is concerned with examples, showing off the code's range of applications with an emphasis on verification and validation. The program agrees with well with analytic solutions of a single reaction, as well as existing codes such as CEA. The program also performs well in a validation exercise against experimental data collected from a reflected shock tunnel.\\

In summary then, though much of the technical capability of equilibrium-c is not new, it fills an important gap in the ecosystem of computational research tools that should be of interest to the wider computational physics community. This paper is a part of that promise, and though it has been long, and it times, somewhat intense, I hope that it has helped the reader in understanding chemical equilibrium a little better.

\newpage

\bibliography{refs}

\section*{Acknowledgements}
\noindent In preparing this manuscript, the author benefitted greatly from Leonard Susskind's ``The Theoretical Minimum'' lecture series, which is freely available online.






% \end{thebibliography}

\appendix

%------------------------------------------------------------------------------------
\section{Formula for Dice Roll Microstates}
\label{dice_formula}
For n six-sided dice the formula for generating the number of microstates is:
\begin{equation*}
N_S = \sum^M_{k=0} (-1)^k \frac{n !}{k!(n-k)!} \times \frac{(S-6k-1)!}{(n-1)! (S-6k-n)!}
\end{equation*}
\begin{equation*}
M = \textrm{floor}\left(\frac{S-n}{6} \right)
\end{equation*}
This expression is related to the Irwin-Hall distribution and is courtesy of reference \cite{stack_exchange_dicerolls}.

% --------------------------------------------------------
\section{Partial Derivative of Entropy w.r.t Temperature}
\label{S_wrt_T}
Beginning with equation \ref{entropy_Q}.

\begin{equation*}
S = k N \left[ln \frac{Q(V,T)}{N} + 1 \right] + \frac{E(V,T)}{T}
\end{equation*}

We first differentiate with respect to T. 

\begin{equation*}
\frac{\partial S}{\partial T}\Bigr|_{\substack{V}} = k N \frac{ \partial ln Q} {\partial T} +  \frac{\frac{\partial E}{\partial T} T - E \frac{\partial T}{\partial T}}{T^2}
\end{equation*}

\begin{equation}
\frac{\partial S}{\partial T}\Bigr|_{\substack{V}} = \frac{k N}{Q} \frac{ \partial Q} {\partial T}\Bigr|_{\substack{V}} +  \frac{\partial E}{\partial T}\Bigr|_{\substack{V}} \frac{1}{T} - \frac{E}{T^2}
\label{ST_from_Q}
\end{equation}

We now compute the partial derivative of the partition function Q, noting that the energy levels depend on volume $V$, but not the temperature.

\begin{equation*}
Q = \sum_j g_j e^{\frac{-\varepsilon_j}{kT}}
\end{equation*}

\begin{equation*}
\frac{\partial Q}{\partial T}\Bigr|_{\substack{V}} = \frac{1}{k T^2} \sum_j \varepsilon_j g_j e^{\frac{-\varepsilon_j}{kT}}
\end{equation*}

Recall that the most probable set of occupation numbers $N_j^*$ has the following expression.

\begin{equation*}
N^*_j = \frac{N}{Q} g_j e^{\frac{-\varepsilon_j}{kT}}
\end{equation*}

Recall also that the energy at equilibrium is the number of particles in each level $N^*_j$, multiplied by the energy of that level, summed over the whole collection: $E = \sum_j \varepsilon_j N^*_j$. Putting these two pieces together, we get.

\begin{equation*}
E = \sum_j \varepsilon_j \frac{N}{Q} g_j e^{\frac{-\varepsilon_j}{kT}}
\end{equation*}

\begin{equation*}
\frac{EQ}{N} = \sum_j \varepsilon_j g_j e^{\frac{-\varepsilon_j}{kT}}
\end{equation*}

Notice that the left hand side of this expression matches the derivative of Q we are looking for, which gives us the expression as follows.

\begin{equation*}
\frac{\partial Q}{\partial T}\Bigr|_{\substack{V}} = \frac{1}{k T^2} \frac{EQ}{N}
\end{equation*}

Substituting this expression into \ref{ST_from_Q}, we get.

\begin{equation*}
\frac{\partial S}{\partial T}\Bigr|_{\substack{V}} = \frac{\bcancel{k} \bcancel{N}}{\bcancel{Q}} \frac{1}{\bcancel{k} T^2} \frac{E\bcancel{Q}}{\bcancel{N}}
 +  \frac{\partial E}{\partial T}\Bigr|_{\substack{V}} \frac{1}{T} - \frac{E}{T^2}
\end{equation*}

\begin{equation*}
\frac{\partial S}{\partial T}\Bigr|_{\substack{V}} = \bcancel{\frac{E}{T^2}}
 +  \frac{\partial E}{\partial T}\Bigr|_{\substack{V}} \frac{1}{T} - \bcancel{\frac{E}{T^2}}
\end{equation*}

Leaving us with the target expression.
\begin{equation*}
\frac{\partial S}{\partial T}\Bigr|_{\substack{V}} = \frac{\partial E}{\partial T}\Bigr|_{\substack{V}} \frac{1}{T}
\end{equation*}

% --------------------------------------------------------
\section{Partial Derivative of Entropy w.r.t Volume}
\label{S_wrt_V}

The derivation in this section proceeds in a slightly strange way, by beginning with the statistical mechanics definition of pressure and working backwards toward the expression we are interested in. To recap, the pressure is defined as minus the partial derivative of energy with respect to volume, \emph{at constant entropy}.

\begin{equation}
p = - \frac{\partial E}{\partial V}\Bigr|_{\substack{S}}
\label{SV_from_Q}
\end{equation}

This expression makes some intuitive sense; the energy change of a box of gas as it is compressed is going to be a force, in this gas the pressure, integrated over the change in volume. But in mathematical terms it is a somewhat strange formula, because it is a partial derivative of a dependant variable (which is okay) but with \emph{another dependant variable}, S, held constant.

This is a strange thing to do, but it can be done using a lemma derived by Susskind \cite{susskind_ttm}, which is reproduced here. Consider two functions, $E(V,T)$ and $S(V,T)$, which are sketched in the figure below. We wish to move a small distance $\Delta E$, from $E_1$ to $E_2$, along a line of constant $S$.

\begin{figure}[h]
    \centering
    \def\svgwidth{0.90\columnwidth}
    \input{leonards_theorum.pdf_tex}
    \label{leonards_theorum}
    \caption{Two independent functions $E$ and $S$ which depend on $V$ and $T$.}
\end{figure}
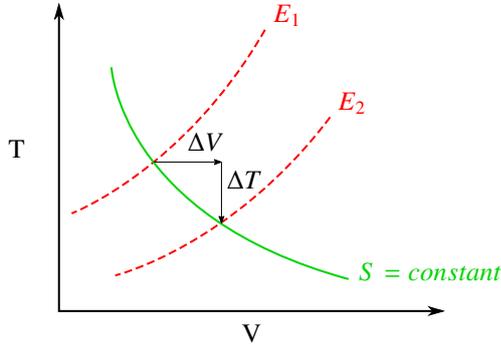

Using the multivariate difference expression, the change in $E$ can be written as follows.

\begin{equation*}
\Delta E = \frac{\partial E}{\partial V}\Bigr|_{\substack{T}} \Delta V
         + \frac{\partial E}{\partial T}\Bigr|_{\substack{V}} \Delta T
\end{equation*}

The second partial derivative can be expanded using the chain rule, to introduce $S$.

\begin{equation*}
\Delta E = \frac{\partial E}{\partial V}\Bigr|_{\substack{T}} \Delta V
         +   \frac{\partial E}{\partial S}\Bigr|_{\substack{V}}
             \frac{\partial S}{\partial T}\Bigr|_{\substack{V}} \Delta T
\end{equation*}

\begin{equation}
\frac{\Delta E}{\Delta V} = \frac{\partial E}{\partial V}\Bigr|_{\substack{T}}
                          +   \frac{\partial E}{\partial S}\Bigr|_{\substack{V}}
                              \frac{\partial S}{\partial T}\Bigr|_{\substack{V}}
                              \frac{\Delta T}{\Delta V}
\label{E_V_diffs}
\end{equation}

We now consider moving along a line of constant $S$. Once again, the multivariate difference theorem gives:
\begin{equation*}
\Delta S = \frac{\partial S}{\partial V}\Bigr|_{\substack{T}} \Delta V
         + \frac{\partial S}{\partial T}\Bigr|_{\substack{V}} \Delta T
\end{equation*}

Setting $\Delta S$ equal to zero, this gives a constraint equation which guides the ratio of $\Delta V$ and $\Delta T$ we need to ensure we travel along the green line in the figure.

\begin{equation}
\frac{\Delta T}{\Delta V} =
    - \frac{
        \frac{\partial S}{\partial V}\Bigr|_{\substack{T}}
    }
    {
         \frac{\partial S}{\partial T}\Bigr|_{\substack{V}}
    }
\label{S_constant}
\end{equation}

This expression is important because it encodes the constraint of $\Delta S$ being zero. Accordingly, we can substitute it into \ref{E_V_diffs} to get the following.
\begin{equation}
\frac{\Delta E}{\Delta V} = \frac{\partial E}{\partial V}\Bigr|_{\substack{T}}
                          + \frac{\partial E}{\partial S}\Bigr|_{\substack{V}} \bcancel{\frac{\partial S}{\partial T}\Bigr|_{\substack{V}}} \times - \frac{\frac{\partial S}{\partial V}\Bigr|_{\substack{T}}}
                                 {\bcancel{\frac{\partial S}{\partial T}\Bigr|_{\substack{V}}}}
\end{equation}

\begin{equation*}
\frac{\Delta E}{\Delta V} = \frac{\partial E}{\partial V}\Bigr|_{\substack{T}}
                          - \frac{\partial E}{\partial S}\Bigr|_{\substack{V}} \frac{\partial S}{\partial V}\Bigr|_{\substack{T}}
\end{equation*}

Letting the $\Delta$ differences on the left hand side go to zero, the fraction there becomes a partial derivative. But a partial derivative with which variable held constant? The answer is $S$, because of the constraint implied by equation \ref{S_constant}.

\begin{equation}
\frac{\partial E}{\partial V}\Bigr|_{\substack{S}} = \frac{\partial E}{\partial V}\Bigr|_{\substack{T}}
                          - \frac{\partial E}{\partial S}\Bigr|_{\substack{V}} \frac{\partial S}{\partial V}\Bigr|_{\substack{T}}
\end{equation}

This is the unusual expression derived by Susskind that relates a pair of partial derivatives with their independent variables held constant. The choice of variables is no accident; Note that partial derivative on the left hand side is just minus the pressure, and also, that one of the terms on the right hand side is the partial derivative of entropy with respect to volume.

\begin{equation*}
- p = \frac{\partial E}{\partial V}\Bigr|_{\substack{T}}
    - \frac{\partial E}{\partial S}\Bigr|_{\substack{V}}
      \frac{\partial S}{\partial V}\Bigr|_{\substack{T}}
\end{equation*}

\begin{equation*}
\frac{\partial S}{\partial V}\Bigr|_{\substack{T}} = 
\frac{p + \frac{\partial E}{\partial V}\Bigr|_{\substack{T}}}
     {\frac{\partial E}{\partial S}\Bigr|_{\substack{V}}}
\end{equation*}

In statistical mechanics, the temperature $T$ is defined as the partial derivative of energy with respect to entropy, which can be used to simplify this equation to get to the target expression we have been looking for.

\begin{equation*}
\frac{\partial S}{\partial V}\Bigr|_{\substack{T}} = 
\frac{p}{T} + \frac{\partial E}{\partial V}\Bigr|_{\substack{T}} \frac{1}{T}
\end{equation*}

% --------------------------------------------------------
\section{Partial Derivatives of the Gibbs Energy wrt. Composition}
\label{dGdNs_appendix}

Recall that the Gibbs energy G for a mixture of species is given by:

\begin{equation*}
G = \sum_s N_s G_s
\end{equation*}

Or, expanding $G_s$ into its explicit form:

\begin{equation*}
G = \sum_s N_s \left[E_s + R_u T - T S_s^\circ - R_u T ln \left( \frac{N_s}{N} \right) - R_u T ln \left( \frac{p}{p^\circ} \right) \right]
\end{equation*}

Differentiating this expression with respect to a given species amount $N_p$ is a actually a nontrivial exercise, because of the sum over species $s$ and the presence of both $N_s$ and the total $N$ in the expression for species molar entropy $S_s$, which appear in the log terms in the large bracket. This creates extra terms because of the product rule that must be accounted for very carefully. 

We are going to differentiate each term in the summation over species one at a time, considering two cases, one where the species index $s$ matches the index differentiating variable $p$, and a second case where it does not. We begin by writing the derivative as follows, using the product rule.

\begin{equation}
\frac{\partial G}{\partial N_p}
  =  \sum_s \frac{\partial N_s}{\partial N_p} G_s
  + \sum_s N_s \frac{\partial G_s}{\partial N_p}
\label{G_deriv_stage_one}
\end{equation}

The first term contains $\partial N_s/\partial N_p$, which is 1 if $s=p$ and zero otherwise, leaving just $G_p$. The second term can be split up into one term for $s=p$, and a summation that includes all of the terms where $s \ne p$. 

\begin{equation*}
\frac{\partial G}{\partial N_p} = 1 \times G_p
  + N_p \frac{\partial G_p}{\partial N_p} + \sum_s^{s \ne p} \frac{\partial G_s}{\partial N_p}
\end{equation*}

Using the full expression for $G_p$, the first term is differentiated as follows.

\begin{equation*}
G_p  = \left[E_p + R_u T - T S^\circ_p + R_u T ln\left(\frac{N_p}{N}\right) + R_u T ln \left(\frac{p}{p^\circ}\right) \right]
\end{equation*}

\begin{equation*}
\frac{\partial G_p}{\partial N_p}  = \left[0 + R_u T \frac{\partial}{\partial N_p} ln\left(\frac{N_p}{N}\right) + 0 \right]
\end{equation*}

\begin{equation*}
\frac{\partial G_p}{\partial N_p}  = R_u T\left[ \frac{\partial}{\partial N_p} ln\left(N_p\right) \frac{\partial}{\partial N_p} ln\left(N\right) \right]
\end{equation*}

\begin{equation*}
\frac{\partial G_p}{\partial N_p}  = R_u T\left[ \frac{1}{N_p} - \frac{1}{N} \frac{\partial N}{\partial N_p} \right]
\end{equation*}

Note that $N$, the total particle count of the system in moles, is the sum of the individual species amounts $N_s$, $\partial N/\partial N_p$ is equal to 1 leaving:

\begin{equation}
\frac{\partial G_p}{\partial N_p}  = R_u T\left[ \frac{1}{N_p} - \frac{1}{N} \right]
\label{s_eq_p}
\end{equation}

For $s \ne p$, we differentiate in a similar manner, but note some minor differences, since $\partial N_s/\partial N_p=0$.

\begin{equation*}
G_s  = \left[E_s + R_u T - T S^\circ_s + R_u T ln\left(\frac{N_s}{N}\right) + R_u T ln \left(\frac{p}{p^\circ}\right) \right]
\end{equation*}

\begin{equation*}
\frac{\partial G_s}{\partial N_p}  = \left[0 + R_u T \frac{\partial}{\partial N_p} ln\left(\frac{N_s}{N}\right) + 0 \right]
\end{equation*}

\begin{equation*}
\frac{\partial G_s}{\partial N_p}  = R_u T\left[ \frac{\partial}{\partial N_p} ln\left(N_s\right) \frac{\partial}{\partial N_p} ln\left(N\right) \right]
\end{equation*}

\begin{equation*}
\frac{\partial G_s}{\partial N_p}  = R_u T\left[ 0 - \frac{1}{N} \frac{\partial N}{\partial N_p} \right]
\end{equation*}

\begin{equation}
\frac{\partial G_s}{\partial N_p}  = -R_u T \frac{1}{N}
\label{s_ne_p}
\end{equation}

Putting the pieces back together, we pick up equation \ref{G_deriv_stage_one}, and substitute in equations \ref{s_eq_p} and \ref{s_ne_p}.

\begin{equation*}
\frac{\partial G}{\partial N_p} = G_p
  + N_p \frac{\partial G_p}{\partial N_p} + \sum_s^{s \ne p} N_s \frac{\partial G_s}{N_p}
\end{equation*}

\begin{equation*}
\frac{\partial G}{\partial N_p} = G_p
  + N_p R_u T\left[ \frac{1}{N_p} - \frac{1}{N} \right]
 + \sum_s^{s \ne p} N_s \times -R_u T \frac{1}{N}
\end{equation*}

\begin{equation*}
\frac{\partial G}{\partial N_p} = G_p
  + R_u T\left[ \frac{N_p}{N_p} - \frac{N_p}{N} \right]
 - \sum_s^{s \ne p} R_u T \frac{N_s}{N}
\end{equation*}

\begin{equation*}
\frac{\partial G}{\partial N_p} = G_p
  + R_u T\left[ 1 - \frac{N_p}{N} - \sum_s^{s \ne p} \frac{N_s}{N}\right]
\end{equation*}

\begin{equation*}
\frac{\partial G}{\partial N_p} = G_p
  + R_u T \left[ 1 - \frac{\sum_s N_s}{N}\right]
\end{equation*}

\begin{equation*}
\frac{\partial G}{\partial N_p} = G_p
  + R_u T \left[ 1 - 1 \right]
\end{equation*}

\begin{equation*}
\frac{\partial G}{\partial N_p} = G_p
\end{equation*}

This is a very tidy and interesting result. Note that, although the Gibbs energy of species $p$ has nonlinear dependencies on the composition $N_s$, when we differentiate, the extra terms from the product rule cancel out exactly to zero. More concretely, we can see that if $G = N_s G_s$, it is true that $\partial G/\partial N_p = G_p$, almost as if $G$ were not a function of $N_p$, even though it certainly is.

% --------------------------------------------------------
\section{Equation for Entropy at Nonstandard Pressure}
\label{appendix_entropy_derivation}

Many thermodynamics textbooks differentiate between the specific (per kilogram) entropy $s$ and the molar (per mole) entropy $S$ using capital and lower case letters. Unfortunately, the upper-case notation is also used in many places for the total, extrinsic entropy $S$. This notation is a source of much confusion. Unfortunately this paper too has been guilty of abusing this notation in several places, which has not caused us any problems until now, where we must be very careful not to mix the two quantities. To that end, we begin by designating the total entropy using a calligraphic font $\mathcal{S}$, to distinguish it from the molar entropy $S$, and the total energy $\mathcal{E}$ as opposed to the energy per mole $E$.

\begin{equation*}
S = \frac{\mathcal{S}}{N} \quad E = \frac{\mathcal{E}}{N}
\end{equation*}

Where N is the number of particles in the system, measured in moles. The differential form of the second law is actually about the extrinsic entropy, so it should be properly written as:

\begin{equation*}
d \mathcal{S} = \frac{d \mathcal{E} + p dV}{T}
\end{equation*}

Or, using the fact that $d\mathcal{S} = N d S$ and $d\mathcal{E} = N d E$:

\begin{equation*}
d S = \frac{N d E + p dV}{NT}
\end{equation*}

We are interested in finding an expression for the entropy per mole at a given pressure and temperature, given that we can look up the entropy at a reference pressure $S^\circ(T)$ using our thermodynamic tables. 

\begin{equation*}
S(p) = S^\circ + \int_{p^\circ}^p dS
\end{equation*}

\begin{equation*}
S(p) = S^\circ + \int_{p^\circ}^p \frac{N d E + p dV}{NT}
\end{equation*}

We can proceed by integrating the entropy over a change in volume; starting at whatever the volume would be at the reference pressure, to whatever the volume would be at our target pressure. We do this because of the handy $d V$ term in the second law, which can be replaced later on by pressure. Also, note that the temperature is fixed in this integral, as we have already accounted for it in the calculation of $S^\circ$, which means that $dE=0$.

\begin{equation*}
S(p) = S^\circ(T) + \int_{V^\circ}^V \frac{p dV}{NT}
\end{equation*}

Using the ideal gas law, $p V = N R_u T$:

\begin{equation*}
S(p) = S^\circ(T) + R_u \int_{V^\circ}^V \frac{dV}{V}
\end{equation*}

\begin{equation*}
S(p) = S^\circ(T) + R_u ln \left(\frac{V}{V^\circ} \right)
\end{equation*}

Now we use the ideal gas law to reintroduce the pressure. Note again that the temperature at our target state and the temperature at the reference state are the same.

\begin{equation*}
S(p) = S^\circ(T) + R_u ln \left( \frac{N R_u T p^\circ}{N R_u T p} \right)
\end{equation*}

\begin{equation*}
S(p) = S^\circ(T) - R_u ln \left(\frac{p}{p^\circ} \right)
\end{equation*}

This is the equation for a single component gas. For a mixture of species, we simply add the subscript to the molar quantities $S$, and change the pressure p to the partial pressure $p_s$.

\begin{equation*}
S_s = S_s^\circ - R_u ln \left(\frac{p_s}{p^\circ} \right)
\end{equation*}

Using the expression for the mole fraction in terms of partial pressure $X_s = p_s/p$ and the mole fraction in terms of the total amounts $X_s = N_s/N$, we can reintroduce the ordinary pressure $p$. 

\begin{equation*}
S_s = S_s^\circ - R_u ln \left( \frac{p}{p^\circ} \frac{N_s}{N} \right)
\end{equation*}

A slight rearrangement gives the species molar entropy in the standard form. 

\begin{equation}
S_s = S_s^\circ - R_u ln \left( \frac{N_s}{N} \right) - R_u ln \left( \frac{p}{p^\circ} \right)
\label{entropy_per_N}
\end{equation}

In \citet{nasacea_I}, a slightly different form of the entropy is presented. This uses the specific molarities $n_s$, which are related to the total amounts $N_s$ as follows.

\begin{equation*}
X_s = \frac{N_s}{N} = \frac{n_s}{n}
\end{equation*}

This gives the entropy in the form that is used in section \ref{governing_equations}, equation \ref{species_entropy}.

\begin{equation}
S_s = S_s^\circ - R_u ln \left( \frac{n_s}{n} \right) - R_u ln \left( \frac{p}{p^\circ} \right)
\label{entropy_per_n}
\end{equation}

\section{Differentiation of the pt Lagrangian}
\label{pt_lagrangian_appendix}
The fixed pressure-temperature Lagrangian (equation \ref{lagrangian}) is as follows.
\begin{equation*}
\begin{split}
\mathcal{L} = &\sum_s n_s \left[ G_s^\circ + R_u T ln \left(\frac{n_s}{n}\right) + R_u T ln \left(\frac{p}{p^\circ}\right) \right]  \\
   + &\sum_j \lambda_j \left( \sum_s a_{js} n_s - a_{js} n^0_s \right) 
\end{split}
\end{equation*}

Differentiating this expression with respect to each composition unknown $n_p$ is actually a tricky problem, because of the sum over species $s$ present in the expression and the existence of $n$ in the Gibbs energy. Importantly, $n$ is currently not an unknown, at this stage of the process it is the sum of $n_s$, which introduces crucial extra terms in the chain rule.

Thankfully, this problem is actually the same tricky problem as differentiating the Gibbs energy, which we have already worked through. To advantage of this prior work, begin by applying the product rule.
\begin{equation*}
\mathcal{L} = \sum_s n_s G_s + \sum_j \lambda_j \left( \sum_s a_{js} n_s - a_{js} n^0_s \right) 
\end{equation*}

\begin{equation*}
\frac{\partial \mathcal{L}}{\partial n_p} =
\sum_s \frac{\partial n_s}{\partial n_p} G_s +
\sum_s n_s \frac{\partial G_s}{\partial n_p} +
\sum_j \lambda_j \sum_s a_{js} \frac{\partial n_s}{\partial n_p}
\end{equation*}

The derivative $\partial n_s/\partial n_p$ is 1 when $s=p$ and zero otherwise, which eliminates two of the summation operators.
\begin{equation}
\frac{\partial \mathcal{L}}{\partial n_p} =
 G_p +
\sum_s n_s \frac{\partial G_s}{\partial n_p} +
\sum_j \lambda_j a_{jp} n_p 
\label{dG_dnp_halfwaydone}
\end{equation}

The simplest way to evaluate $\partial G_s/\partial n_p$ is to take advantage of the work in \ref{dGdNs_appendix}. Recall that there we began with the following.

\begin{equation*}
G_s  = \left[G_s^\circ + R_u T ln\left(\frac{N_s}{N}\right) + R_u T ln \left(\frac{p}{p^\circ}\right) \right]
\end{equation*}

Recall also that the specific molarities $n_s$ are defined as follows.

\begin{equation*}
n_s = \frac{N_s}{\rho V}
\end{equation*}

An irritating detail of this expression is that the mixture total mass $\rho V$ actually depends on $N_s$: $\rho V=\sum_p N_p M_p$, so differentiating it directly creates something of a mess. A simple way to get around this problem is to rewrite the molar Gibbs energy to be in terms of $n_s$.

\begin{equation*}
G_s  = \left[G^\circ_s + R_u T ln\left(\frac{n_s}{n} \frac{\cancel{\rho V}}{\cancel{\rho V}} \right) + R_u T ln \left(\frac{p}{p^\circ}\right) \right]
\end{equation*}

\begin{equation*}
G_s  = \left[G_s^\circ + R_u T ln\left(\frac{n_s}{n} \right) + R_u T ln \left(\frac{p}{p^\circ}\right) \right]
\end{equation*}

Notice that this expression is precisely identical to the one in terms of $N_s$, which means we can simply recapitulate exactly the same algebraic operations as in \ref{dGdNs_appendix}, to get.

\begin{equation*}
\frac{\partial G_p}{\partial n_p} = R_u T \left[\frac{1}{n_p} - \frac{1}{n} \right]
\end{equation*}

\begin{equation*}
\frac{\partial G_s}{\partial n_p} = -R_u T  \frac{1}{n}
\end{equation*}

Putting these pieces into the summation from equation \ref{dG_dnp_halfwaydone}, we get the following.

\begin{equation*}
\frac{\partial \mathcal{L}}{\partial n_p} =
 G_p -
\sum_s^{s \ne p} n_s R_u T \frac{1}{n} + n_p  R_u T \left[\frac{1}{n_p} - \frac{1}{n} \right]+
\sum_j \lambda_j a_{jp} n_p 
\end{equation*}

\begin{equation*}
\frac{\partial \mathcal{L}}{\partial n_p} =
 G_p - R_u T
\sum_s^{s \ne p} \frac{n_s}{n} + R_u T \left[\frac{n_p}{n_p} - \frac{n_p}{n} \right]+
\sum_j \lambda_j a_{jp} n_p 
\end{equation*}

\begin{equation*}
\frac{\partial \mathcal{L}}{\partial n_p} =
 G_p - R_u T
\left[\sum_s^{s \ne p} \frac{n_s}{n} + \frac{n_p}{n}  \right]+ R_u T \cancel{\frac{n_p}{n_p}}+
\sum_j \lambda_j a_{jp} n_p 
\end{equation*}

\begin{equation*}
\frac{\partial \mathcal{L}}{\partial n_p} =
 G_p - R_u T
\left[\sum_s \frac{n_s}{n} \right] + R_u T +
\sum_j \lambda_j a_{jp} n_p 
\end{equation*}

\begin{equation*}
\frac{\partial \mathcal{L}}{\partial n_p} =
 G_p - R_u T
\left[\cancel{\frac{n}{n}} \right] + R_u T +
\sum_j \lambda_j a_{jp} n_p 
\end{equation*}

\begin{equation*}
\frac{\partial \mathcal{L}}{\partial n_p} =
 G_p \cancel{- R_u T + R_u T} +
\sum_j \lambda_j a_{jp} n_p 
\end{equation*}

\begin{equation}
\frac{\partial \mathcal{L}}{\partial n_p} =
 G_p +
\sum_j \lambda_j a_{jp} n_p 
\label{end_appendix_dGdns}
\end{equation}

Once again we see that the additional terms from the chain rule cancel out, giving $\partial g/\partial n_p = G_p$. Equation \ref{end_appendix_dGdns} is the target expression we are looking for.

\section{Reduced Iteration Equations for known Density/Energy}
\label{rhoe_equations}
The main body of this paper introduces the equations for a fixed temperature and pressure equilibrium problem. In this section, I present those needed for fixed density and internal energy $\rho e$, and in the next appendix, those for fixed pressure and entropy $ps$ problems. $\rho e$ problems are useful in constant volume reactors where energy and mass are conserved, and $p s$ mode is helpful for fluid mechanics problems featuring isentropic flow. For $\rho e$ problems, the key theoretical difference is the minimisation of Helmholtz energy, a thermodynamic potential defined as:

\begin{equation*}
F = E - TS
\end{equation*}

Note the similarity to the Gibbs energy $E + p V - T S$; the pressure term has been removed. Before we begin, it will be helpful to write the Helmholtz energy per kilogram $f$ in a specific way.

\begin{equation*}
f = \frac{F}{\rho V} = n_s F_s = n_s [E_s - T S_s]
\end{equation*}

\begin{equation*}
f = n_s [E_s - T S_s^\circ + R_u T ln\frac{n_s}{n} + R_u T ln\frac{p}{p^\circ} ]
\end{equation*}

\begin{equation*}
f = n_s [E_s + R_u T - R_u T -  T S_s^\circ + R_u T ln\frac{n_s}{n} + R_u T ln\frac{p}{p^\circ} ]
\end{equation*}

\begin{equation*}
f = n_s [G_s^\circ + R_u T ln\frac{n_s}{n} + R_u T ln\frac{p}{p^\circ}  - R_u T]
\end{equation*}

\begin{equation*}
f = n_s [G_s^\circ + R_u T ln\frac{n_s p}{n p^\circ} - R_u T]
\end{equation*}

We then introduce the ideal gas equation, $p = \rho R_u T/ M_{mix}$, and remembering that $n=1/M_{mix}$, remove the unknown pressure from the log term, in favour of the density, which is known.

\begin{equation*}
f = n_s [G_s^\circ + R_u T ln\frac{\rho n_s R_u T}{p^\circ} - R_u T]
\end{equation*}

The $\rho e$ Lagrangian that needs to be minimised is then.

\begin{equation*}
L = \sum_s n_s \left[ G_s^\circ + R_u T ln\frac{\rho n_s R_u T}{p^\circ} - R_u T \right] + \sum_j \left(\sum_s a_{js} n_s - b^0_k\right)
\end{equation*}

With no $n$ term, differentiating this expression is much easier than the $pt$ version, though note the chain rule still applies.

\begin{equation*}
\begin{split}
\frac{\partial L}{\partial n_s} = &
G_s^\circ + R_u T ln\frac{\rho n_s R_u T}{p^\circ} - R_u T \\
&+ n_s \left[R_u T \frac{\partial}{\partial n_s} ln\frac{\rho n_s R_u T}{p^\circ} + 0 \right] \\
& + \sum_j \lambda_j a_{js} 
\end{split}
\end{equation*}

\begin{equation*}
\begin{split}
\frac{\partial L}{\partial n_s} = &
G_s^\circ + R_u T ln\frac{\rho n_s R_u T}{p^\circ} - R_u T \\
&+ n_s \left[R_u T \frac{1}{n_s}\right] \\
& + \sum_j \lambda_j a_{js} 
\end{split}
\end{equation*}

\begin{equation*}
\frac{\partial L}{\partial n_s} =
G_s^\circ + R_u T ln\frac{\rho n_s R_u T}{p^\circ} - \cancel{R_u T} + \cancel{R_u T}
 + \sum_j \lambda_j a_{js} 
\end{equation*}

\begin{equation}
\frac{\partial L}{\partial n_s} =
G_s^\circ + R_u T ln\frac{\rho n_s R_u T}{p^\circ} + \sum_j \lambda_j a_{js}
\label{rhou_Fs}
\end{equation}

And the constraint equations are the same as before.

\begin{equation}
\frac{\partial L}{\partial \lambda_j} =
\sum_s a_{js} n_s - a_{js} n^0_s
\label{rhou_Fj}
\end{equation}

At this point we encounter something new. What we have just derived are the equations that maximise the equilibrium disorder of the system: Any composition that evaluates to zero on the right hand side of \ref{rhou_Fs} and \ref{rhou_Fj} is an equilibrium state. But note that the temperature appears in these equations, and the temperature is now an unknown. We solve this problem by introducing another nonlinear equation to the set being solved, an apparently trivial equation that checks whether the internal energy $e(T)$ at our current guessed temperature matches the target $e_0$.

\begin{equation}
0 = e_0 - e(T, n_s) = e_0 - \sum_s n_s E_s(T)
\end{equation}

We can now proceed to derive analytic update equations for the Newton step, following \citet{nasacea_I}. Our unknowns are the logarithms of the species $ln(n_s)$ (but not $ln(n)$, this time), the Lagrange multipliers, which are converted to $\pi_j$ in exactly the same way as before, and the logarithm of the temperature $ln(T)$. The iteration equations are then as follows.

\begin{equation}
\Delta ln(n_s) - \sum_j a_{sj} \pi_j - \frac{E_j}{R_u T} \Delta ln T = -\frac{G_s}{R_u T} ~~~ ~~~ (s=0,...,n_s)
\label{helm_ds}
\end{equation}

\begin{equation}
\sum_s a_{sj} n_s \Delta ln(n_s) = \sum_s a_{js} n^0_s - a_{js} n_s  ~~~ ~~~ (j=0,...,n_e)
\label{helm_dj}
\end{equation}

\begin{equation}
\begin{split}
\sum_s \frac{n_s E_s}{R_u T} \Delta ln(n_s) + \sum_s \frac{n_s C_{vs}}{R_u} \Delta ln T = -\frac{e_0 - n_s E_s}{R_u T}\\ ~~~ ~~~ (s=0,...,n_s)
\label{helm_de}
\end{split}
\end{equation}

Once again the species update equations contain only $\Delta ln(n_s)$, which means they can be eliminated analytically to produce what \cite{nasacea_I} call the Reduced Helmholtz Iteration Equations.

\begin{equation}
\begin{split}
\sum_i \sum_s a_{js} a_{is} n_s \pi_i
+ \left(\sum_s \frac{a_{js} n_s E_s}{R_u T}\right) \Delta ln T =\\
b^0_j - a_{js} n_s + \sum_s \frac{a_{js} n_s G_s}{R_u T}\\ ~~~ ~~~ (j=0,...,n_e)
\end{split}
\label{reduced_helm_Fj}
\end{equation}

\begin{equation}
\begin{split}
\sum_i \left(\sum_s \frac{a_{is} n_s E_s}{R_u T} \right) pi_i +
\left[\sum_s \frac{n_s C_{v,s}}{R_u} + \frac{n_s E_s^2}{R_u^2 T^2} \right] \Delta ln T=\\
\frac{e_0 - \sum_s n_s E_s}{R_u T} + \sum_s \frac{n_s E_s G_s}{R_u^2 T^2}
\label{reduced_helm_Fn}
\end{split}
\end{equation}

These are the equations that are iterated on and solved, using the exact same methodology as from section \ref{convergence}, to solve for a known density and internal energy.
\newpage

%------------------------------------------------------------------------------------
\section{Reduced Iteration Equations for known Pressure/Entropy}
\label{ps_equations}

The equations for known pressure and entropy, or $ps$ problems, are based on the Gibbs free energy and are very similar to the $pt$ problem in the main text. However, just like in the $\rho e$ derivation in the previous appendix, the temperature is now an unknown we must add to the set of nonlinear equations needed to solve for the equilibrium state. In this case, we add an apparently trivial equation that the entropy at the current guess must match the target entropy $s_0$.

\begin{equation}
0 = s_0 - n_s S_s(T,p)
\end{equation}

Following the same process as in the $\rho e$ appendix, we get the following Reduced Iteration Equations.

\begin{equation}
\begin{split}
\sum_i \sum_s a_{js} a_{is} n_s \pi_i
+ \left(\sum_s a_{js} n_s\right) \Delta ln(n)
+ \left(\sum_s \frac{a_{js} n_s H_s}{R_u T}\right) \Delta ln T\\
=
\sum_s a_{js} n^0_s - a_{js} n_s + \sum_s \frac{a_{js} n_s G_s}{R_u T}\\ ~~~ ~~~ (j=0,...,n_e)
\end{split}
\label{reduced_gibbs_Fj}
\end{equation}

\begin{equation}
\begin{split}
\sum_i \sum_s a_{is} n_s \pi_i +
\left( \sum_s n_s - n \right) \Delta ln(n) +
\left( \sum_s \frac{n_s H_s}{R_u T} \right) \Delta ln T\\
=
n - \sum_s n_s + \sum_s \frac{n_s G_s}{R_u T}
\label{reduced_gibbs_Fn}
\end{split}
\end{equation}

\begin{equation}
\begin{split}
\sum_i \left( \sum_s \frac{a_{is} n_s S_s}{R_u} \right) \pi_i
+ \left( \sum_s \frac{n_s S_s}{R_u} \right) \Delta ln(n)\\
+ \left( \sum_s \frac{n_s C_{p,s}}{R_u} + \frac{n_s H_s S_s}{R_u^2 T} \right) \Delta ln T =\\
\frac{s_0 - \sum_s n_s S_s}{R_u} + n - \sum_s n_s  + \sum_s \frac{n_s S_s G_s}{R_u^2 T}
\label{reduced_gibbs_Fn}
\end{split}
\end{equation}
\newpage



%% References
%%   \cite[chap. ]{key} ==>> [#, chap. 2]
%%

\section{Derivation of Single Reaction Analytic Solution}
\label{co2_derivation}

Recall that the equilibrium condition is given by equation \ref{equilibrium_condition_2}, reproduced below.
\begin{equation*}
0 =  \sum_s \left[G^{\circ}_s + R_u T ln\left(\frac{p_s}{p^{\circ}} \right) \right] d N_s
\end{equation*}

The changes in species mole densities can be written in terms of alpha as follows.
\begin{equation*}
\begin{split}
d N_{CO_2} & = -d \alpha \\
d N_{CO}   & = d \alpha \\
d N_{O_2}  & = \frac{1}{2} d \alpha \\
\end{split}
\end{equation*}

Substituting these expressions in equation \ref{equilibrium_condition_2} results in a single equation in terms of the partial pressures.
\begin{equation*}
\begin{split}
0 =
  & \left[G^{\circ}_{CO_2} + R_u T ln\left(\frac{p_{CO_2}}{p^{\circ}} \right) \right] -d \alpha\\
 +& \left[G^{\circ}_{CO} + R_u T ln\left(\frac{p_{CO}}{p^{\circ}} \right) \right] d \alpha\\
 +& \left[G^{\circ}_{O_2} + R_u T ln\left(\frac{p_{O_2}}{p^{\circ}} \right) \right] \frac{1}{2} d \alpha
\end{split}
\end{equation*}

Dividing by $d \alpha$ removes the reaction progress variable, leaving just the partial pressures.
\begin{equation*}
\begin{split}
0 = &  \left[-G^{\circ}_{CO_2} + G^{\circ}_{CO} + \frac{1}{2}G^{\circ}_{O_2} \right]\\
  + & R_u T \left[
  - ln \left(\frac{p_{CO_2}}{p^{\circ}} \right)
  + ln \left(\frac{p_{CO}}{p^{\circ}} \right)
  +\frac{1}{2} ln \left(\frac{p_{O_2}}{p^{\circ}} \right)
\right]
\end{split}
\end{equation*}

\begin{equation*}
0 =  \frac{\Delta G}{R_u T}
  + ln \left[
             \left(\frac{p_{CO_2}}{p^{\circ}}\right)^{-1}
      \times \left(\frac{p_{CO}}{p^{\circ}}\right)
      \times \left(\frac{p_{O_2}}{p^{\circ}}\right)^{\frac{1}{2}}
   \right]
\end{equation*}

Where $\Delta G$ is the sum of the stoichiometric coefficients multiplied by the individual species energies $\Delta G = -G_{CO_2} + G_{CO} + \sfrac{1}{2} G_{O_2}$. We can remove the partial pressures from this expression by using the definition of the mole fractions $X_s = p_s/p$, which results in a more convenient result since the pressure $p$ is known.
\begin{equation*}
0 =  \frac{\Delta G}{R_u T}
  + ln \left[
             \left(\frac{p_{CO_2}}{p p^{\circ}}\right)^{-1}
      \times \left(\frac{p_{CO}}{p p^{\circ}}\right)
      \times \left(\frac{p_{O_2}}{p p^{\circ}}\right)^{\frac{1}{2}}
      \times \left(\frac{p p^{\frac{1}{2}}}{p}\right)
   \right]
\end{equation*}

\begin{equation*}
0 =  \frac{\Delta G}{R_u T}
  + ln \left[
             \frac{X_{CO} X_{O_2}^{\frac{1}{2}}}{X_{CO_2}}
      \times \left(\frac{p}{p^{\circ}}\right)^{\frac{1}{2}}
\right]
\end{equation*}

\begin{equation*}
\exp\left(\frac{-\Delta G}{R_u T}\right)
=
     \frac{X_{CO} X_{O_2}^{\frac{1}{2}}}{X_{CO_2}}
      \times \left(\frac{p}{p^{\circ}}\right)^{\frac{1}{2}}
\end{equation*}

The left hand side of this expression contains the exponential of $\Delta G$ divided by the thermal energy. This quantity is frequently referred to as the equilibrium constant $K_p$ for the reaction, which gives the following formula, found in many textbooks.
\begin{equation*}
K_p
=
     \frac{X_{CO} X_{O_2}^{\frac{1}{2}}}{X_{CO_2}}
      \times \left(\frac{p}{p^{\circ}}\right)^{\frac{1}{2}}
\label{equilibrium_constant}
\end{equation*}

Reduction to a single unknown can be performed once again using the reaction progress variable $\alpha$. If we begin with $N^0_{CO_2}$ moles of carbon dioxide, for a given amount of reaction progress $\alpha$, the amount of each species is given by.
\begin{equation*}
\begin{split}
N_{CO_2} & = N^0_{CO_2} (1 -\alpha) \\
N_{CO}   & = N^0_{CO_2} \alpha \\
N_{O_2}  & = N^0_{CO_2} \frac{1}{2} \alpha \\
\end{split}
\end{equation*}

With the total moles $N_T$ equal to $N_{CO_2} + N_{CO} + N_{O_2}$ or $N^0_{CO_2} (1 + \sfrac{1}{2} \alpha)$, and the mole fractions written as:
\begin{equation*}
\begin{split}
X_{CO_2} & = \frac{N_{CO_2}}{N_T} = \frac{1 -\alpha}{1 + \sfrac{1}{2} \alpha} \\
X_{CO}   & = \frac{N_{CO}}{N_T}   = \frac{\alpha}{1 + \sfrac{1}{2} \alpha}   \\
X_{O_2}  & = \frac{N_{O_2}}{N_T}  = \frac{\sfrac{1}{2} \alpha}{1 + \sfrac{1}{2} \alpha}  \\
\label{X_from_alpha}
\end{split}
\end{equation*}

Substituting this expression into equation \ref{equilibrium_constant} gives an expression entirely in terms of $\alpha$, which can be rearranged into a cubic polynomial as follows.
\begin{equation*}
K_p
=
     \frac{\alpha}{1 + \sfrac{1}{2} \alpha} \times \frac{\left(\sfrac{1}{2} \alpha\right)^{\sfrac{1}{2}}}{1 + \sfrac{1}{2} \alpha} \times \frac{1 + \sfrac{1}{2} \alpha}{1-\alpha}
      \times \left(\frac{p}{p^{\circ}}\right)^{\frac{1}{2}}
\end{equation*}

\begin{equation*}
K_p \sqrt{\frac{p^{\circ}}{p}}
=
     \frac{\alpha^{1.5}}{\sqrt{2}} \times \frac{1}{1 - \alpha} \times \frac{1}{\left(1 + \sfrac{1}{2}\alpha\right)^{\sfrac{1}{2}}}
\end{equation*}

\begin{equation*}
K_p^2 \frac{p^{\circ}}{p}
=
     \frac{\alpha^{3}}{2} \times \frac{1}{\left(1 - \alpha\right)^2} \times \frac{1}{1 + \sfrac{1}{2}\alpha}
\end{equation*}

\begin{equation}
\alpha^3 \left( 1 - \frac{p}{p^{\circ}}\frac{1}{K_p^2} \right) - 3 \alpha + 2 = 0
\end{equation}

This expression is readily solved using numpy's \texttt{roots} library for $\alpha$.

% Put this an an appendix probably.
%\begin{equation}
%M_1 = \frac{v_1}{\sqrt{\gamma R T_1}} 
%\end{equation}
%\begin{equation}
%T^{guess}_2 = T_1 \frac{\left(2 \gamma M_1^2 - (\gamma - 1)\right) \left((\gamma - 1) M_1^2 + 2\right)}{(\gamma+1)^2*M_1^2}
%\end{equation}
%\begin{equation}
%M^{guess}_2 = \sqrt{\frac{M_1^2 (\gamma-1) + 2)}{2 \gamma M_1^2 - (\gamma - 1)}} 
%\end{equation}
%\begin{equation}
%v^{guess}_2 = \frac{1}{2} M^{guess}_2 \sqrt{\gamma R T_1}
%\end{equation}

\end{document}

%% file: images/equilibrium.pdf_tex
%% Creator: Inkscape 1.1.2 (0a00cf5339, 2022-02-04), www.inkscape.org
%% Accompanies image file 'equilibrium.pdf' (pdf, eps, ps)
\begingroup%
  \makeatletter%
  \providecommand\color[2][]{%
    \errmessage{(Inkscape) Color is used for the text in Inkscape, but the package 'color.sty' is not loaded}%
    \renewcommand\color[2][]{}%
  }%
  \providecommand\transparent[1]{%
    \errmessage{(Inkscape) Transparency is used (non-zero) for the text in Inkscape, but the package 'transparent.sty' is not loaded}%
    \renewcommand\transparent[1]{}%
  }%
  \providecommand\rotatebox[2]{#2}%
  \newcommand*\fsize{\dimexpr\f@size pt\relax}%
  \newcommand*\lineheight[1]{\fontsize{\fsize}{#1\fsize}\selectfont}%
  \ifx\svgwidth\undefined%
    \setlength{\unitlength}{333.34204488bp}%
    \ifx\svgscale\undefined%
      \relax%
    \else%
      \setlength{\unitlength}{\unitlength * \real{\svgscale}}%
    \fi%
  \else%
    \setlength{\unitlength}{\svgwidth}%
  \fi%
  \global\let\svgwidth\undefined%
  \global\let\svgscale\undefined%
  \makeatother%
  \begin{picture}(1,0.73048461)%
    \lineheight{1}%
    \setlength\tabcolsep{0pt}%
    \put(0,0){\includegraphics[width=\unitlength,page=1]{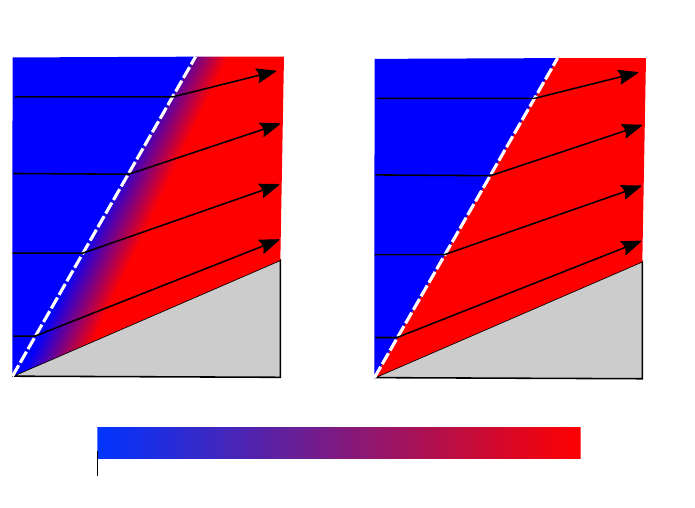}}%
    \put(0.021943,0.69895531){\color[rgb]{0,0,0}\makebox(0,0)[lt]{\lineheight{1.25}\smash{\begin{tabular}[t]{l}Finite-Rate Reactions\end{tabular}}}}%
    \put(0.54360894,0.70313379){\color[rgb]{0,0,0}\makebox(0,0)[lt]{\lineheight{1.25}\smash{\begin{tabular}[t]{l}Chemical Equilibrium\end{tabular}}}}%
    \put(0,0){\includegraphics[width=\unitlength,page=2]{equilibrium.pdf}}%
    \put(-0.00312882,0.00748808){\color[rgb]{0,0,0}\makebox(0,0)[lt]{\lineheight{1.25}\smash{\begin{tabular}[t]{l}Unreacted Flow\end{tabular}}}}%
    \put(0.66707928,0.00748808){\color[rgb]{0,0,0}\makebox(0,0)[lt]{\lineheight{1.25}\smash{\begin{tabular}[t]{l}Fully Reacted Flow\end{tabular}}}}%
  \end{picture}%
\endgroup%

%% file: images/box.pdf_tex
%% Creator: Inkscape 1.1.2 (0a00cf5339, 2022-02-04), www.inkscape.org
%% Accompanies image file 'box.pdf' (pdf, eps, ps)
\begingroup%
  \makeatletter%
  \providecommand\color[2][]{%
    \errmessage{(Inkscape) Color is used for the text in Inkscape, but the package 'color.sty' is not loaded}%
    \renewcommand\color[2][]{}%
  }%
  \providecommand\transparent[1]{%
    \errmessage{(Inkscape) Transparency is used (non-zero) for the text in Inkscape, but the package 'transparent.sty' is not loaded}%
    \renewcommand\transparent[1]{}%
  }%
  \providecommand\rotatebox[2]{#2}%
  \newcommand*\fsize{\dimexpr\f@size pt\relax}%
  \newcommand*\lineheight[1]{\fontsize{\fsize}{#1\fsize}\selectfont}%
  \ifx\svgwidth\undefined%
    \setlength{\unitlength}{440.23742243bp}%
    \ifx\svgscale\undefined%
      \relax%
    \else%
      \setlength{\unitlength}{\unitlength * \real{\svgscale}}%
    \fi%
  \else%
    \setlength{\unitlength}{\svgwidth}%
  \fi%
  \global\let\svgwidth\undefined%
  \global\let\svgscale\undefined%
  \makeatother%
  \begin{picture}(1,0.37881344)%
    \lineheight{1}%
    \setlength\tabcolsep{0pt}%
    \put(0,0){\includegraphics[width=\unitlength,page=1]{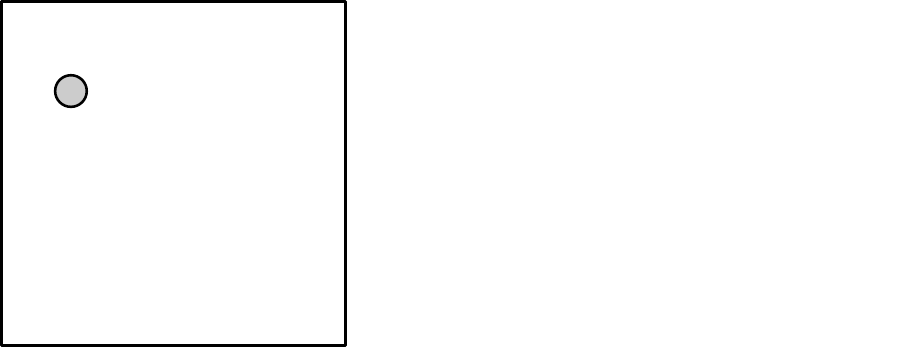}}%
    \put(0.06930465,0.27030779){\color[rgb]{0,0,0}\makebox(0,0)[lt]{\lineheight{1.25}\smash{\begin{tabular}[t]{l}1\end{tabular}}}}%
    \put(0,0){\includegraphics[width=\unitlength,page=2]{box.pdf}}%
    \put(0.27546019,0.33152971){\color[rgb]{0,0,0}\rotatebox{-0.72763267}{\makebox(0,0)[lt]{\lineheight{1.25}\smash{\begin{tabular}[t]{l}3\end{tabular}}}}}%
    \put(0,0){\includegraphics[width=\unitlength,page=3]{box.pdf}}%
    \put(0.03920482,0.08069959){\color[rgb]{0,0,0}\rotatebox{-1.2216976}{\makebox(0,0)[lt]{\lineheight{1.25}\smash{\begin{tabular}[t]{l}6\end{tabular}}}}}%
    \put(0,0){\includegraphics[width=\unitlength,page=4]{box.pdf}}%
    \put(0.15907846,0.1642679){\color[rgb]{0,0,0}\rotatebox{0.06554205}{\makebox(0,0)[lt]{\lineheight{1.25}\smash{\begin{tabular}[t]{l}2\end{tabular}}}}}%
    \put(0,0){\includegraphics[width=\unitlength,page=5]{box.pdf}}%
    \put(0.17098809,0.01224562){\color[rgb]{0,0,0}\rotatebox{0.05426039}{\makebox(0,0)[lt]{\lineheight{1.25}\smash{\begin{tabular}[t]{l}4\end{tabular}}}}}%
    \put(0,0){\includegraphics[width=\unitlength,page=6]{box.pdf}}%
    \put(0.33864072,0.09975041){\color[rgb]{0,0,0}\rotatebox{-1.6361913}{\makebox(0,0)[lt]{\lineheight{1.25}\smash{\begin{tabular}[t]{l}7\end{tabular}}}}}%
    \put(0,0){\includegraphics[width=\unitlength,page=7]{box.pdf}}%
    \put(0.25673944,0.22247524){\color[rgb]{0,0,0}\rotatebox{-1.9918119}{\makebox(0,0)[lt]{\lineheight{1.25}\smash{\begin{tabular}[t]{l}5\end{tabular}}}}}%
    \put(0,0){\includegraphics[width=\unitlength,page=8]{box.pdf}}%
    \put(0.04890386,0.23868266){\color[rgb]{0,0,0}\rotatebox{1.1488033}{\makebox(0,0)[lt]{\lineheight{1.25}\smash{\begin{tabular}[t]{l}8\end{tabular}}}}}%
    \put(0,0){\includegraphics[width=\unitlength,page=9]{box.pdf}}%
    \put(0.70250724,0.09103509){\color[rgb]{0,0,0}\makebox(0,0)[lt]{\lineheight{1.25}\smash{\begin{tabular}[t]{l}1\end{tabular}}}}%
    \put(0,0){\includegraphics[width=\unitlength,page=10]{box.pdf}}%
    \put(0.55821822,0.0931431){\color[rgb]{0,0,0}\rotatebox{-0.72763267}{\makebox(0,0)[lt]{\lineheight{1.25}\smash{\begin{tabular}[t]{l}3\end{tabular}}}}}%
    \put(0,0){\includegraphics[width=\unitlength,page=11]{box.pdf}}%
    \put(0.60488143,0.09096891){\color[rgb]{0,0,0}\rotatebox{0.06554205}{\makebox(0,0)[lt]{\lineheight{1.25}\smash{\begin{tabular}[t]{l}2\end{tabular}}}}}%
    \put(0,0){\includegraphics[width=\unitlength,page=12]{box.pdf}}%
    \put(0.70142691,0.1338285){\color[rgb]{0,0,0}\rotatebox{-1.2216976}{\makebox(0,0)[lt]{\lineheight{1.25}\smash{\begin{tabular}[t]{l}6\end{tabular}}}}}%
    \put(0,0){\includegraphics[width=\unitlength,page=13]{box.pdf}}%
    \put(0.70225206,0.17636431){\color[rgb]{0,0,0}\rotatebox{0.05426039}{\makebox(0,0)[lt]{\lineheight{1.25}\smash{\begin{tabular}[t]{l}4\end{tabular}}}}}%
    \put(0,0){\includegraphics[width=\unitlength,page=14]{box.pdf}}%
    \put(0.79405631,0.09194264){\color[rgb]{0,0,0}\rotatebox{-1.6361913}{\makebox(0,0)[lt]{\lineheight{1.25}\smash{\begin{tabular}[t]{l}7\end{tabular}}}}}%
    \put(0,0){\includegraphics[width=\unitlength,page=15]{box.pdf}}%
    \put(0.79527489,0.13249416){\color[rgb]{0,0,0}\rotatebox{1.1488033}{\makebox(0,0)[lt]{\lineheight{1.25}\smash{\begin{tabular}[t]{l}8\end{tabular}}}}}%
    \put(0,0){\includegraphics[width=\unitlength,page=16]{box.pdf}}%
    \put(0.74708608,0.09068935){\color[rgb]{0,0,0}\rotatebox{-1.9918119}{\makebox(0,0)[lt]{\lineheight{1.25}\smash{\begin{tabular}[t]{l}5\end{tabular}}}}}%
    \put(0.63834415,0.01308228){\color[rgb]{0,0,0}\makebox(0,0)[lt]{\lineheight{1.25}\smash{\begin{tabular}[t]{l}Energy levels\end{tabular}}}}%
    \put(0,0){\includegraphics[width=\unitlength,page=17]{box.pdf}}%
    \put(0.46196848,0.04575306){\color[rgb]{0,0,0}\makebox(0,0)[lt]{\lineheight{1.25}\smash{\begin{tabular}[t]{l}0\end{tabular}}}}%
    \put(0.50734733,0.04575306){\color[rgb]{0,0,0}\makebox(0,0)[lt]{\lineheight{1.25}\smash{\begin{tabular}[t]{l}1\end{tabular}}}}%
    \put(0.56028936,0.04575306){\color[rgb]{0,0,0}\makebox(0,0)[lt]{\lineheight{1.25}\smash{\begin{tabular}[t]{l}2\end{tabular}}}}%
    \put(0.60769403,0.04575306){\color[rgb]{0,0,0}\makebox(0,0)[lt]{\lineheight{1.25}\smash{\begin{tabular}[t]{l}3\end{tabular}}}}%
    \put(0.65442367,0.04575306){\color[rgb]{0,0,0}\makebox(0,0)[lt]{\lineheight{1.25}\smash{\begin{tabular}[t]{l}4\end{tabular}}}}%
    \put(0.70358392,0.04575306){\color[rgb]{0,0,0}\makebox(0,0)[lt]{\lineheight{1.25}\smash{\begin{tabular}[t]{l}5\end{tabular}}}}%
    \put(0.74896262,0.04575306){\color[rgb]{0,0,0}\makebox(0,0)[lt]{\lineheight{1.25}\smash{\begin{tabular}[t]{l}6\end{tabular}}}}%
    \put(0.79636685,0.04575306){\color[rgb]{0,0,0}\makebox(0,0)[lt]{\lineheight{1.25}\smash{\begin{tabular}[t]{l}7\end{tabular}}}}%
    \put(0.84552665,0.04575306){\color[rgb]{0,0,0}\makebox(0,0)[lt]{\lineheight{1.25}\smash{\begin{tabular}[t]{l}8\end{tabular}}}}%
    \put(0.89468636,0.04575306){\color[rgb]{0,0,0}\makebox(0,0)[lt]{\lineheight{1.25}\smash{\begin{tabular}[t]{l}9\end{tabular}}}}%
    \put(0.9335818,0.04575306){\color[rgb]{0,0,0}\makebox(0,0)[lt]{\lineheight{1.25}\smash{\begin{tabular}[t]{l}10\end{tabular}}}}%
  \end{picture}%
\endgroup%

%% file: images/dice_pdfs.pdf_tex
%% Creator: Inkscape 1.1.2 (0a00cf5339, 2022-02-04), www.inkscape.org
%% Accompanies image file 'dice_pdfs.pdf' (pdf, eps, ps)
\begingroup%
  \makeatletter%
  \providecommand\color[2][]{%
    \errmessage{(Inkscape) Color is used for the text in Inkscape, but the package 'color.sty' is not loaded}%
    \renewcommand\color[2][]{}%
  }%
  \providecommand\transparent[1]{%
    \errmessage{(Inkscape) Transparency is used (non-zero) for the text in Inkscape, but the package 'transparent.sty' is not loaded}%
    \renewcommand\transparent[1]{}%
  }%
  \providecommand\rotatebox[2]{#2}%
  \newcommand*\fsize{\dimexpr\f@size pt\relax}%
  \newcommand*\lineheight[1]{\fontsize{\fsize}{#1\fsize}\selectfont}%
  \ifx\svgwidth\undefined%
    \setlength{\unitlength}{421.454271bp}%
    \ifx\svgscale\undefined%
      \relax%
    \else%
      \setlength{\unitlength}{\unitlength * \real{\svgscale}}%
    \fi%
  \else%
    \setlength{\unitlength}{\svgwidth}%
  \fi%
  \global\let\svgwidth\undefined%
  \global\let\svgscale\undefined%
  \makeatother%
  \begin{picture}(1,0.67438697)%
    \lineheight{1}%
    \setlength\tabcolsep{0pt}%
    \put(0,0){\includegraphics[width=\unitlength,page=1]{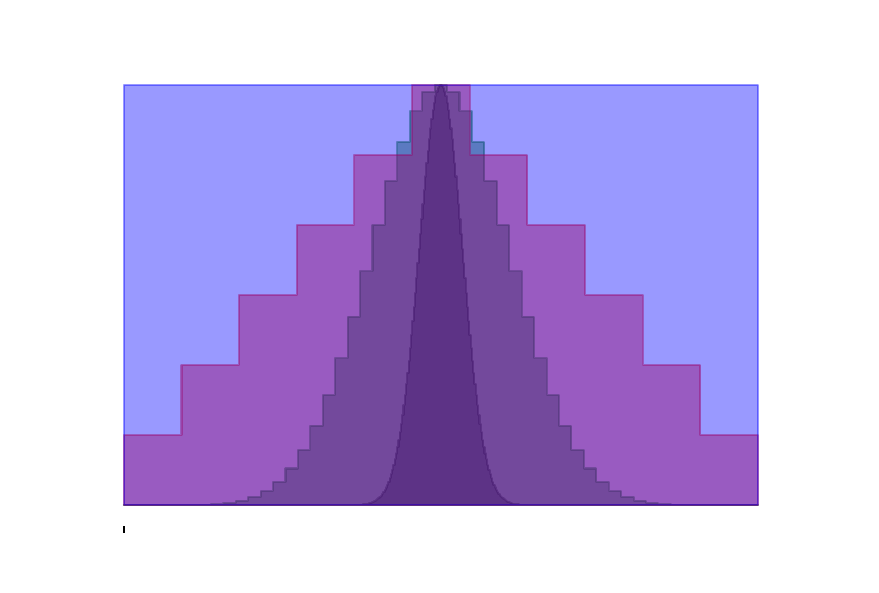}}%
    \put(0.14102353,0.04053008){\makebox(0,0)[t]{\lineheight{0}\smash{\begin{tabular}[t]{c}0\end{tabular}}}}%
    \put(0,0){\includegraphics[width=\unitlength,page=2]{dice_pdfs.pdf}}%
    \put(0.86319831,0.04053008){\makebox(0,0)[t]{\lineheight{0}\smash{\begin{tabular}[t]{c}1\end{tabular}}}}%
    \put(0.50211088,0.00807548){\makebox(0,0)[t]{\lineheight{0}\smash{\begin{tabular}[t]{c}x=(S-minroll)/(maxroll-minroll)\end{tabular}}}}%
    \put(0,0){\includegraphics[width=\unitlength,page=3]{dice_pdfs.pdf}}%
    \put(0.08830563,0.09007092){\makebox(0,0)[rt]{\lineheight{0}\smash{\begin{tabular}[t]{r}0.0\end{tabular}}}}%
    \put(0,0){\includegraphics[width=\unitlength,page=4]{dice_pdfs.pdf}}%
    \put(0.08830563,0.18573964){\makebox(0,0)[rt]{\lineheight{0}\smash{\begin{tabular}[t]{r}0.2\end{tabular}}}}%
    \put(0,0){\includegraphics[width=\unitlength,page=5]{dice_pdfs.pdf}}%
    \put(0.08830563,0.2814084){\makebox(0,0)[rt]{\lineheight{0}\smash{\begin{tabular}[t]{r}0.4\end{tabular}}}}%
    \put(0,0){\includegraphics[width=\unitlength,page=6]{dice_pdfs.pdf}}%
    \put(0.08830563,0.37707714){\makebox(0,0)[rt]{\lineheight{0}\smash{\begin{tabular}[t]{r}0.6\end{tabular}}}}%
    \put(0,0){\includegraphics[width=\unitlength,page=7]{dice_pdfs.pdf}}%
    \put(0.08830563,0.4727459){\makebox(0,0)[rt]{\lineheight{0}\smash{\begin{tabular}[t]{r}0.8\end{tabular}}}}%
    \put(0,0){\includegraphics[width=\unitlength,page=8]{dice_pdfs.pdf}}%
    \put(0.08830563,0.56841463){\makebox(0,0)[rt]{\lineheight{0}\smash{\begin{tabular}[t]{r}1.0\end{tabular}}}}%
    \put(0.03614621,0.3382573){\rotatebox{90}{\makebox(0,0)[t]{\lineheight{0}\smash{\begin{tabular}[t]{c}Normalised P(x)/max(P(x))\end{tabular}}}}}%
    \put(0,0){\includegraphics[width=\unitlength,page=9]{dice_pdfs.pdf}}%
    \put(0.88839127,0.63742455){\makebox(0,0)[lt]{\lineheight{0}\smash{\begin{tabular}[t]{l}100d6\end{tabular}}}}%
    \put(0,0){\includegraphics[width=\unitlength,page=10]{dice_pdfs.pdf}}%
    \put(0.88839127,0.59214904){\makebox(0,0)[lt]{\lineheight{0}\smash{\begin{tabular}[t]{l}10d6\end{tabular}}}}%
    \put(0,0){\includegraphics[width=\unitlength,page=11]{dice_pdfs.pdf}}%
    \put(0.88839127,0.54687351){\makebox(0,0)[lt]{\lineheight{0}\smash{\begin{tabular}[t]{l}2d6\end{tabular}}}}%
    \put(0,0){\includegraphics[width=\unitlength,page=12]{dice_pdfs.pdf}}%
    \put(0.88839127,0.501598){\makebox(0,0)[lt]{\lineheight{0}\smash{\begin{tabular}[t]{l}1d6\end{tabular}}}}%
  \end{picture}%
\endgroup%

%% file: images/nuclear_matrix2.pdf_tex
%% Creator: Inkscape 1.1.2 (0a00cf5339, 2022-02-04), www.inkscape.org
%% Accompanies image file 'nuclear_matrix2.pdf' (pdf, eps, ps)
\begingroup%
  \makeatletter%
  \providecommand\color[2][]{%
    \errmessage{(Inkscape) Color is used for the text in Inkscape, but the package 'color.sty' is not loaded}%
    \renewcommand\color[2][]{}%
  }%
  \providecommand\transparent[1]{%
    \errmessage{(Inkscape) Transparency is used (non-zero) for the text in Inkscape, but the package 'transparent.sty' is not loaded}%
    \renewcommand\transparent[1]{}%
  }%
  \providecommand\rotatebox[2]{#2}%
  \newcommand*\fsize{\dimexpr\f@size pt\relax}%
  \newcommand*\lineheight[1]{\fontsize{\fsize}{#1\fsize}\selectfont}%
  \ifx\svgwidth\undefined%
    \setlength{\unitlength}{447.53826925bp}%
    \ifx\svgscale\undefined%
      \relax%
    \else%
      \setlength{\unitlength}{\unitlength * \real{\svgscale}}%
    \fi%
  \else%
    \setlength{\unitlength}{\svgwidth}%
  \fi%
  \global\let\svgwidth\undefined%
  \global\let\svgscale\undefined%
  \makeatother%
  \begin{picture}(1,0.32672753)%
    \lineheight{1}%
    \setlength\tabcolsep{0pt}%
    \put(0,0){\includegraphics[width=\unitlength,page=1]{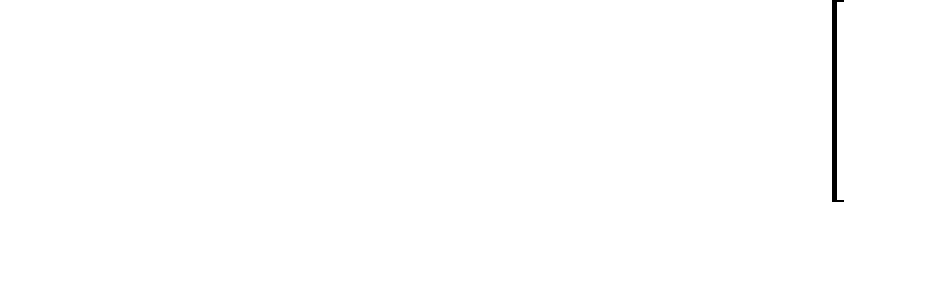}}%
    \put(0.94633734,0.25521073){\makebox(0,0)[t]{\lineheight{1.25}\smash{\begin{tabular}[t]{c}C\end{tabular}}}}%
    \put(0.94866779,0.13790376){\makebox(0,0)[t]{\lineheight{1.25}\smash{\begin{tabular}[t]{c}O\end{tabular}}}}%
    \put(0,0){\includegraphics[width=\unitlength,page=2]{nuclear_matrix2.pdf}}%
    \put(0.20960366,0.20905059){\makebox(0,0)[rt]{\lineheight{1.25}\smash{\begin{tabular}[t]{r}$a_{js}=$\end{tabular}}}}%
    \put(0,0){\includegraphics[width=\unitlength,page=3]{nuclear_matrix2.pdf}}%
    \put(0.32215633,0.25521073){\makebox(0,0)[t]{\lineheight{1.25}\smash{\begin{tabular}[t]{c}1\end{tabular}}}}%
    \put(0.44616819,0.25521073){\makebox(0,0)[t]{\lineheight{1.25}\smash{\begin{tabular}[t]{c}1\end{tabular}}}}%
    \put(0.56199675,0.25452385){\makebox(0,0)[t]{\lineheight{1.25}\smash{\begin{tabular}[t]{c}0\end{tabular}}}}%
    \put(0.32215633,0.13790376){\makebox(0,0)[t]{\lineheight{1.25}\smash{\begin{tabular}[t]{c}2\end{tabular}}}}%
    \put(0.44616819,0.13790376){\makebox(0,0)[t]{\lineheight{1.25}\smash{\begin{tabular}[t]{c}1\end{tabular}}}}%
    \put(0.56199675,0.13719543){\makebox(0,0)[t]{\lineheight{1.25}\smash{\begin{tabular}[t]{c}2\end{tabular}}}}%
    \put(0,0){\includegraphics[width=\unitlength,page=4]{nuclear_matrix2.pdf}}%
    \put(0.31399966,0.02730008){\makebox(0,0)[t]{\lineheight{1.25}\smash{\begin{tabular}[t]{c}CO$_2$\end{tabular}}}}%
    \put(0.45458663,0.02730008){\makebox(0,0)[t]{\lineheight{1.25}\smash{\begin{tabular}[t]{c}CO\end{tabular}}}}%
    \put(0.58352087,0.02730008){\makebox(0,0)[t]{\lineheight{1.25}\smash{\begin{tabular}[t]{c}O$_2$\end{tabular}}}}%
    \put(0,0){\includegraphics[width=\unitlength,page=5]{nuclear_matrix2.pdf}}%
    \put(0.22702085,0.02969499){\makebox(0,0)[rt]{\lineheight{1.25}\smash{\begin{tabular}[t]{r}species = \end{tabular}}}}%
    \put(0.88175593,0.21740283){\makebox(0,0)[rt]{\lineheight{1.25}\smash{\begin{tabular}[t]{r}elem=\end{tabular}}}}%
  \end{picture}%
\endgroup%

%% file: images/newton_2d.pdf_tex
%% Creator: Inkscape 1.1.2 (0a00cf5339, 2022-02-04), www.inkscape.org
%% Accompanies image file 'newton_2d.pdf' (pdf, eps, ps)
\begingroup%
  \makeatletter%
  \providecommand\color[2][]{%
    \errmessage{(Inkscape) Color is used for the text in Inkscape, but the package 'color.sty' is not loaded}%
    \renewcommand\color[2][]{}%
  }%
  \providecommand\transparent[1]{%
    \errmessage{(Inkscape) Transparency is used (non-zero) for the text in Inkscape, but the package 'transparent.sty' is not loaded}%
    \renewcommand\transparent[1]{}%
  }%
  \providecommand\rotatebox[2]{#2}%
  \newcommand*\fsize{\dimexpr\f@size pt\relax}%
  \newcommand*\lineheight[1]{\fontsize{\fsize}{#1\fsize}\selectfont}%
  \ifx\svgwidth\undefined%
    \setlength{\unitlength}{555.05228bp}%
    \ifx\svgscale\undefined%
      \relax%
    \else%
      \setlength{\unitlength}{\unitlength * \real{\svgscale}}%
    \fi%
  \else%
    \setlength{\unitlength}{\svgwidth}%
  \fi%
  \global\let\svgwidth\undefined%
  \global\let\svgscale\undefined%
  \makeatother%
  \begin{picture}(1,0.7689792)%
    \lineheight{1}%
    \setlength\tabcolsep{0pt}%
    \put(0,0){\includegraphics[width=\unitlength,page=1]{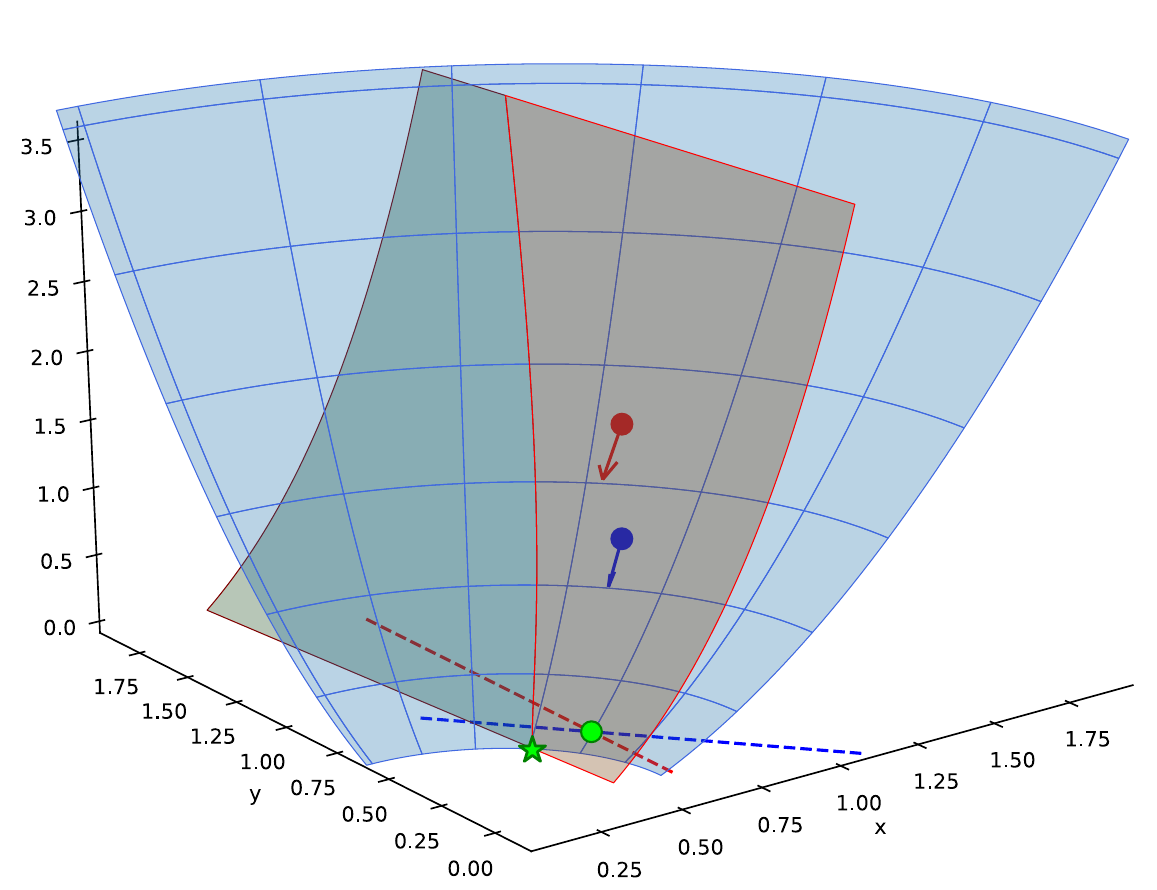}}%
  \end{picture}%
\endgroup%

%% file: images/co2.pdf_tex
%% Creator: Inkscape inkscape 0.92.5, www.inkscape.org
%% Accompanies image file 'co2.pdf' (pdf, eps, ps)
\begingroup%
  \makeatletter%
  \providecommand\color[2][]{%
    \errmessage{(Inkscape) Color is used for the text in Inkscape, but the package 'color.sty' is not loaded}%
    \renewcommand\color[2][]{}%
  }%
  \providecommand\transparent[1]{%
    \errmessage{(Inkscape) Transparency is used (non-zero) for the text in Inkscape, but the package 'transparent.sty' is not loaded}%
    \renewcommand\transparent[1]{}%
  }%
  \providecommand\rotatebox[2]{#2}%
  \newcommand*\fsize{\dimexpr\f@size pt\relax}%
  \newcommand*\lineheight[1]{\fontsize{\fsize}{#1\fsize}\selectfont}%
  \ifx\svgwidth\undefined%
    \setlength{\unitlength}{831.00249576bp}%
    \ifx\svgscale\undefined%
      \relax%
    \else%
      \setlength{\unitlength}{\unitlength * \real{\svgscale}}%
    \fi%
  \else%
    \setlength{\unitlength}{\svgwidth}%
  \fi%
  \global\let\svgwidth\undefined%
  \global\let\svgscale\undefined%
  \makeatother%
  \begin{picture}(1,0.31720177)%
    \lineheight{1}%
    \setlength\tabcolsep{0pt}%
    \put(0,0){\includegraphics[width=\unitlength,page=1]{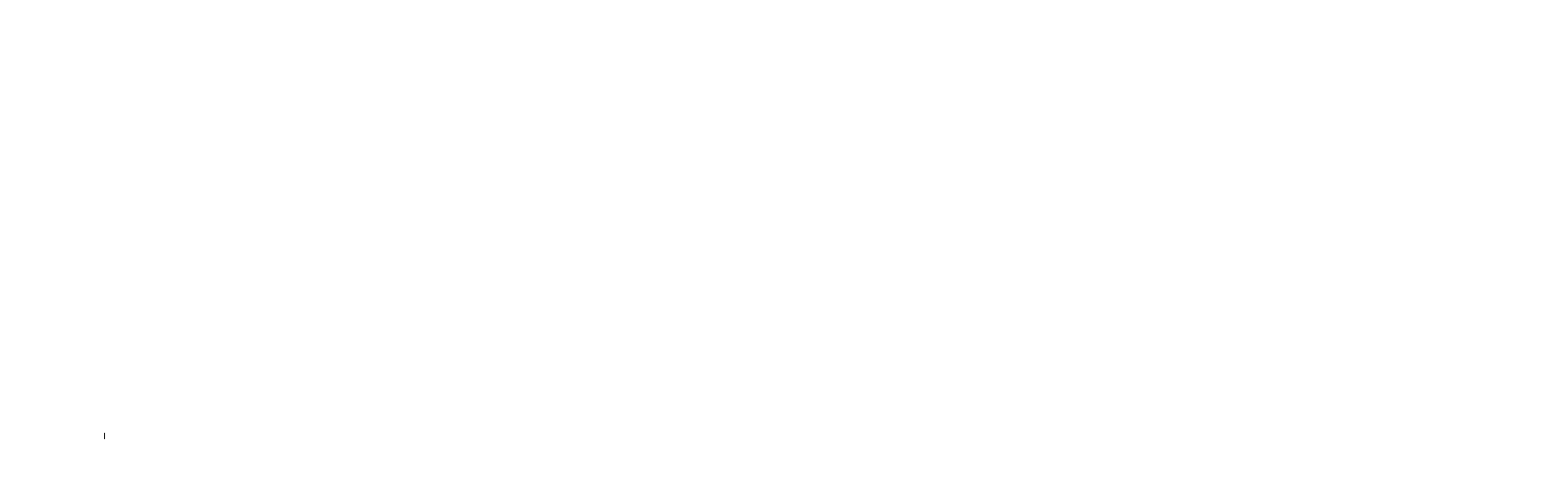}}%
    \put(0.06666318,0.02179275){\makebox(0,0)[t]{\lineheight{1.25}\smash{\begin{tabular}[t]{c}1500\end{tabular}}}}%
    \put(0,0){\includegraphics[width=\unitlength,page=2]{co2.pdf}}%
    \put(0.14828719,0.02179275){\makebox(0,0)[t]{\lineheight{1.25}\smash{\begin{tabular}[t]{c}2000\end{tabular}}}}%
    \put(0,0){\includegraphics[width=\unitlength,page=3]{co2.pdf}}%
    \put(0.2299112,0.02179275){\makebox(0,0)[t]{\lineheight{1.25}\smash{\begin{tabular}[t]{c}2500\end{tabular}}}}%
    \put(0,0){\includegraphics[width=\unitlength,page=4]{co2.pdf}}%
    \put(0.31153522,0.02179275){\makebox(0,0)[t]{\lineheight{1.25}\smash{\begin{tabular}[t]{c}3000\end{tabular}}}}%
    \put(0.1890992,0.00300371){\makebox(0,0)[t]{\lineheight{1.25}\smash{\begin{tabular}[t]{c}Temperature (K)\end{tabular}}}}%
    \put(0,0){\includegraphics[width=\unitlength,page=5]{co2.pdf}}%
    \put(0.04599602,0.05013204){\makebox(0,0)[rt]{\lineheight{1.25}\smash{\begin{tabular}[t]{r}$10^{-1}$\end{tabular}}}}%
    \put(0,0){\includegraphics[width=\unitlength,page=6]{co2.pdf}}%
    \put(0.04599602,0.28554739){\makebox(0,0)[rt]{\lineheight{1.25}\smash{\begin{tabular}[t]{r}$10^{0}$\end{tabular}}}}%
    \put(0,0){\includegraphics[width=\unitlength,page=7]{co2.pdf}}%
    \put(0.01097131,0.16915468){\rotatebox{90}{\makebox(0,0)[t]{\lineheight{1.25}\smash{\begin{tabular}[t]{c}Mole Fraction\end{tabular}}}}}%
    \put(0,0){\includegraphics[width=\unitlength,page=8]{co2.pdf}}%
    \put(0.1890992,0.30434079){\makebox(0,0)[t]{\lineheight{1.25}\smash{\begin{tabular}[t]{c}CO2\end{tabular}}}}%
    \put(0,0){\includegraphics[width=\unitlength,page=9]{co2.pdf}}%
    \put(0.06741593,0.1419714){\makebox(0,0)[lt]{\lineheight{1.25}\smash{\begin{tabular}[t]{l}Turns\end{tabular}}}}%
    \put(0,0){\includegraphics[width=\unitlength,page=10]{co2.pdf}}%
    \put(0.17571884,0.12077561){\makebox(0,0)[lt]{\lineheight{1.25}\smash{\begin{tabular}[t]{l}0.1 Atm\end{tabular}}}}%
    \put(0,0){\includegraphics[width=\unitlength,page=11]{co2.pdf}}%
    \put(0.17571884,0.09957984){\makebox(0,0)[lt]{\lineheight{1.25}\smash{\begin{tabular}[t]{l}1 Atm\end{tabular}}}}%
    \put(0,0){\includegraphics[width=\unitlength,page=12]{co2.pdf}}%
    \put(0.17571884,0.07838405){\makebox(0,0)[lt]{\lineheight{1.25}\smash{\begin{tabular}[t]{l}10 Atm\end{tabular}}}}%
    \put(0,0){\includegraphics[width=\unitlength,page=13]{co2.pdf}}%
    \put(0.17571884,0.05718826){\makebox(0,0)[lt]{\lineheight{1.25}\smash{\begin{tabular}[t]{l}100 Atm\end{tabular}}}}%
    \put(0,0){\includegraphics[width=\unitlength,page=14]{co2.pdf}}%
    \put(0.40218762,0.02179275){\makebox(0,0)[t]{\lineheight{1.25}\smash{\begin{tabular}[t]{c}1500\end{tabular}}}}%
    \put(0,0){\includegraphics[width=\unitlength,page=15]{co2.pdf}}%
    \put(0.48381164,0.02179275){\makebox(0,0)[t]{\lineheight{1.25}\smash{\begin{tabular}[t]{c}2000\end{tabular}}}}%
    \put(0,0){\includegraphics[width=\unitlength,page=16]{co2.pdf}}%
    \put(0.56543567,0.02179275){\makebox(0,0)[t]{\lineheight{1.25}\smash{\begin{tabular}[t]{c}2500\end{tabular}}}}%
    \put(0,0){\includegraphics[width=\unitlength,page=17]{co2.pdf}}%
    \put(0.64705966,0.02179275){\makebox(0,0)[t]{\lineheight{1.25}\smash{\begin{tabular}[t]{c}3000\end{tabular}}}}%
    \put(0.52462366,0.00300371){\makebox(0,0)[t]{\lineheight{1.25}\smash{\begin{tabular}[t]{c}Temperature (K)\end{tabular}}}}%
    \put(0,0){\includegraphics[width=\unitlength,page=18]{co2.pdf}}%
    \put(0.38152048,0.03936003){\makebox(0,0)[rt]{\lineheight{1.25}\smash{\begin{tabular}[t]{r}$10^{-5}$\end{tabular}}}}%
    \put(0,0){\includegraphics[width=\unitlength,page=19]{co2.pdf}}%
    \put(0.38152048,0.09054639){\makebox(0,0)[rt]{\lineheight{1.25}\smash{\begin{tabular}[t]{r}$10^{-4}$\end{tabular}}}}%
    \put(0,0){\includegraphics[width=\unitlength,page=20]{co2.pdf}}%
    \put(0.38152048,0.14173277){\makebox(0,0)[rt]{\lineheight{1.25}\smash{\begin{tabular}[t]{r}$10^{-3}$\end{tabular}}}}%
    \put(0,0){\includegraphics[width=\unitlength,page=21]{co2.pdf}}%
    \put(0.38152048,0.19291914){\makebox(0,0)[rt]{\lineheight{1.25}\smash{\begin{tabular}[t]{r}$10^{-2}$\end{tabular}}}}%
    \put(0,0){\includegraphics[width=\unitlength,page=22]{co2.pdf}}%
    \put(0.38152048,0.24410549){\makebox(0,0)[rt]{\lineheight{1.25}\smash{\begin{tabular}[t]{r}$10^{-1}$\end{tabular}}}}%
    \put(0,0){\includegraphics[width=\unitlength,page=23]{co2.pdf}}%
    \put(0.38152048,0.29529186){\makebox(0,0)[rt]{\lineheight{1.25}\smash{\begin{tabular}[t]{r}$10^{0}$\end{tabular}}}}%
    \put(0,0){\includegraphics[width=\unitlength,page=24]{co2.pdf}}%
    \put(0.52462366,0.30434079){\makebox(0,0)[t]{\lineheight{1.25}\smash{\begin{tabular}[t]{c}CO\end{tabular}}}}%
    \put(0,0){\includegraphics[width=\unitlength,page=25]{co2.pdf}}%
    \put(0.73771209,0.02179275){\makebox(0,0)[t]{\lineheight{1.25}\smash{\begin{tabular}[t]{c}1500\end{tabular}}}}%
    \put(0,0){\includegraphics[width=\unitlength,page=26]{co2.pdf}}%
    \put(0.81933612,0.02179275){\makebox(0,0)[t]{\lineheight{1.25}\smash{\begin{tabular}[t]{c}2000\end{tabular}}}}%
    \put(0,0){\includegraphics[width=\unitlength,page=27]{co2.pdf}}%
    \put(0.90096007,0.02179275){\makebox(0,0)[t]{\lineheight{1.25}\smash{\begin{tabular}[t]{c}2500\end{tabular}}}}%
    \put(0,0){\includegraphics[width=\unitlength,page=28]{co2.pdf}}%
    \put(0.9825841,0.02179275){\makebox(0,0)[t]{\lineheight{1.25}\smash{\begin{tabular}[t]{c}3000\end{tabular}}}}%
    \put(0.8601481,0.00300371){\makebox(0,0)[t]{\lineheight{1.25}\smash{\begin{tabular}[t]{c}Temperature (K)\end{tabular}}}}%
    \put(0,0){\includegraphics[width=\unitlength,page=29]{co2.pdf}}%
    \put(0.71704489,0.03936003){\makebox(0,0)[rt]{\lineheight{1.25}\smash{\begin{tabular}[t]{r}$10^{-5}$\end{tabular}}}}%
    \put(0,0){\includegraphics[width=\unitlength,page=30]{co2.pdf}}%
    \put(0.71704489,0.09054639){\makebox(0,0)[rt]{\lineheight{1.25}\smash{\begin{tabular}[t]{r}$10^{-4}$\end{tabular}}}}%
    \put(0,0){\includegraphics[width=\unitlength,page=31]{co2.pdf}}%
    \put(0.71704489,0.14173277){\makebox(0,0)[rt]{\lineheight{1.25}\smash{\begin{tabular}[t]{r}$10^{-3}$\end{tabular}}}}%
    \put(0,0){\includegraphics[width=\unitlength,page=32]{co2.pdf}}%
    \put(0.71704489,0.19291914){\makebox(0,0)[rt]{\lineheight{1.25}\smash{\begin{tabular}[t]{r}$10^{-2}$\end{tabular}}}}%
    \put(0,0){\includegraphics[width=\unitlength,page=33]{co2.pdf}}%
    \put(0.71704489,0.24410549){\makebox(0,0)[rt]{\lineheight{1.25}\smash{\begin{tabular}[t]{r}$10^{-1}$\end{tabular}}}}%
    \put(0,0){\includegraphics[width=\unitlength,page=34]{co2.pdf}}%
    \put(0.71704489,0.29529186){\makebox(0,0)[rt]{\lineheight{1.25}\smash{\begin{tabular}[t]{r}$10^{0}$\end{tabular}}}}%
    \put(0,0){\includegraphics[width=\unitlength,page=35]{co2.pdf}}%
    \put(0.8601481,0.30434079){\makebox(0,0)[t]{\lineheight{1.25}\smash{\begin{tabular}[t]{c}O2\end{tabular}}}}%
    \put(0.13239769,0.1419714){\makebox(0,0)[lt]{\lineheight{1.25}\smash{\begin{tabular}[t]{l}ceq\end{tabular}}}}%
  \end{picture}%
\endgroup%

%% file: images/end.pdf_tex
%% Creator: Inkscape inkscape 0.92.5, www.inkscape.org
%% Accompanies image file 'end.pdf' (pdf, eps, ps)
\begingroup%
  \makeatletter%
  \providecommand\color[2][]{%
    \errmessage{(Inkscape) Color is used for the text in Inkscape, but the package 'color.sty' is not loaded}%
    \renewcommand\color[2][]{}%
  }%
  \providecommand\transparent[1]{%
    \errmessage{(Inkscape) Transparency is used (non-zero) for the text in Inkscape, but the package 'transparent.sty' is not loaded}%
    \renewcommand\transparent[1]{}%
  }%
  \providecommand\rotatebox[2]{#2}%
  \newcommand*\fsize{\dimexpr\f@size pt\relax}%
  \newcommand*\lineheight[1]{\fontsize{\fsize}{#1\fsize}\selectfont}%
  \ifx\svgwidth\undefined%
    \setlength{\unitlength}{613.66126036bp}%
    \ifx\svgscale\undefined%
      \relax%
    \else%
      \setlength{\unitlength}{\unitlength * \real{\svgscale}}%
    \fi%
  \else%
    \setlength{\unitlength}{\svgwidth}%
  \fi%
  \global\let\svgwidth\undefined%
  \global\let\svgscale\undefined%
  \makeatother%
  \begin{picture}(1,0.52722731)%
    \lineheight{1}%
    \setlength\tabcolsep{0pt}%
    \put(0,0){\includegraphics[width=\unitlength,page=1]{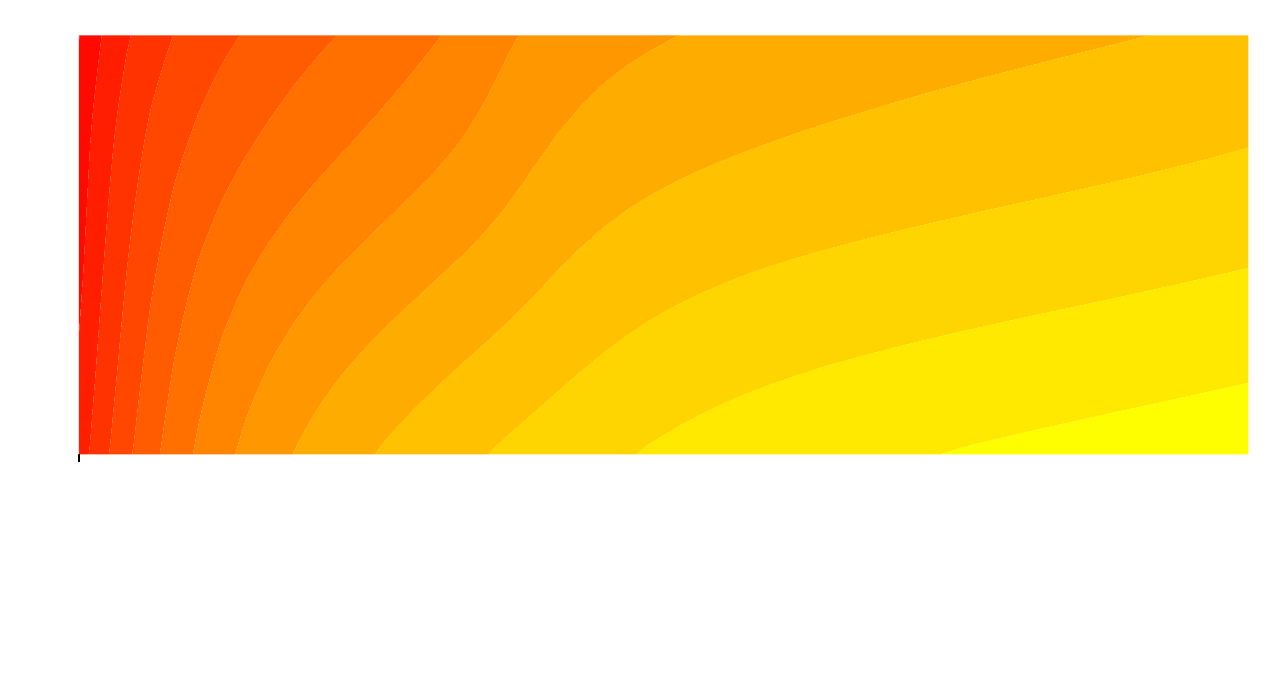}}%
    \put(0.06173245,0.14550014){\makebox(0,0)[t]{\lineheight{1.25}\smash{\begin{tabular}[t]{c}2000\end{tabular}}}}%
    \put(0,0){\includegraphics[width=\unitlength,page=2]{end.pdf}}%
    \put(0.21417969,0.14550014){\makebox(0,0)[t]{\lineheight{1.25}\smash{\begin{tabular}[t]{c}3000\end{tabular}}}}%
    \put(0,0){\includegraphics[width=\unitlength,page=3]{end.pdf}}%
    \put(0.36662693,0.14550014){\makebox(0,0)[t]{\lineheight{1.25}\smash{\begin{tabular}[t]{c}4000\end{tabular}}}}%
    \put(0,0){\includegraphics[width=\unitlength,page=4]{end.pdf}}%
    \put(0.51907417,0.14550014){\makebox(0,0)[t]{\lineheight{1.25}\smash{\begin{tabular}[t]{c}5000\end{tabular}}}}%
    \put(0,0){\includegraphics[width=\unitlength,page=5]{end.pdf}}%
    \put(0.67152141,0.14550014){\makebox(0,0)[t]{\lineheight{1.25}\smash{\begin{tabular}[t]{c}6000\end{tabular}}}}%
    \put(0,0){\includegraphics[width=\unitlength,page=6]{end.pdf}}%
    \put(0.8239687,0.14550014){\makebox(0,0)[t]{\lineheight{1.25}\smash{\begin{tabular}[t]{c}7000\end{tabular}}}}%
    \put(0,0){\includegraphics[width=\unitlength,page=7]{end.pdf}}%
    \put(0.97641589,0.14550014){\makebox(0,0)[t]{\lineheight{1.25}\smash{\begin{tabular}[t]{c}8000\end{tabular}}}}%
    \put(0.51907417,0.12005656){\makebox(0,0)[t]{\lineheight{1.25}\smash{\begin{tabular}[t]{c}Velocity (m/s)\end{tabular}}}}%
    \put(0,0){\includegraphics[width=\unitlength,page=8]{end.pdf}}%
    \put(0.0503255,0.16433637){\makebox(0,0)[rt]{\lineheight{1.25}\smash{\begin{tabular}[t]{r}10\end{tabular}}}}%
    \put(0,0){\includegraphics[width=\unitlength,page=9]{end.pdf}}%
    \put(0.0503255,0.22990753){\makebox(0,0)[rt]{\lineheight{1.25}\smash{\begin{tabular}[t]{r}20\end{tabular}}}}%
    \put(0,0){\includegraphics[width=\unitlength,page=10]{end.pdf}}%
    \put(0.0503255,0.29547869){\makebox(0,0)[rt]{\lineheight{1.25}\smash{\begin{tabular}[t]{r}30\end{tabular}}}}%
    \put(0,0){\includegraphics[width=\unitlength,page=11]{end.pdf}}%
    \put(0.0503255,0.36104986){\makebox(0,0)[rt]{\lineheight{1.25}\smash{\begin{tabular}[t]{r}40\end{tabular}}}}%
    \put(0,0){\includegraphics[width=\unitlength,page=12]{end.pdf}}%
    \put(0.0503255,0.42662103){\makebox(0,0)[rt]{\lineheight{1.25}\smash{\begin{tabular}[t]{r}50\end{tabular}}}}%
    \put(0,0){\includegraphics[width=\unitlength,page=13]{end.pdf}}%
    \put(0.0503255,0.4921922){\makebox(0,0)[rt]{\lineheight{1.25}\smash{\begin{tabular}[t]{r}60\end{tabular}}}}%
    \put(0.01485704,0.33569358){\rotatebox{90}{\makebox(0,0)[t]{\lineheight{1.25}\smash{\begin{tabular}[t]{c}Altitude (km)\end{tabular}}}}}%
    \put(0,0){\includegraphics[width=\unitlength,page=14]{end.pdf}}%
    \put(0.07470798,0.40359621){\rotatebox{87.414955}{\makebox(0,0)[lt]{\lineheight{1.25}\smash{\begin{tabular}[t]{l}12\end{tabular}}}}}%
    \put(0.07436127,0.16936966){\rotatebox{86.736012}{\makebox(0,0)[lt]{\lineheight{1.25}\smash{\begin{tabular}[t]{l}13\end{tabular}}}}}%
    \put(0.09021118,0.16934114){\rotatebox{86.274542}{\makebox(0,0)[lt]{\lineheight{1.25}\smash{\begin{tabular}[t]{l}14\end{tabular}}}}}%
    \put(0.14489683,0.38981472){\rotatebox{78.399839}{\makebox(0,0)[lt]{\lineheight{1.25}\smash{\begin{tabular}[t]{l}15\end{tabular}}}}}%
    \put(0.13488772,0.20270106){\rotatebox{84.601576}{\makebox(0,0)[lt]{\lineheight{1.25}\smash{\begin{tabular}[t]{l}16\end{tabular}}}}}%
    \put(0.27466896,0.40356173){\rotatebox{60.854394}{\makebox(0,0)[lt]{\lineheight{1.25}\smash{\begin{tabular}[t]{l}17\end{tabular}}}}}%
    \put(0.21064047,0.22703907){\rotatebox{72.353231}{\makebox(0,0)[lt]{\lineheight{1.25}\smash{\begin{tabular}[t]{l}18\end{tabular}}}}}%
    \put(0.44144776,0.41682382){\rotatebox{62.895038}{\makebox(0,0)[lt]{\lineheight{1.25}\smash{\begin{tabular}[t]{l}19\end{tabular}}}}}%
    \put(0.70723319,0.44213464){\rotatebox{25.021364}{\makebox(0,0)[lt]{\lineheight{1.25}\smash{\begin{tabular}[t]{l}20\end{tabular}}}}}%
    \put(0.77667213,0.35709821){\rotatebox{19.987115}{\makebox(0,0)[lt]{\lineheight{1.25}\smash{\begin{tabular}[t]{l}21\end{tabular}}}}}%
    \put(0.77661818,0.26686425){\rotatebox{19.662006}{\makebox(0,0)[lt]{\lineheight{1.25}\smash{\begin{tabular}[t]{l}22\end{tabular}}}}}%
    \put(0.83265371,0.18944624){\rotatebox{19.870107}{\makebox(0,0)[lt]{\lineheight{1.25}\smash{\begin{tabular}[t]{l}23\end{tabular}}}}}%
    \put(0.51907417,0.50939886){\makebox(0,0)[t]{\lineheight{1.25}\smash{\begin{tabular}[t]{c}Postshock Ionisation Map: Electron Number Density\end{tabular}}}}%
    \put(0,0){\includegraphics[width=\unitlength,page=15]{end.pdf}}%
    \put(0.06441962,0.02951112){\makebox(0,0)[t]{\lineheight{1.25}\smash{\begin{tabular}[t]{c}11\end{tabular}}}}%
    \put(0,0){\includegraphics[width=\unitlength,page=16]{end.pdf}}%
    \put(0.13680814,0.02951112){\makebox(0,0)[t]{\lineheight{1.25}\smash{\begin{tabular}[t]{c}12\end{tabular}}}}%
    \put(0,0){\includegraphics[width=\unitlength,page=17]{end.pdf}}%
    \put(0.20919666,0.02951112){\makebox(0,0)[t]{\lineheight{1.25}\smash{\begin{tabular}[t]{c}13\end{tabular}}}}%
    \put(0,0){\includegraphics[width=\unitlength,page=18]{end.pdf}}%
    \put(0.2815852,0.02951112){\makebox(0,0)[t]{\lineheight{1.25}\smash{\begin{tabular}[t]{c}14\end{tabular}}}}%
    \put(0,0){\includegraphics[width=\unitlength,page=19]{end.pdf}}%
    \put(0.35397372,0.02951112){\makebox(0,0)[t]{\lineheight{1.25}\smash{\begin{tabular}[t]{c}15\end{tabular}}}}%
    \put(0,0){\includegraphics[width=\unitlength,page=20]{end.pdf}}%
    \put(0.42636226,0.02951112){\makebox(0,0)[t]{\lineheight{1.25}\smash{\begin{tabular}[t]{c}16\end{tabular}}}}%
    \put(0,0){\includegraphics[width=\unitlength,page=21]{end.pdf}}%
    \put(0.49875077,0.02951112){\makebox(0,0)[t]{\lineheight{1.25}\smash{\begin{tabular}[t]{c}17\end{tabular}}}}%
    \put(0,0){\includegraphics[width=\unitlength,page=22]{end.pdf}}%
    \put(0.57113929,0.02951112){\makebox(0,0)[t]{\lineheight{1.25}\smash{\begin{tabular}[t]{c}18\end{tabular}}}}%
    \put(0,0){\includegraphics[width=\unitlength,page=23]{end.pdf}}%
    \put(0.64352781,0.02951112){\makebox(0,0)[t]{\lineheight{1.25}\smash{\begin{tabular}[t]{c}19\end{tabular}}}}%
    \put(0,0){\includegraphics[width=\unitlength,page=24]{end.pdf}}%
    \put(0.71591632,0.02951112){\makebox(0,0)[t]{\lineheight{1.25}\smash{\begin{tabular}[t]{c}20\end{tabular}}}}%
    \put(0,0){\includegraphics[width=\unitlength,page=25]{end.pdf}}%
    \put(0.78830484,0.02951112){\makebox(0,0)[t]{\lineheight{1.25}\smash{\begin{tabular}[t]{c}21\end{tabular}}}}%
    \put(0,0){\includegraphics[width=\unitlength,page=26]{end.pdf}}%
    \put(0.8606933,0.02951112){\makebox(0,0)[t]{\lineheight{1.25}\smash{\begin{tabular}[t]{c}22\end{tabular}}}}%
    \put(0,0){\includegraphics[width=\unitlength,page=27]{end.pdf}}%
    \put(0.93308187,0.02951112){\makebox(0,0)[t]{\lineheight{1.25}\smash{\begin{tabular}[t]{c}23\end{tabular}}}}%
    \put(0.51907417,0.00406754){\makebox(0,0)[t]{\lineheight{1.25}\smash{\begin{tabular}[t]{c}log10(Electron Number Density)\end{tabular}}}}%
    \put(0,0){\includegraphics[width=\unitlength,page=28]{end.pdf}}%
  \end{picture}%
\endgroup%

%% file: images/nozzle_exit_state.pdf_tex
%% Creator: Inkscape 1.1.2 (0a00cf5339, 2022-02-04), www.inkscape.org
%% Accompanies image file 'nozzle_exit_state.pdf' (pdf, eps, ps)
\begingroup%
  \makeatletter%
  \providecommand\color[2][]{%
    \errmessage{(Inkscape) Color is used for the text in Inkscape, but the package 'color.sty' is not loaded}%
    \renewcommand\color[2][]{}%
  }%
  \providecommand\transparent[1]{%
    \errmessage{(Inkscape) Transparency is used (non-zero) for the text in Inkscape, but the package 'transparent.sty' is not loaded}%
    \renewcommand\transparent[1]{}%
  }%
  \providecommand\rotatebox[2]{#2}%
  \newcommand*\fsize{\dimexpr\f@size pt\relax}%
  \newcommand*\lineheight[1]{\fontsize{\fsize}{#1\fsize}\selectfont}%
  \ifx\svgwidth\undefined%
    \setlength{\unitlength}{142.03498838bp}%
    \ifx\svgscale\undefined%
      \relax%
    \else%
      \setlength{\unitlength}{\unitlength * \real{\svgscale}}%
    \fi%
  \else%
    \setlength{\unitlength}{\svgwidth}%
  \fi%
  \global\let\svgwidth\undefined%
  \global\let\svgscale\undefined%
  \makeatother%
  \begin{picture}(1,0.64927124)%
    \lineheight{1}%
    \setlength\tabcolsep{0pt}%
    \put(0,0){\includegraphics[width=\unitlength,page=1]{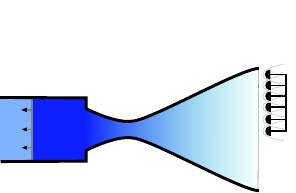}}%
    \put(0.01518855,0.19670854){\makebox(0,0)[lt]{\lineheight{1.25}\smash{\begin{tabular}[t]{l}2\end{tabular}}}}%
    \put(0.22646594,0.19670854){\makebox(0,0)[lt]{\lineheight{1.25}\smash{\begin{tabular}[t]{l}5\end{tabular}}}}%
    \put(0.41235622,0.19670854){\makebox(0,0)[lt]{\lineheight{1.25}\smash{\begin{tabular}[t]{l}6\end{tabular}}}}%
    \put(0.76786079,0.19522869){\makebox(0,0)[lt]{\lineheight{1.25}\smash{\begin{tabular}[t]{l}7\end{tabular}}}}%
    \put(0,0){\includegraphics[width=\unitlength,page=2]{nozzle_exit_state.pdf}}%
    \put(0.08639307,0.61615692){\makebox(0,0)[lt]{\lineheight{1.25}\smash{\begin{tabular}[t]{l}Shock\\Tube\end{tabular}}}}%
    \put(0,0){\includegraphics[width=\unitlength,page=3]{nozzle_exit_state.pdf}}%
    \put(0.50978324,0.61756687){\makebox(0,0)[lt]{\lineheight{1.25}\smash{\begin{tabular}[t]{l}Laval\\Nozzle\end{tabular}}}}%
    \put(0,0){\includegraphics[width=\unitlength,page=4]{nozzle_exit_state.pdf}}%
    \put(0.88988864,0.61428378){\makebox(0,0)[lt]{\lineheight{1.25}\smash{\begin{tabular}[t]{l}Pitot\\Rake\end{tabular}}}}%
  \end{picture}%
\endgroup%

%% file: images/nozzle_results.pdf_tex
%% Creator: Inkscape 1.1.2 (0a00cf5339, 2022-02-04), www.inkscape.org
%% Accompanies image file 'nozzle_results.pdf' (pdf, eps, ps)
\begingroup%
  \makeatletter%
  \providecommand\color[2][]{%
    \errmessage{(Inkscape) Color is used for the text in Inkscape, but the package 'color.sty' is not loaded}%
    \renewcommand\color[2][]{}%
  }%
  \providecommand\transparent[1]{%
    \errmessage{(Inkscape) Transparency is used (non-zero) for the text in Inkscape, but the package 'transparent.sty' is not loaded}%
    \renewcommand\transparent[1]{}%
  }%
  \providecommand\rotatebox[2]{#2}%
  \newcommand*\fsize{\dimexpr\f@size pt\relax}%
  \newcommand*\lineheight[1]{\fontsize{\fsize}{#1\fsize}\selectfont}%
  \ifx\svgwidth\undefined%
    \setlength{\unitlength}{131.09327193bp}%
    \ifx\svgscale\undefined%
      \relax%
    \else%
      \setlength{\unitlength}{\unitlength * \real{\svgscale}}%
    \fi%
  \else%
    \setlength{\unitlength}{\svgwidth}%
  \fi%
  \global\let\svgwidth\undefined%
  \global\let\svgscale\undefined%
  \makeatother%
  \begin{picture}(1,1.16132339)%
    \lineheight{1}%
    \setlength\tabcolsep{0pt}%
    \put(0,0){\includegraphics[width=\unitlength,page=1]{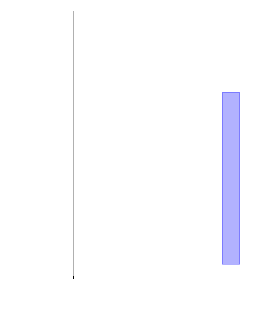}}%
    \put(0.26973421,0.0977709){\makebox(0,0)[t]{\lineheight{0}\smash{\begin{tabular}[t]{c}0.002\end{tabular}}}}%
    \put(0,0){\includegraphics[width=\unitlength,page=2]{nozzle_results.pdf}}%
    \put(0.40548032,0.0977709){\makebox(0,0)[t]{\lineheight{0}\smash{\begin{tabular}[t]{c}0.004\end{tabular}}}}%
    \put(0,0){\includegraphics[width=\unitlength,page=3]{nozzle_results.pdf}}%
    \put(0.54122627,0.0977709){\makebox(0,0)[t]{\lineheight{0}\smash{\begin{tabular}[t]{c}0.006\end{tabular}}}}%
    \put(0,0){\includegraphics[width=\unitlength,page=4]{nozzle_results.pdf}}%
    \put(0.67697223,0.0977709){\makebox(0,0)[t]{\lineheight{0}\smash{\begin{tabular}[t]{c}0.008\end{tabular}}}}%
    \put(0,0){\includegraphics[width=\unitlength,page=5]{nozzle_results.pdf}}%
    \put(0.81271834,0.0977709){\makebox(0,0)[t]{\lineheight{0}\smash{\begin{tabular}[t]{c}0.010\end{tabular}}}}%
    \put(0,0){\includegraphics[width=\unitlength,page=6]{nozzle_results.pdf}}%
    \put(0.94846429,0.0977709){\makebox(0,0)[t]{\lineheight{0}\smash{\begin{tabular}[t]{c}0.012\end{tabular}}}}%
    \put(0.58522681,0.04740239){\makebox(0,0)[t]{\lineheight{0}\smash{\begin{tabular}[t]{c}Pressure $p_7/p_5$\end{tabular}}}}%
    \put(0,0){\includegraphics[width=\unitlength,page=7]{nozzle_results.pdf}}%
    \put(0.16358531,0.1791388){\makebox(0,0)[rt]{\lineheight{0}\smash{\begin{tabular}[t]{r}0\end{tabular}}}}%
    \put(0,0){\includegraphics[width=\unitlength,page=8]{nozzle_results.pdf}}%
    \put(0.16358531,0.30508046){\makebox(0,0)[rt]{\lineheight{0}\smash{\begin{tabular}[t]{r}20\end{tabular}}}}%
    \put(0,0){\includegraphics[width=\unitlength,page=9]{nozzle_results.pdf}}%
    \put(0.16358531,0.4310218){\makebox(0,0)[rt]{\lineheight{0}\smash{\begin{tabular}[t]{r}40\end{tabular}}}}%
    \put(0,0){\includegraphics[width=\unitlength,page=10]{nozzle_results.pdf}}%
    \put(0.16358531,0.55696315){\makebox(0,0)[rt]{\lineheight{0}\smash{\begin{tabular}[t]{r}60\end{tabular}}}}%
    \put(0,0){\includegraphics[width=\unitlength,page=11]{nozzle_results.pdf}}%
    \put(0.16358531,0.68290481){\makebox(0,0)[rt]{\lineheight{0}\smash{\begin{tabular}[t]{r}80\end{tabular}}}}%
    \put(0,0){\includegraphics[width=\unitlength,page=12]{nozzle_results.pdf}}%
    \put(0.1635853,0.80884615){\makebox(0,0)[rt]{\lineheight{0}\smash{\begin{tabular}[t]{r}100\end{tabular}}}}%
    \put(0,0){\includegraphics[width=\unitlength,page=13]{nozzle_results.pdf}}%
    \put(0.1635853,0.93478781){\makebox(0,0)[rt]{\lineheight{0}\smash{\begin{tabular}[t]{r}120\end{tabular}}}}%
    \put(0,0){\includegraphics[width=\unitlength,page=14]{nozzle_results.pdf}}%
    \put(0.1635853,1.06072916){\makebox(0,0)[rt]{\lineheight{0}\smash{\begin{tabular}[t]{r}140\end{tabular}}}}%
    \put(0.06874191,0.63464123){\rotatebox{90}{\makebox(0,0)[t]{\lineheight{0}\smash{\begin{tabular}[t]{c}Radial Distance from Nozzle Axis (mm)\end{tabular}}}}}%
    \put(0,0){\includegraphics[width=\unitlength,page=15]{nozzle_results.pdf}}%
    \put(0.3293965,0.30629703){\makebox(0,0)[lt]{\lineheight{0}\smash{\begin{tabular}[t]{l}eqc (0mm)\end{tabular}}}}%
    \put(0,0){\includegraphics[width=\unitlength,page=16]{nozzle_results.pdf}}%
    \put(0.3293965,0.24947699){\makebox(0,0)[lt]{\lineheight{0}\smash{\begin{tabular}[t]{l}eqc (6mm)\end{tabular}}}}%
    \put(0,0){\includegraphics[width=\unitlength,page=17]{nozzle_results.pdf}}%
    \put(0.3293965,0.19265663){\makebox(0,0)[lt]{\lineheight{0}\smash{\begin{tabular}[t]{l}Shot 11310\end{tabular}}}}%
  \end{picture}%
\endgroup%

%% file: images/leonards_theorum.pdf_tex
%% Creator: Inkscape 1.1.2 (0a00cf5339, 2022-02-04), www.inkscape.org
%% Accompanies image file 'leonards_theorum.pdf' (pdf, eps, ps)
\begingroup%
  \makeatletter%
  \providecommand\color[2][]{%
    \errmessage{(Inkscape) Color is used for the text in Inkscape, but the package 'color.sty' is not loaded}%
    \renewcommand\color[2][]{}%
  }%
  \providecommand\transparent[1]{%
    \errmessage{(Inkscape) Transparency is used (non-zero) for the text in Inkscape, but the package 'transparent.sty' is not loaded}%
    \renewcommand\transparent[1]{}%
  }%
  \providecommand\rotatebox[2]{#2}%
  \newcommand*\fsize{\dimexpr\f@size pt\relax}%
  \newcommand*\lineheight[1]{\fontsize{\fsize}{#1\fsize}\selectfont}%
  \ifx\svgwidth\undefined%
    \setlength{\unitlength}{198.33454245bp}%
    \ifx\svgscale\undefined%
      \relax%
    \else%
      \setlength{\unitlength}{\unitlength * \real{\svgscale}}%
    \fi%
  \else%
    \setlength{\unitlength}{\svgwidth}%
  \fi%
  \global\let\svgwidth\undefined%
  \global\let\svgscale\undefined%
  \makeatother%
  \begin{picture}(1,0.6424215)%
    \lineheight{1}%
    \setlength\tabcolsep{0pt}%
    \put(0,0){\includegraphics[width=\unitlength,page=1]{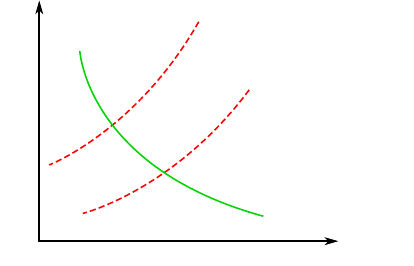}}%
    \put(0.00018478,0.3487866){\makebox(0,0)[lt]{\lineheight{1.25}\smash{\begin{tabular}[t]{l}T\end{tabular}}}}%
    \put(0.4381516,-0){\makebox(0,0)[lt]{\lineheight{1.25}\smash{\begin{tabular}[t]{l}V\end{tabular}}}}%
    \put(0,0){\includegraphics[width=\unitlength,page=2]{leonards_theorum.pdf}}%
    \put(0.33637492,0.36135403){\makebox(0,0)[lt]{\lineheight{1.25}\smash{\begin{tabular}[t]{l}$\Delta V$\end{tabular}}}}%
    \put(0.40993567,0.29018308){\makebox(0,0)[lt]{\lineheight{1.25}\smash{\begin{tabular}[t]{l}$\Delta T$\end{tabular}}}}%
    \put(0.49622158,0.60524036){\color[rgb]{1,0,0}\makebox(0,0)[lt]{\lineheight{1.25}\smash{\begin{tabular}[t]{l}$E_1$\end{tabular}}}}%
    \put(0.61753435,0.43977568){\color[rgb]{1,0,0}\makebox(0,0)[lt]{\lineheight{1.25}\smash{\begin{tabular}[t]{l}$E_2$\end{tabular}}}}%
    \put(0.6569825,0.11749938){\color[rgb]{0,0.83137255,0}\makebox(0,0)[lt]{\lineheight{1.25}\smash{\begin{tabular}[t]{l}$S=constant$\end{tabular}}}}%
  \end{picture}%
\endgroup%

%% file: article.bbl
\begin{thebibliography}{10}
\expandafter\ifx\csname natexlab\endcsname\relax\def\natexlab#1{#1}\fi
\providecommand{\url}[1]{\texttt{#1}}
\providecommand{\href}[2]{#2}
\providecommand{\path}[1]{#1}
\providecommand{\DOIprefix}{doi:}
\providecommand{\ArXivprefix}{arXiv:}
\providecommand{\URLprefix}{URL: }
\providecommand{\Pubmedprefix}{pmid:}
\providecommand{\doi}[1]{\href{http://dx.doi.org/#1}{\path{#1}}}
\providecommand{\Pubmed}[1]{\href{pmid:#1}{\path{#1}}}
\providecommand{\bibinfo}[2]{#2}
\ifx\xfnm\relax \def\xfnm[#1]{\unskip,\space#1}\fi
%Type = Techreport
\bibitem[{Reynolds(1986)}]{reynolds_stanjan86}
\bibinfo{author}{W.~C. Reynolds}, \bibinfo{title}{The Element Potential Method
  for Chemical Equilibrium Analysis: Implementation in the Interactive Program
  STANJAN}, \bibinfo{type}{Technical Report}, Department of Mechanical
  Engineering, Stanford University, \bibinfo{year}{1986}.
%Type = Article
\bibitem[{Pope(2004)}]{pope_ceq04}
\bibinfo{author}{S.~B. Pope},
\newblock \bibinfo{title}{Gibbs function continuation for the stable
  computation of chemical equilibrium},
\newblock \bibinfo{journal}{Combustion and flame} \bibinfo{volume}{139}
  (\bibinfo{year}{2004}) \bibinfo{pages}{222--226}.
%Type = Techreport
\bibitem[{Gordon and McBride(1994)}]{nasacea_I}
\bibinfo{author}{S.~Gordon}, \bibinfo{author}{B.~J. McBride},
  \bibinfo{title}{Computer Program for Calculation of Complex Chemical
  Equilibrium Compositions and Applications}, \bibinfo{type}{Technical Report}
  \bibinfo{number}{Reference Publication 1311}, National Aeronautics and Space
  Administration, \bibinfo{year}{1994}.
%Type = Misc
\bibitem[{Goodwin et~al.(2023)Goodwin, Moffat, Schoegl, Speth, and
  Weber}]{cantera}
\bibinfo{author}{D.~G. Goodwin}, \bibinfo{author}{H.~K. Moffat},
  \bibinfo{author}{I.~Schoegl}, \bibinfo{author}{R.~L. Speth},
  \bibinfo{author}{B.~W. Weber}, \bibinfo{title}{Cantera: An object-oriented
  software toolkit for chemical kinetics, thermodynamics, and transport
  processes}, \bibinfo{howpublished}{\url{https://www.cantera.org}},
  \bibinfo{year}{2023}. \DOIprefix\doi{10.5281/zenodo.8137090},
  \bibinfo{note}{version 3.0.0}.
%Type = Article
\bibitem[{Dirks et~al.(2007)Dirks, Bois, Schaeffer, Winfree, and
  Pierce}]{dirks2007}
\bibinfo{author}{R.~M. Dirks}, \bibinfo{author}{J.~S. Bois},
  \bibinfo{author}{J.~M. Schaeffer}, \bibinfo{author}{E.~Winfree},
  \bibinfo{author}{N.~A. Pierce},
\newblock \bibinfo{title}{Thermodynamic analysis of interacting nucleic acid
  strands},
\newblock \bibinfo{journal}{SIAM Review} \bibinfo{volume}{49}
  (\bibinfo{year}{2007}) \bibinfo{pages}{64--88}. \URLprefix
  \url{https://doi.org/10.1137/060651100}. \DOIprefix\doi{10.1137/060651100}.
%Type = Book
\bibitem[{Turns and Haworth(2020)}]{turns}
\bibinfo{author}{S.~R. Turns}, \bibinfo{author}{D.~C. Haworth},
  \bibinfo{title}{An Introduction to Combustion: Concepts and Applications},
  \bibinfo{edition}{4th} ed., \bibinfo{publisher}{McGraw-Hill Education},
  \bibinfo{year}{2020}.
%Type = Book
\bibitem[{Anderson(1989)}]{andersonhypersonics}
\bibinfo{author}{J.~Anderson}, \bibinfo{title}{Hypersonic and High Temperature
  Gas Dynamics}, \bibinfo{publisher}{McGraw Hill}, \bibinfo{year}{1989}.
%Type = Article
\bibitem[{Chan et~al.(2018)Chan, Jacobs, Smart, Grieve, Craddock, and
  Doherty}]{chan_nozzles18}
\bibinfo{author}{W.~Y.~K. Chan}, \bibinfo{author}{P.~A. Jacobs},
  \bibinfo{author}{M.~K. Smart}, \bibinfo{author}{S.~Grieve},
  \bibinfo{author}{C.~S. Craddock}, \bibinfo{author}{L.~J. Doherty},
\newblock \bibinfo{title}{Aerodynamic design of nozzles with uniform outflow
  for hypervelocity ground-test facilities},
\newblock \bibinfo{journal}{Journal of Propulsion and Power}
  \bibinfo{volume}{34} (\bibinfo{year}{2018}) \bibinfo{pages}{1467--1478}.
  \DOIprefix\doi{10.2514/1.B36938}.
%Type = Misc
\bibitem[{Anonymous(2024)}]{stack_exchange_dicerolls}
\bibinfo{author}{Anonymous}, \bibinfo{title}{Probability that the sum of k dice
  is n},
  \bibinfo{howpublished}{\url{https://math.stackexchange.com/questions/2290090/probability-that-the-sum-of-k-dice-is-n}},
  \bibinfo{year}{2024}. \bibinfo{note}{Accessed 22/08/24}.
%Type = Misc
\bibitem[{Susskind(2013)}]{susskind_ttm}
\bibinfo{author}{L.~Susskind}, \bibinfo{title}{The theoretical minimum:
  Statistical mechanics},
  \bibinfo{howpublished}{\url{https://theoreticalminimum.com/courses/statistical-mechanics/2013/spring}},
  \bibinfo{year}{2013}. \bibinfo{note}{Accessed 27/08/24}.

\end{thebibliography}
